\documentclass[superscriptaddress,prd,aps,amssymb,nofootinbib]{revtex4}
\usepackage{graphicx, wrapfig}

\newcommand{\beq}{\begin{equation}} 
\newcommand{\eeq}{\end{equation}} 
\newcommand{\beqn}{\begin{eqnarray}} 
\newcommand{\eeqn}{\end{eqnarray}} 
\newcommand{\pa}{\partial}
\newcommand{\na}{\nabla}
\newcommand{\gab}{g_{\alpha\beta}}
\newcommand{\gabu}{g^{\alpha\beta}}
\newcommand{\gabd}{g_{\alpha\beta}}

\newcommand{\Tabu}{T^{\alpha\beta}}

\newcommand{\zD}{{\raise1.0ex\hbox{${}^{\ \circ}$}}\!\!\!\!\!D}
\newcommand{\alone}{{\raise0.5ex\hbox{${}^{\ 1}$}}\!\!\!\!\alpha}

\newcommand{\dl}{\delta}

\newcommand{\nalam}{\mathrel{\raise0.9ex\hbox{$^\lambda$}\mkern-14mu
\lower0.0ex\hbox{$\nabla$}}}

\newcommand{\be}{\begin{equation}}
\newcommand{\ee}{\end{equation}}
\newcommand{\bea}{\begin{eqnarray}}
\newcommand{\eea}{\end{eqnarray}}
\newcommand{\beaa}{\begin{eqnarray*}}
\newcommand{\eeaa}{\end{eqnarray*}}
\newcommand{\ben}{\begin{enumerate}}
\newcommand{\een}{\end{enumerate}}
\newcommand{\dis}{$\displaystyle}
\newcommand{\ep}{\epsilon}
\newcommand{\btau}{\bar{\tau}}
\newcommand{\bx}{\bar{x}}
\newcommand{\bmm}{\bar{m}}
\newcommand{\bR}{\bar{R}}
\newcommand{\dtx}{{\dot{x}}}
\newcommand{\dtbx}{{\dot{\bar{x}}}{}}
\newcommand{\ddtx}{\ddot{x}}
\newcommand{\ddtbx}{\ddot{\bar{x}}{}}
\newcommand{\bt}{\bar{t}}
\newcommand{\bphi}{\bar{\phi}}
\newcommand{\bz}{\bar{z}}
\newcommand{\bpi}{\bar{\varpi}}
\newcommand{\ba}{\bar{a}}
\newcommand{\bv}{\bar{v}}
\newcommand{\bk}{\bar{k}}
\newcommand{\bhh}{{\bar{h}}}
\newcommand{\bT}{{\bar{T}}}

\newcommand{\bgamma}{{\bar{\gamma}}}
\newcommand{\bareta}{\bar{\eta}}

\newcommand{\bX}{\bar{X}}
\newcommand{\dtX}{\dot{X}}
\newcommand{\dtbX}{\dot{\bar{X}}{}}
\newcommand{\tPhi}{\tilde{\Phi}}
\newcommand{\albe}{\alpha\beta}
\newcommand{\bega}{\beta\gamma}
\newcommand{\gade}{{\gamma\delta}}

\newcommand{\hphi}{{\widehat\phi}}
\newcommand{\hbphi}{\widehat{\bar{\phi}}{}}
\newcommand{\intinf}{\int_{-\infty}^{\infty}}
\newcommand{\cUm}{{\cal U}_m}
\newcommand{\bcUm}{{\cal U}_{\bmm}}
\newcommand{\bA}{\bar{A}}
\newcommand{\dtphi}{{\dot{\phi}}}
\newcommand{\dtbphi}{{\dot{\bphi}}{}}

\begin{document} 
\preprint{}

\title{Post-Minkowski action for point particles and a helically 
symmetric binary solution} 
 
\author{John L. Friedman} 
%\email{}
\affiliation{
Department of Physics, University of Wisconsin-Milwaukee, P.O. Box 413,  
Milwaukee, WI 53201}
\author{K\=oji Ury\=u}
%\email{}
\affiliation{
Department of Physics, University of Wisconsin-Milwaukee, P.O. Box 413,  
Milwaukee, WI 53201}
\affiliation{
SISSA, via Beirut 4, 34014 Trieste, Italy}

\begin{abstract} 

Two Fokker actions and corresponding equations of motion are
obtained for two point particles in a post-Minkowski framework, in
which the field of each particle is given by the half-retarded +
half-advanced solution to the linearized Einstein equations. 
The first action is parametrization invariant, the second a
generalization of the affinely parametrized quadratic action for 
a relativistic particle. Expressions for a conserved 4-momentum and 
angular momentum tensor are obtained in terms of the particles' 
trajectories in this post-Minkowski approximation.  A formal solution 
to the equations of motion is found for a binary system with circular 
orbits.  For a bound system of this 
kind, the post-Minkowski solution is a toy model that omits nonlinear
terms of relevant post-Newtonian order; and we also obtain a Fokker
action that is accurate to first post-Newtonian order, by adding to the
post-Minkowski action a term cubic in the particle masses.  Curiously, 
the conserved energy and angular momentum associated with the 
Fokker action are each finite and well-defined for this bound 2-particle 
system despite the fact that the total energy and angular momentum 
of the radiation field diverge.  Corresponding solutions and  
conserved quantities are found for two scalar charges (for electromagnetic 
charges we exhibit the solution found by Schild). 
For a broad class of parametrization-invariant Fokker actions 
and for the affinely parametrized action, 
binary systems with circular orbits satisfy the relation $dE = \Omega dL$ 
(a form of the first law of thermodynamics), relating the energy, angular 
velocity and angular momentum of nearby equilibrium configurations.  

\end{abstract} 

%\maketitle 
%%\pacs{PACS numbers: 04.20.Fy,04.30.Db,04.40.Dg,04.70.Dy}
\pacs{PACS numbers: }
\maketitle  
 
\section{Introduction} 
\label{intro} 

In constructing numerical initial data for compact binary systems, it
may be useful to find exact solutions to the Einstein equations with
helical symmetry, solutions that are stationary in a corotating
frame \cite{bd92,bct98,whelan00,whelan02,fus02,price04,andrade04,
torre03,bromley05,klein05}.   In the case of two charged particles, an exact
solution of this kind was found by Schild\cite{sc63}: The force on each
particle is associated with the half-retarded plus half-advanced field
of the other. (An earlier but less explicit version is given by
Sch\"{o}nberg\cite{schonberg46}.)  We obtain here an analogous solution for
point masses in a post-Minkowski framework. 

A time-symmetric interaction allows one to derive the equations of
motion for point-particles from a single action integral that is
written solely in terms of the dynamical variables of each particle
(without mediating field variables).  This kind of {\sl action at a
distance theory} has been formulated by Fokker and by Wheeler and
Feynman; for their treatments of the electromagnetic field, see
\cite{fo29,wf4549}.  We exhibit an analogous Fokker action for linearized 
gravity.  The equations of motion are those of a post-Minkowski
approximation, in which, as in the electromagnetic case, each 
particle moves in the half-advanced plus half-retarded field of the
other.  

A Fokker action is not a true action: Its variation leads to equations in 
which the endpoints of the action integral explicitly appear in integrals 
that should yield the field of each particle.  Only by taking a limit 
of the varied action as the endpoints go to infinity does one recover 
the correct equations of motion. Invariance of the Fokker action under 
Poincar\'e transformations nevertheless leads to expressions for a conserved 
energy and angular momentum of an n-particle system.  For helically symmetric
binary systems
(in, e.g., models with 
scalar, electromagnetic, or gravitational interactions), the energy and 
angular momentum of the field is infinite, because the particles have radiated
for infinite time in past and future.  The conserved energy-momentum and angular 
momentum associated with the Fokker action, however, are finite. For circular 
orbits of particles of mass $m$ and $\bar m$, we find the surprising result 
that the form of this energy, 
\be
 E = \frac m\gamma + \frac{\bar m}{\bar\gamma},
\ee
is identical for scalar, electromagnetic, and linearized gravitational 
interactions, described by a parametrization invariant Fokker action. 
The angular momentum is in each case proportional to the value of 
the interaction field dotted on each free index into the helical 
Killing vector.  We show that nearby circular orbits satisfy the familiar 
first law, $dE=\Omega dL$. 

Although one can construct formal solutions to the post-Minkowski equations 
in which the source is gravitationally bound, care is needed in their 
interpretation.  A conceptual problem is related to the fact that, 
at zeroth order in the perturbed metric, particles move on geodesics of 
flat space, not in bound orbits. Because the matter density vanishes 
at zeroth order, the stress-energy tensor $\Tabu = \rho u^\alpha u^\beta$
has at first order the form $\delta\Tabu = (\delta\rho) u^\alpha u^\beta$,
where $u^\alpha$ is the zeroth-order velocity field describing straight-line 
motion.  If one then finds the metric from the perturbed equation 
\be 
0=-2\delta\left[G^{\alpha\beta} - 8\pi T^{\alpha\beta}
\right] 
= \Box(h^{\alpha\beta}-\frac12\eta^{\alpha\beta}h) +16\pi 
(\delta\rho) u^\alpha u^\beta,
\ee
the perturbed metric will not yield perturbed bound trajectories.

A resolution to this problem is to look at families of solutions, 
$\gabd(s),T^{\alpha\beta}(s) $ that 
describe finite fluid masses in bound orbits for all nonzero values of $s$, despite the fact that 
$s=0$ is flat space.  An example can be easily understood by first 
looking at Newtonian gravity, with a family of solutions corresponding
to the scaling $v\sim \epsilon$, $m/r\sim \epsilon^2$ of the post-Newtonian
approximation: That is, consider a family of solutions for 
two point masses $m=m^{(1)}s$ and $\bar m=\bar m^{(1)} s$ 
in circular orbits of radii $a$ and $ \bar a$ (independent of $s$), 
with speeds given approximately by $v^{(1)}\sqrt s$ and $\bar v^{(1)} \sqrt s$. 
At $s=0$, the solution is 
flat space with no matter.  For each nonzero $s$ the solution describes 
masses in circular orbit, and as $s\rightarrow 0$, the period of the orbit 
increases without bound. An exact solution is given 
for each mass by equations in which the perturbed particle trajectory is 
found {\em self-consistently}.  That is, with 
$\Phi = s\Phi^{(1)} $, the center-of-mass 
trajectories ${\bf r}$ and $\bar{\bf r}$ satisfy 
\beqn
\na^2\Phi^{(1)}({\bf r}) &=& 4\pi \bar m^{(1)}\delta({\bf r}-\bar{\bf r}(t)), \qquad 
\ddot{\bf r} = -\nabla\Phi,\nonumber\\
\bar\na^2\bar\Phi^{(1)}(\bar {\bf r}) &=& 4\pi m^{(1)}\delta(\bar{\bf r}-{\bf r}(t)), \qquad 
\ddot{\bf \bar r} = -\bar\nabla\bar\Phi.
\eeqn
Here $\Phi({\bf r})$ is the potential at ${\bf r}$ due to the mass 
$\bar m$.  Because $v\rightarrow 0$ as $s\rightarrow 0$, the orbit remains 
close to a straight line for increasingly long times, but for each finite $s$
the orbit is circular .

In general relativity, one can presumably construct 
a similar family of solutions with bound orbits whose energy-momentum tensor, 
$T^{\alpha\beta}(s) 
= \epsilon(s) u^\alpha(s) u^\beta(s)+p(s)[\gabu(s)+u^\alpha(s) u^\beta(s)]$, 
and metric $\gabd(s)$ are pointwise continuous in the parameter $s$ and for which one 
has flat, empty space at $s=0$, 
\[ 
   \gabd(0)=\eta_{\alpha\beta}, \qquad \epsilon(0)=p(0)=0.
\]
The equations for the first-order metric in a radiation gauge, 
\be
-2 G^{(1)}_{\alpha\beta}
\equiv \Box(h_{\alpha\beta}-\frac12\eta_{\alpha\beta}) h 
= - 16\pi T^{(1)}_{\alpha\beta},
\ee
have as their source $T^{(1)}$ a stress-energy tensor constructed from 
the {\em perturbed} velocity field and density and from the unperturbed metric. With the 
masses shrinking to zero as $s\rightarrow 0$,
the motion of each mass is given to linear order by the linear 
field of the other: The self-force at linear order serves only to 
renormalize the mass. 

The resulting ``post-Minkowski'' solution in which 
two masses move in bound orbit, each responding to the linear field of 
the other has the following features:\\
(i) It is correct to Newtonian order.\\ 
(ii) The radiation field of the linearized metric is correct 
to lowest nonvanishing post-Newtonian order (2 1/2 post-Newtonian order).\\
(iii) The orbit is not correct to first post-Newtonian order.\\
Using the linearized metric leads to equations of motion that are missing 
a term quadratic in the particle masses that enters
the first post-Newtonian equations. In this sense, the first post-Minkowski 
approximation for bound orbits is a toy model, 
keeping terms of all orders in $v$ but discarding nonlinear terms 
in $h_{\alpha\beta}$ that, for bound states, give corrections of order $v^2$
to the orbit.  

Following recent codes that obtain helically symmetric solutions to nonlinear wave 
equations \cite{yo06,bbp06}, Uryu \cite{yo06,uryu06} has obtained a neutron star 
code that solves the full Einstein-perfect fluid equations on an asymptotically flat 
initial hypersurface S, obtaining a solution with exact helical symmetry in the near zone. 
By solving the full Einstein equation, one expects more accurately to enforce the
balance of gravity and centripetal acceleration in circular motion.  The error 
associated with ignoring the radial motion associated with radiation reaction, however, 
remains. In future work, we anticipate using the point-particle model developed here 
to estimate the accuracy of such helically symmetric initial data sets and 
corresponding quasiequilibrium sequences, by comparing outgoing point-particle 
solution (in a post-Minkowski framework) to helically symmetric models and sequences.
A further problem is related to the fact that the codes mentioned above typically 
diverge when helical symmetry is enforced in a region larger than the near 
zone (beyond one wavelength). Extending the present work to second order in the 
post-Minkowski approximation may be useful in developing codes with larger 
domains of convergence. Finally, existence of helically 
symmetric binaries is unproved and the second-order (nonlinear) extension 
may also help in understanding existence and asymptotic behavior of models of helically 
symmetric binaries.

In Sect.~\ref{sec:II}, we review Fokker actions, obtaining the 
equations of motion and the form of the conserved 4-momentum and 
corresponding angular momentum.  In Sect.~\ref{sec:PMinv}, we introduce a 
parametrization-invariant Fokker action that describes point particles 
in the first post-Minkowski approximation.  The formalism of the previous 
section is used to obtain equations of motion and conserved quantities;
derivations of the explicit forms of the conserved quantities are relegated 
to Appendix~\ref{sec:eqshij}.  
Sect.~\ref{sec:affine} then introduces the Fokker analog of the quadratic 
action, \dis\int d\tau \frac m2 \dot x^\alpha \dot x_\alpha$, for affinely 
parametrized particles, again obtaining equations of motion, momentum,
and angular momentum.
In Sect.~\ref{sec:orbit} and \ref{sec:cirint} we consider two particles 
of masses $m$ and $\bar m$ in circular orbit, finding the equations 
governing the orbit, and computing the system's energy and angular 
momentum.  The corresponding conserved quantities 
for scalar and electromagnetic interactions are summarized.
Sect.~\ref{sec:1stlaw} is devoted to proving the relation 
$dE=\Omega dL$ for a class of parametrization-invariant Fokker actions
and the affinely parametrized action.
In Sect.~\ref{sec:pn} we introduce two forms of a post-Newtonian correction
term to make the post-Minkowski action accurate to first post-Newtonian
order. We obtain the corresponding corrections to the conserved 
momentum and angular momentum.  Sect.~\ref{sec:numer} displays the 
results of a numerical solution to the post-Minkowski equations of motion,
with and without the added post-Newtonian term. Sect.~\ref{sec:discussion}
discusses features of conserved quantities associated with Fokker actions, 
proposing, in particular, an explanation for their finite behavior when the 
field energy is infinite.  A description of interacting scalar 
charges and many of the details of our calculations are presented in 
appendices.

\section{Action at a distance theory}
\label{sec:II}

An action-at-a-distance theory of interacting classical point charges
was formulated by Fokker and by Wheeler and
Feynman.\cite{fo29,wf4549}.  In these treatments, one obtains the
equations of motion by varying an action integral that is a function
only of the trajectories (worldlines) of the charges.  The price one pays for
eliminating the electromagnetic field is that the action integral is
not a genuine action: One must vary an integral $I$ that involves only
a finite segment of each particle's trajectory, and the equations of
motion emerge from the limit of $\delta I$ as the trajectories are
extended to infinity.  In other words, the limit must be taken {\em
after} the variation of the action integral. In addition, the action is
not an integral over a single parameter time, but instead involves
integrals over parameter times associated with each particle.
\footnote{For parametrization-invariant actions, one can choose to 
parametrize the trajectory of each particle by Minkowski time, 
but interaction terms involve double integrals over the parameter 
time of each particle.}  We will
call an action integral having these properties a {\em Fokker action}.
  
In Dettman and Schild \cite{ds54}, a derivation of the equation of
motion, as well as expressions for the conserved quantities, the energy
momentum and the angular momentum, is presented for a generic Fokker
action that may include self-action terms and variable mass
parameters.

In this paper, we obtain a similar action for two self-gravitating 
point particles, in a post-Minkowski approximation. We describe 
the particles by their constant masses $m,\ \bar m$ and by their 
trajectories written in terms of position vectors 
$x^\alpha(\tau),\ \bx^\alpha(\btau)$ on a flat background spacetime, 
with arbitrary parameter times $\tau$ and $\btau$. 
In this approximation, the motion of each particle is determined by a 
gravitational interaction with the other particle, and there is no 
self-interaction term (any contribution from gravitational self-energy 
is accounted for in the mass of each particle).  The two-particle 
system is then described by a Fokker action of the form
\be 
I(\tau_1, \tau_2, \btau_{1},\btau_{2})
=-m\int_{\tau_1}^{\tau_2}d\tau
(-\dtx_{\alpha}\dtx^{\alpha})^{\frac{1}{2}}
-\bar{m}\int_{\btau_{1}}^{\btau_{2}}d\btau
(-\dtbx_{\alpha}\dtbx^{\alpha})^{\frac{1}{2}}
+\int_{\tau_1}^{\tau_2}d\tau\int_{\btau_{1}}^{\btau_{2}}d\btau
\,\Lambda(x-\bx,\dot{x},\dtbx),
\label{eq:action}
\ee
where the Fokker analog $\Lambda$ of an interaction Lagrangian has the 
property, 
\be
\Lambda(x-\bx,\dtx,\dtbx)=\Lambda(\bx-x,\dtbx,\dtx), 
\label{eq:Lambdasym}
\ee
and where
\be
\dtx^{\alpha}:=\frac{dx^{\alpha}}{d\tau},\ \ \ \mbox{and}\ \ \ 
\dtbx^{\alpha}:=\frac{d\bx^{\alpha}}{d\btau}.
\ee
For given values of $\tau, \bar \tau$, $\Lambda(\tau,\bar\tau)$ 
is a scalar constructed only from $\eta_{\alpha\beta}, x^\alpha(\tau), 
\dot x^\alpha(\tau), \bar x^\alpha(\bar \tau), \dot{\bar x}^\alpha(\bar\tau)$.
It follows that $\Lambda$ and $I$ are Poincar\'e invariant -- invariant
under simultaneous Poincar\'e transformations of the particles' paths.

As noted earlier, to obtain the equations of motion, one must compute 
the variation of the action integral before taking the limit as 
$\tau_1, \btau_1 \rightarrow - \infty$ and 
$\tau_2, \btau_2 \rightarrow\infty$. 

The variation of the action integral (\ref{eq:action}) is given by
\bea
\dl I(\tau_1, \tau_2, \btau_{1},\btau_{2})
&=&\left[\frac{m \dtx_{\alpha}}{(-\dtx_{\gamma}\dtx^{\gamma})^{\frac{1}{2}}}
+\int_{\btau_{1}}^{\btau_{2}}d\btau
\frac{\pa\Lambda}{\pa \dtx^{\alpha}}\right]\dl x^{\alpha}
\left.\phantom{\frac12}\!\!\!\!\!\right|_{\tau_{1}}^{\tau_{2}} 
%%%{\Big |}_{\tau_{1}}^{\tau_{2}}
\nonumber\\
&&
+\int_{\tau_1}^{\tau_2}d\tau\,\dl x^{\alpha}
\left\{-\frac{d}{d\tau}\frac{m\dtx_{\alpha}}
{(-\dtx_{\gamma}\dtx^{\gamma})^{\frac{1}{2}}}
+\int_{\btau_{1}}^{\btau_{2}}d\btau
%%koji\left[\frac{\pa\Lambda}{\pa(x^{\alpha}-\bx^{\alpha})}
\left[\frac{\pa\Lambda}{\pa R^{\alpha}}
-\frac{d}{d\tau}\frac{\pa\Lambda}{\pa\dtx^{\alpha}}\right]\right\} 
\nonumber\\
&&
+\left[\frac{\bar{m}\dtbx_{\alpha}}
{(-\dtbx_{\gamma}\dtbx^{\gamma})^{\frac{1}{2}}}
+\int_{\tau_1}^{\tau_2}d\tau\frac{\pa\Lambda}{\pa\dtbx^{\alpha}}\right]
\dl\bx^{\alpha}
\left.\phantom{\frac12}\!\!\!\!\!\right|_{\btau_{1}}^{\btau_{2}} 
%%%{\Big |}_{\btau_{1}}^{\btau_{2}}
\nonumber\\
&&
+\int_{\btau_{1}}^{\btau_{2}}d\btau\,\dl \bx^{\alpha}
\left\{-\frac{d}{d\btau}
\frac{\bar{m}\dtbx_{\alpha}}{(-\dtbx_{\gamma}\dtbx^{\gamma})^{\frac{1}{2}}}
+\int_{\tau_1}^{\tau_2}d\tau
%%koji\left[\frac{\pa\Lambda}{\pa(\bx^{\alpha}-x^{\alpha})}-\frac{d}{d\btau}
\left[\frac{\pa\Lambda}{\pa\bR^{\alpha}}-\frac{d}{d\btau}
\frac{\pa\Lambda}{\pa\dtbx^{\alpha}}\right]
\right\},  
\label{eq:dlaction}
\eea
where 
$R^\alpha :=x^\alpha-\bx^\alpha =:-\bR^\alpha. $
	
Requiring that the limit of the variation vanish when    
$\dl x^\alpha |_{\pm\infty}=0$ and $\dl\bx^{\alpha}|_{\pm\infty}=0$,  
\beq
  \lim\delta I:= \  
\lim{}_{\hspace{-8mm}\begin{array}{l}
\smash{{}_{(\tau_1,\tau_2)\rightarrow(-\infty,+\infty)}}\\[-1mm]
\smash{{}_{(\btau_1,\btau_2)\rightarrow(-\infty,+\infty)}}
\end{array}} \delta I
=0,
 \label{limI}\eeq
yields the equation of motion for each particle, 
\be
\frac{d}{d\tau}\frac{m\dtx_{\alpha}}
{(-\dtx_{\gamma}\dtx^{\gamma})^{\frac{1}{2}}}
=\intinf d\btau
%%koji\left[\frac{\pa\Lambda}{\pa(x^{\alpha}-\bx^{\alpha})}
\left[\frac{\pa\Lambda}{\pa R^{\alpha}}
-\frac{d}{d\tau}\frac{\pa\Lambda}{\pa \dtx^{\alpha}}\right]
\label{eq:eom1}
\ee
\be
\frac{d}{d\btau}\frac{\bmm\dtbx_{\alpha}}
{(-\dtbx_{\gamma}\dtbx^{\gamma})^{\frac{1}{2}}}
=\intinf d\tau
%%koji\left[\frac{\pa\Lambda}{\pa(\bx^{\alpha}-x^{\alpha})}
\left[\frac{\pa\Lambda}{\pa\bR^{\alpha}}
-\frac{d}{d\btau}\frac{\pa\Lambda}{\pa\dtbx^{\alpha}}\right].
\label{eq:eom2}
\ee

The action integral $I$, with {\em finite} values of $\tau_i, \bar\tau_i$, is 
invariant under Poincar\'e transformations of the paths that leave 
the path parameters fixed.  Invariance of $I$ under 
the infinitesimal spacetime translation of each path by a constant vector 
$a^{\alpha}$, 
\be
  \dl x^{\alpha}=\dl \bx^{\alpha}=a^{\alpha},
\label{eq:infsp}
\ee
implies conservation of 4-momentum:
That is, assuming the equations of motion, (\ref{eq:eom1}) and 
(\ref{eq:eom2}), and substituting 
Eq.~(\ref{eq:infsp}) in Eq.~(\ref{eq:dlaction}),
we obtain 
\beq
\frac{\dl I}{\dl a^\alpha}
=P_\alpha(\tau_2,\btau_2) - P_\alpha(\tau_1,\btau_1), 
\eeq
where
\bea
P_\alpha(\tau,\btau)
=&&\left[\frac{m\dtx_{\alpha}}{(-\dtx_{\gamma}\dtx^{\gamma})^{\frac{1}{2}}}
+\intinf d\btau
\frac{\pa\Lambda}{\pa \dtx^{\alpha}}\right](\tau)
+\left[\frac{\bar{m}\dtbx_{\alpha}}
{(-\dtbx_{\gamma}\dtbx^{\gamma})^{\frac{1}{2}}}
+ \intinf d\tau
\frac{\pa\Lambda}{\pa\dtbx^{\alpha}}\right](\btau)
\nonumber \\
&&+\left(\int_{\tau}^{\infty}\int_{-\infty}^{\btau}
-\int_{-\infty}^{\tau}\int_{\btau}^{\infty}\right)
\frac{\pa\Lambda}{\pa R^{\alpha}}d\tau d\btau.
\label{eq:linmom}
\eea 
Translation invariance, $\delta I =0$, implies that $P_\alpha$ is 
independent of $\tau$ and $\bar\tau$.

Similarly, invariance of $I$ under 
an infinitesimal Lorentz transformation of each path, 
\beq
\dl x^{\alpha}=\epsilon^{\albe} x_{\beta}\ \ \  \mbox{and} \ \ \ 
\dl \bx^{\alpha}=\epsilon^{\albe} \bx_{\beta},
\label{eq:infrot}
\eeq
where $\epsilon^{\albe}=-\epsilon^{\beta\alpha}$
is a constant antisymmetric tensor, implies 
conservation of angular momentum. Again  
assuming the equations of motion, (\ref{eq:eom1}) and 
(\ref{eq:eom2}), and substituting 
Eq.~(\ref{eq:infrot}) in Eq.~(\ref{eq:dlaction}),
we have
\beq
2\frac{\dl I}{\dl \epsilon^{\beta\alpha}}
=L_{\albe}(\tau_2,\btau_2) 
- L_{\albe}(\tau_1,\btau_1), 
\eeq
where
\bea
L_{\albe}(\tau,\btau)
&=&\left[
\frac{m\left(x_\alpha\dtx_\beta-x_\beta\dtx_\alpha\right)}
{(-\dtx_\gamma\dtx^\gamma)^{\frac12}}
+\intinf d\btau
\left(x_\alpha \frac{\pa\Lambda}{\pa \dtx^\beta}
-x_\beta\frac{\pa\Lambda}{\pa\dtx^\alpha}\right)
\right](\tau) 
\nonumber\\
&+&\left[
\frac{\bmm\left(\bx_\alpha\dtbx_\beta-\bx_\beta\dtbx_\alpha\right)}
{(-\dtbx_\gamma\dtbx^\gamma)^{\frac12}}
+\intinf d\tau
\left(\bx_\alpha\frac{\pa\Lambda}{\pa\dtbx^\beta}
-\bx_\beta\frac{\pa\Lambda}{\pa\dtbx^\alpha}\right)
\right](\btau)
\nonumber\\
&+& \left(\int_{\tau}^{\infty}\int_{-\infty}^{\btau}
-\int_{-\infty}^{\tau}\int_{\btau}^{\infty}\right)
\left[
\left(x_\alpha\frac{\pa\Lambda}{\pa R^\beta}
- x_{\beta}\frac{\pa\Lambda}{\pa R^\alpha}\right)
+\left(\dtx_{\alpha}\frac{\pa\Lambda}{\pa \dtx^{\beta}}
-\dtx_{\beta}\frac{\pa\Lambda}{\pa \dtx^{\alpha}}\right)\right]
d\tau d\btau  
\label{eq:angmom}
\eea 
Here, Lorenz invariance implies 
\be
\epsilon^{\albe}
\left(x_\beta\frac{\pa\Lambda}{\pa R^\alpha}
+\bx_\beta\frac{\pa\Lambda}{\pa \bR^\alpha}
+\dtx_{\beta}\frac{\pa\Lambda}{\pa \dtx^{\alpha}}
+\dtbx_{\beta}\frac{\pa\Lambda}{\pa\dtbx^{\alpha}}\right)=0.
\ee
Finally, $\delta I =0$ implies that $L_{\alpha\beta}$ is 
independent of $\tau$ and $\bar\tau$.

With a definition 
\beq
w:=R^\alpha R_\alpha=(x-\bx)^2, 
\eeq
an interaction $\Lambda$ that depends on 
the positions only through $w$, $\Lambda=\Lambda(w,\dtx,\dtbx)$, 
is a restricted form of interactions that satisfies 
the property (\ref{eq:Lambdasym}).  
For such an interaction term, since  
$\pa\Lambda/\pa R^{\alpha} =2R_{\alpha} \pa\Lambda/\pa w$, 
the last line of 
Eq.~(\ref{eq:angmom}) is written 
\beq
2\left(\int_{\tau}^{\infty}\int_{-\infty}^{\btau}
-\int_{-\infty}^{\tau}\int_{\btau}^{\infty}\right)
\left[\frac{\pa\Lambda}{\pa w}
(x_{\beta}\bx_{\alpha}-x_{\alpha}\bx_{\beta})
-\frac{1}{2}\left(\dtx_{\beta}\frac{\pa\Lambda}
{\pa \dtx^{\alpha}}-\dtx_{\alpha}\frac{\pa\Lambda}{\pa \dtx^{\beta}}\right)\right]
d\tau d\btau .
\label{eq:angmomw}
\eeq

\section{Action at a distance theory for post-Minkowskian gravity}
\label{sec:gr}

As usual in a linearized framework, all tensor indices will be raised 
and lowered by the flat metric $\eta_{\alpha\beta}$ of the background 
spacetime.  

\subsection{Fokker actions for point-particles in post-Minkowskian gravity}
\label{sec:PMinv}

Havas and Goldberg \cite{hg62} derived equations of motion for 
point particles in general relativity 
by expanding the metric and demanding that the covariant 
conservation law for the stress-energy tensor be satisfied to first order
in the perturbation, with a time-symmetric, 
half-advanced +  half-retarded field for the first-order metric.  
They found a Fokker action, an action integral $I$ 
for which $\lim\delta I = 0$ (in the sense of Eq. (\ref{limI})) gives the 
the equation of motion to the same order.  In our notation, their 
interaction term is 
\beq
\Lambda(w,\dot{x},\dtbx)
=2m\bmm\,\dl(w)\frac{(\dtx_\alpha\dtbx^\alpha)^2
+\frac12\dtx_\alpha\dtx^\alpha
+\frac12\dtbx_\beta\dtbx^\beta
+\frac12}
{(-\dtx_\gamma\dtx^\gamma)^{\frac12}(-\dtbx_\delta\dtbx^\delta)^{\frac12}}.
\eeq

Ramond \cite{ra73} subsequently formulated a general action-at-a-distance 
theory that is invariant under reparametrization and includes 
tensorial interactions.  Following Ramond's argument, we find that the 
following interaction term is reparametrization invariant: 
\beq
\Lambda(w,\dot{x},\dtbx)
=2m\bmm\,\dl(w)\frac{(\dtx_\alpha\dtbx^\alpha)^2
-\frac12\dtx_\alpha\dtx^\alpha\,\dtbx_\beta\dtbx^\beta}
{(-\dtx_\gamma\dtx^\gamma)^{\frac12}(-\dtbx_\delta\dtbx^\delta)^{\frac12}},
\label{eq:intgeo}
\eeq
When parameters $\tau$ and $\btau$ are chosen to satisfy 
$\dtx_\alpha\dtx^\alpha=-1$ and $\dtbx_\alpha\dtbx^\alpha=-1$, 
as in the Havas-Goldberg treatment, the equations of motion 
agree with theirs. Because $\dot x^\alpha$ is normalized with respect to
$\eta_{\albe}$, not $\eta_{\albe}+ h_{\albe}$, the Havas-Goldberg $\tau$ 
is not an affine parameter.  

We can see as follows how, starting from the post-Minkowski equations 
of motion, one arrives at an interaction term with $\Lambda$ given by 
Eq.~(\ref{eq:intgeo}). The derivation also shows why one obtains a 
Fokker action, not a true action. 

To linear order, the metric $\gab$ is a sum of the flat metric, $\eta_{\albe}$, and 
a perturbation $\widetilde h_{\albe}$, 
\beq
g_{\albe}=\eta_{\albe}+\widetilde h_{\albe}. 
\eeq
where the perturbed metric $\widetilde h_{\albe}$ is a sum of the half-advanced 
+half-retarded field of each particle.  
As mentioned earlier, in 
the post-Minkowski treatment of point particles, the  
effect on the motion of particle $m$ from the field of 
$m$ itself is simply a mass renormalization.  That is, the motion 
of $m$ is described to linear order in $\widetilde h_{\albe}$ by the linearized 
geodesic equation with $\widetilde h_{\albe}$ replaced by the value at 
the position of $m$ of the half-advanced+half-retarded field $h_{\albe}$ of 
$\bar m$ alone: When $\tau$ is an affine parameter (proper time with respect to the 
perturbed metric), the geodesic equation to linear order in $h_{\albe}$ 
has the form 
\beq
(\eta_{\albe}+h_{\albe})\ddtx^{\beta}
+C_{\albe\gamma}\dtx^{\beta}\dtx^{\gamma}=0, \quad
\mbox{ where }\quad 
C_{\albe\gamma} :=\frac{1}{2}(\na_{\beta}h_{\alpha\gamma}+\na_{\gamma}
h_{\beta\alpha}-\na_{\alpha}h_{\bega}).   
\label{eq:geodyn}\eeq
Equivalently, 
\beq
\frac{d}{d\tau}[(\eta_{\albe}+h_{\albe})\dtx^\beta]
 - \frac12 \na_\alpha h_{\bega}\dtx^\beta\dtx^\gamma=0. 
\label{eq:lingeo1}
\eeq
To linear order in $h_{..}$, the expression 
$(\eta_{\albe}+h_{\albe})\ddtx^\beta$ in Eq. (\ref{eq:geodyn}) 
can be replaced by $\ddtx^\alpha =\eta_{\albe}\ddtx^\beta$, 
because $\dtx^\alpha$ is already order $h_{..}$.  
The form given in Eq. (\ref{eq:lingeo1}), however, conforms to that of 
the action integrals below.  
For $\tau$ a generic time parameter, the geodesic equation has the 
form given in Eq.~(\ref{eq:geogeo}). This can be directly shown from 
(\ref{eq:lingeo1}), but we derive it below from the action for a 
point particle, written to linear order in $h_{\albe}$. \\ 

In the deDonder (harmonic) gauge, $\na_\beta h^{\albe} = 0$, $h_{\albe}$ is given by 
\beq
\Box\, (h_{\albe}-\frac1{2}\eta_{\albe}h) =-16\pi \bT_{\albe}, 
\label{eq:boxth}\eeq
where the stress-energy tensor $\bT^{\albe}$ of $\bar m$ is defined by 
\beq
\bT^{\albe}(x)=\bmm \int_{-\infty}^\infty d\bar\tau\, 
\delta\big(x-\bar x(\btau)\big)
\frac{\dtbx^\alpha \dtbx^\beta}{(-\dtbx_\gamma \dtbx^\gamma)^{\frac12}}.
\label{eq:bset}
\eeq
Here $\nabla_\alpha$ is the covariant derivative operator of the
flat metric $\eta_{\albe}$, and $\Box=\na_\alpha\na^\alpha$ is the 
corresponding flat D'Alembertian. Using the half-retarded +  
half-advanced Green function, $G(x,\bar x) = \delta(w)$ 
(a solution to $\Box\, G(x,\bar x)=-4\pi\delta(x-\bar x)$), 
the solution to Eq.~(\ref{eq:boxth}) is written, 
\beq
h^{\albe}(x)
= 4\bmm \int_{-\infty}^\infty d\btau\, \dl(w)
\frac{\dtbx^\alpha\dtbx^\beta
-\frac12 \bareta^{\albe}\dtbx_\gamma\dtbx^\gamma}
{(-\dtbx_\delta \dtbx^\delta)^{\frac12}}.  
\label{eq:hsol}
\eeq

The trajectory of the second particle, $(\bmm,\bx^\alpha(\btau))$, 
is similarly a geodesic of a background spacetime with metric 
$g_{\albe}=\eta_{\albe}+\bhh_{\albe}$, given by Eq.~(\ref{eq:lingeo1}) 
or (\ref{eq:geogeo}), with barred and unbarred quantities exchanged. 
The source for $\bhh_{\albe}$ is the particle 
$(m,x^\alpha(\tau))$: 
\beq
\bhh^{\albe}(\bx)
= 4m \int_{-\infty}^\infty d\tau\, \dl(w)
\frac{\dtx^\alpha\dtx^\beta
-\frac12 \eta^{\albe}\dtx_\gamma\dtx^\gamma}
{(-\dtx_\delta \dtx^\delta)^{\frac12}}.  
\label{eq:bhsol}
\eeq
({\em Note}: We use a bar to label the perturbed field acting 
on particle $\bar m$; 
$\bar h_{\alpha\beta}$ is {\em not} the trace-reversed form 
of $h_{\alpha\beta}$.)

To find an interaction term that reproduces the equations of 
motion, we begin with an action $I_m$ for the first particle in the field of 
the second, a linearized geodesic equation on a 
background spacetime with metric $g_{\albe}=\eta_{\albe}+h_{\albe}$.
The action $I_m$ is then the action for geodesic motion, 
\[
 -m\int_{\tau_1}^{\tau_2} d\tau\,(-\gab \dtx^\alpha \dtx^\beta)^{\frac12}
=-m\int_{\tau_1}^{\tau_2} d\tau\,
\left[-(\eta_{\albe}+h_{\albe})\dtx^\alpha \dtx^\beta\right]^{\frac12},
\]
written to linear order in $h_{\albe}$:
\beq
I_{m} = -m \int_{\tau_1}^{\tau_2} d\tau\,
(-\dtx_\alpha \dtx^\alpha)^{\frac12}
+ \frac12 m\int_{\tau_1}^{\tau_2} d\tau\,
h_{\albe}\frac{\dtx^\alpha \dtx^\beta}
{(-\dtx_\gamma \dtx^\gamma)^{\frac12}}.
\label{eq:linac}\eeq

The action $I_m$ is invariant under time reparametrization. From its
variation, $\delta I_m =0$,  we can directly compute the linearized
geodesic equation with arbitrary parameter $\tau$:
\beqn
\dl I_m &=& \left.
m \left[\eta_{\albe}+h_{\albe}
+\frac12\eta_{\albe} h_\gade
\frac{\dtx^\gamma \dtx^\delta}
{(-\dtx_\gamma \dtx^\gamma)^{\frac12}}
\right]
\frac{\dot{x}^\beta}{(-\dtx_\gamma \dtx^\gamma)^{\frac12}}
\dl x^\alpha \right|_{\tau_1}^{\tau_2}
\nonumber \\
&&- m \int_{\tau_1}^{\tau_2}d\tau\left\{
\frac{d}{d\tau}\left[\left(\eta_{\albe}+h_{\albe}
+\frac12\eta_{\albe} h_\gade
\frac{\dtx^\gamma \dtx^\delta}
{(-\dtx_\gamma \dtx^\gamma)^{\frac12}}
\right)
\frac{\dot{x}^\beta}{(-\dtx_\gamma \dtx^\gamma)^{\frac12}}\right]
- \frac12\na_\alpha h_{\bega}
\frac{\dtx^\beta\dtx^\gamma}
{(-\dtx_\gamma \dtx^\gamma)^{\frac12}}
\right\}\dl x^\alpha.
\label{eq:varac}
\eeqn
Then, requiring $\delta I_m = 0$ for variations $\delta x^\alpha$ 
with fixed end points yields the equation of motion, we have
\beq
\frac{d}{d\tau}\left[\left(\eta_{\albe}+h_{\albe}
+\frac12\eta_{\albe} h_\gade
\frac{\dtx^\gamma \dtx^\delta}
{(-\dtx_\gamma \dtx^\gamma)^{\frac12}}
\right)
\frac{\dot{x}^\beta}{(-\dtx_\gamma \dtx^\gamma)^{\frac12}}\right]
= \frac12\na_\alpha h_{\bega}
\frac{\dtx^\beta\dtx^\gamma}
{(-\dtx_\gamma \dtx^\gamma)^{\frac12}}.
\label{eq:geogeo}
\eeq

If in the action $I_m$ of Eq.~(\ref{eq:linac}) we substitute Eq.~(\ref{eq:hsol}),
we obtain 
\beqn
I_{m}
= -m \int_{\tau_1}^{\tau_2} d\tau\,
(-\dtx_\alpha \dtx^\alpha)^{\frac12}
+ 2m\bmm\int_{\tau_1}^{\tau_2} d\tau \intinf d\btau\,
\dl(w)\frac{(\dtx_\alpha\dtbx^\alpha)^2
-\frac12\dtx_\alpha\dtx^\alpha\,\dtbx_\beta\dtbx^\beta}
{(-\dtx_\gamma\dtx^\gamma)^{\frac12}(-\dtbx_\delta\dtbx^\delta)^{\frac12}}.
\label{eq:linac2}
\eeqn

To obtain an action whose variations with respect to both $x^\alpha$ 
and $\bar x^\alpha$ yield the equations of motion for each particle, one 
cannot simply add a kinetic term for $\bar m$ and a second interaction 
term in which the barred and unbarred quantities are interchanged, 
because the new interaction term would involve $x^\alpha$ and  
alter the equations of motion of the first particle. (Roughly speaking, 
one would be double-counting the gravitational binding energy.)  
Instead, one observes that the interaction term becomes symmetric under 
interchange of the two particles $m, x^\alpha(\tau)$ and 
$\bmm, \bx^\alpha(\btau)$ if, in the infinite integral, we make the 
replacements $-\infty\rightarrow\bar\tau_1$ and 
$\infty\rightarrow\bar\tau_2 $.  The resulting action integral, 
symmetric under interchange of two particles, has the form 
of Eq.~(\ref{eq:action}), with interaction term~(\ref{eq:intgeo})
\beqn
I = -m \int_{\tau_1}^{\tau_2} d\tau\, 
(-\dtx_\alpha \dtx^\alpha)^{\frac12}
-\bmm \int_{\btau_1}^{\btau_2} d\btau\,
(-\dtbx_\alpha \dtbx^\alpha)^{\frac12}
+ 2m\bmm\int_{\tau_1}^{\tau_2} d\tau \int_{\btau_1}^{\btau_2} d\btau\,
\dl(w)\frac{(\dtx_\alpha\dtbx^\alpha)^2
-\frac12\dtx_\alpha\dtx^\alpha\,\dtbx_\beta\dtbx^\beta}
{(-\dtx_\gamma\dtx^\gamma)^{\frac12}(-\dtbx_\delta\dtbx^\delta)^{\frac12}}.
\label{eq:actiongeo}
\eeqn
The price for this symmetry is that a variation of $I$ yields equations 
of motion in which the integrals giving $h_{\albe}$ and
$\bar h_{\albe}$ extend only from 
$\bar\tau_1$ to $\bar\tau_2$ (or $\tau_1$ to $\tau_2$). It is for 
this reason that action-at-a-distance theories require a Fokker action, 
whose equations of motion are given by $\lim \delta I = 0$. 
 
As mentioned, the above theory is invariant under reparametrization 
of $\tau$ and $\btau$.  
The obvious choice of a proper-time parametrization, with  
$\eta_{\albe}\dtx^\alpha\dtx^\beta=-1$ and 
$\bareta_{\albe}\dtbx^\alpha\dtbx^\beta=-1$, simplifies 
computations of physical quantities.

\subsection{A Fokker action for the affinely parametrized equations}
\label{sec:affine}

One can, of course, specialize the parametrization-invariant action to 
affinely parametrized trajectories, but affine parametrization also allows 
a generalization to interacting particles of the quadratic action
for geodesic motion,  
$\displaystyle \frac12 m\int_{\tau_1}^{\tau_2}d\tau\,
\dtx_{\alpha}\dtx^{\alpha}$.
To construct an action integral $I$ with quadratic kinetic term, 
for which $\lim\delta I=0$ reproduces
the affinely parametrized equation of motion, Eq.~(\ref{eq:lingeo1}), and 
its barred $\leftrightarrow$ unbarred form, 
we modify the kinetic terms in the action integral of  
Eq.~({\ref{eq:action}), writing 
\be 
I(\tau_1, \tau_2, \btau_{1},\btau_{2})
=\frac12 m\int_{\tau_1}^{\tau_2}d\tau\,
\dtx_{\alpha}\dtx^{\alpha}
+\frac12\bmm\int_{\btau_1}^{\btau_2}d\btau\,
\dtbx_{\alpha}\dtbx^{\alpha}
+\int_{\tau_1}^{\tau_2}d\tau\int_{\btau_{1}}^{\btau_{2}}d\btau
\,\Lambda(w,\dot{x},\dtbx); 
\label{eq:actionlgr}
\ee
and we take as the interaction term 
\beq
\Lambda(w,\dtx,\dtbx) = 
2m\bmm\,\dl(w)\left[(\dtx_\alpha\dtbx^\alpha)^2
-\frac12 \dtx_\alpha\dtx^\alpha \dtbx_\beta\dtbx^\beta\right].  
\label{eq:interac}
\eeq
Affine parametrization, the requirement 
\be 
(\eta_{\alpha\beta}+h_{\alpha\beta})\dot x^\alpha \dot x^\beta = \mbox{ constant},
\qquad 
(\eta_{\alpha\beta}+\bar h_{\alpha\beta})\dot{\bx}^\alpha \dot {\bx}^\beta = \mbox{ constant},
\ee
is enforced by the equations of motion. 

Formulas for the equation of motion Eq.~(\ref{eq:eom1}) and 
Eq.~(\ref{eq:eom2}), 
the 4-momentum Eq.~(\ref{eq:linmom})
and the angular momentum tensor Eq.~(\ref{eq:angmom}) 
are changed as a result of this modification of the kinetic terms in 
Eq.~(\ref{eq:actionlgr}).  The variation of our kinetic terms  
\beqn
\frac12 m\,\dl\int_{\tau_1}^{\tau_2}d\tau\,\dtx_{\alpha}\dtx^{\alpha} 
&&= m \dtx_{\alpha} \dl x^{\alpha} \big|_{\tau_{1}}^{\tau_{2}} 
- \int_{\tau_1}^{\tau_2}d\tau\,m\ddtx_{\alpha}\dl x^{\alpha}, 
\\
\frac12 \bmm\,\dl\int_{\btau_1}^{\btau_2}d\btau\,\dtbx_{\alpha}\dtbx^{\alpha} 
&&= \bmm \dtbx_{\alpha} \dl \bx^{\alpha} \big|_{\btau_{1}}^{\btau_{2}} 
- \int_{\btau_1}^{\btau_2}d\btau\,\bmm\ddtbx_{\alpha}\dl \bx^{\alpha}, 
\eeqn 
results in the following form for the equation of motion, 
\be
m\ddtx_{\alpha}
=\intinf d\btau
\left[\frac{\pa\Lambda}{\pa R^{\alpha}}
-\frac{d}{d\tau}\frac{\pa\Lambda}{\pa \dtx^{\alpha}}\right]
= -\frac{d}{d\tau}(h_{\albe}\dtx^\beta)
 + \frac12 \na_\alpha h_{\bega}\dtx^\beta\dtx^\gamma, 
\label{eq:eomlgr1}
\ee
\be
\bmm\ddtbx_{\alpha}
=\intinf d\tau
\left[\frac{\pa\Lambda}{\pa \bR^{\alpha}}
-\frac{d}{d\btau}\frac{\pa\Lambda}{\pa\dtbx^{\alpha}}\right]
=-\frac{d}{d\tau}(\bar h_{\albe}\dtbx^\beta)
 + \frac12 \na_\alpha \bar h_{\bega}\dtbx^\beta\dtbx^\gamma, 
\label{eq:eomlgr2}
\ee
in agreement with the linearized geodesic equation (\ref{eq:lingeo1}).

The 4-momentum is given by 
\beqn
P_\alpha(\tau,\btau)
=&&\left[m\dtx_{\alpha}
+\intinf d\btau
\frac{\pa\Lambda}{\pa \dtx^{\alpha}}\right](\tau)
+\left[\bar{m}\dtbx_{\alpha}
+ \intinf d\tau
\frac{\pa\Lambda}{\pa\dtbx^{\alpha}}\right](\btau)
\nonumber \\
&&+\left(\int_{\tau}^{\infty}\int_{-\infty}^{\btau}
-\int_{-\infty}^{\tau}\int_{\btau}^{\infty}\right)
\frac{\pa\Lambda}{\pa R^{\alpha}}d\tau d\btau, 
\label{eq:linmomlgr}
\eeqn
and the angular momentum by 
\bea
L_{\albe}(\tau,\btau)
&=&\left[m\left(x_{\alpha}\dtx_{\beta}-x_{\beta}\dtx_{\alpha}\right)
+\intinf d\btau 
\left(x_{\alpha}\frac{\pa\Lambda}{\pa \dtx^{\beta}}
     -x_{\beta}\frac{\pa\Lambda}{\pa \dtx^{\alpha}}
\right)\right](\tau) 
\nonumber\\
&+&\left[\bar{m}\left(\bx_{\alpha}\dtbx_{\beta}-\bx_{\beta}\dtbx_\alpha\right)
+\intinf d\tau
\left(\bx_{\alpha}\frac{\pa\Lambda}{\pa\dtbx^{\beta}}
     -\bx_{\beta}\frac{\pa\Lambda}{\pa\dtbx^{\alpha}}
\right)\right](\btau)
\nonumber\\
&+& 2\left(\int_{\tau}^{\infty}\int_{-\infty}^{\btau}
-\int_{-\infty}^{\tau}\int_{\btau}^{\infty}\right)
\left[\frac{\pa\Lambda}{\pa w}
(x_{\beta}\bx_{\alpha}-x_{\alpha}\bx_{\beta})
-\frac{1}{2}\left(\dtx_{\beta}\frac{\pa\Lambda}
{\pa \dtx^{\alpha}}-\dtx_{\alpha}\frac{\pa\Lambda}{\pa \dtx^{\beta}}
\right)\right]
d\tau d\btau ,
\label{eq:angmomlgr}
\eea 
These expressions for momentum and angular momentum are 
formally identical to Eqs. (\ref{eq:linmom}) and (\ref{eq:angmom}) 
if one sets $(-\eta_{\alpha\beta}{\dot x}^\alpha{\dot x}^\beta)$ to $1$. 
This is misleading: In the present section, ${\dot x}^\alpha$ is a unit vector {\em not} of the Minkowski metric but of the perturbed metric 
$\eta_{\alpha\beta}+h_{\alpha\beta}$, and the altered form 
of $\Lambda$ compensates for the altered normalization, leading to 
identical Newtonian limits of the two expressions.    

\subsection{Circular orbits}
\label{sec:orbit}

We consider now a system of two point particles in circular orbit, for which  
the gravitational field seen by each particle is the linearized 
half-advanced + half-retarded field of the other.  This is a gravitational 
analog of a solution obtained by Schild \cite{sc63} for two point charges.
In the gravitational context, however, linearized gravity is a toy model 
for bound systems: Discarded nonlinear terms are of the same post-Newtonian 
order as linear terms that are kept, and hence of the same magnitude 
as terms of the next post-Minkowskian order.  
  
We introduce a basis $\{t^\alpha,\varpi^\alpha,\widehat\phi^\alpha,z^\alpha\}$ 
of unit vectors of the flat metric 
$\eta_{\albe}$.  
A particle in circular orbit in the $z=0$ plane with constant 
orbital radius $a$ has cylindrical coordinates $\{t, \varpi=a, \phi,z=0\}$ and 
position vector  
\beq
x^\alpha = t\, t^\alpha + a \varpi^\alpha . 
\eeq 
Its spacetime trajectory is tangent to the helical Killing vector 
\beq
k^\alpha=t^\alpha+\Omega\phi^\alpha, 
\eeq
where $\phi^\alpha = \varpi \widehat\phi^\alpha$ is a rotational Killing 
vector of the flat metric $\eta_{\alpha\beta}$ and $\Omega$ is the 
particle's constant angular velocity.  
With $\gamma := dt/d\tau$ and $v:=a \Omega$, the particle's 4-velocity 
and acceleration are given by 
\beq 
\dtx^\alpha = \gamma k^\alpha = \gamma(t^\alpha + v\widehat\phi^\alpha) 
\eeq
and 
\beq
\ddtx^\alpha = - \gamma^2 v \Omega \varpi^\alpha.
\eeq

The second particle $\bmm$ has a circular orbit of radius $\bar a$ 
about the same origin with coordinates \\
$\{\bt,\bpi=\bar a,\bphi,\bz=0\}$ and position vector 
\be
 \bx^{\alpha}=\bt\, t^{\alpha}+ \ba \bpi^{\alpha}. 
\ee   
The particle trajectory is again tangent to the helical Killing vector 
\beq
\bk^\alpha=t^\alpha+\Omega\bphi^\alpha,  
\eeq
with $\bphi^\alpha$ the value of the vector field $\phi^\alpha$ at the position 
of the second particle.  
With $\bgamma = d\bt/d\btau$ and $\bv:=\ba \Omega$, the acceleration and 
4-velocity of the second particle are given by
\beq
\dtbx^\alpha = \bgamma \bk^\alpha = \bgamma(t^\alpha + \bv\hbphi^\alpha)  
\eeq
and
\beq
\ddtbx^\alpha = - \bgamma^2 \bv \Omega \bpi^\alpha.  
\eeq

For the parametrization-invariant formulation, 
it is convenient to parametrize the trajectories by proper time with
respect to the flat metric $\eta_{\alpha\beta}$; their tangent vectors
are then unit vectors of $\eta_{\alpha\beta}$, 
\beq
\eta_{\albe} \dtx^\alpha\dtx^\beta = -1, \ \ \ \ \ \ 
\bareta_{\albe} \dtbx^\alpha\dtbx^\beta = -1.  
\label{eq:norm}
\eeq
The quantities $\gamma$, $v$, $\bgamma$ and $\bv$ obey the familiar 
relations 
\beq
\gamma = (1-v^2)^{-\frac12},\ \ \ \ \ \ 
\bgamma = (1-\bv^2)^{-\frac12}.
\label{eq:normgf}
\eeq
For the affinely parametrized formulation, a correct Newtonian limit is 
reproduced when a normalization with the perturbed metric 
(Sec.~\ref{sec:affine}) is 
used:   
\beq
\left(\eta_{\albe} + h_{\albe}\right)\dtx^\alpha\dtx^\beta = -1, \ \ \ \ \ \ 
\left(\bareta_{\albe} + \bar{h}_{\albe}\right)\dtbx^\alpha\dtbx^\beta = -1,   
\label{eq:normaf}
\eeq
which yield relations 
\beq
\gamma = (1-v^2 - h_{\albe}k^\alpha k^\beta)^{-\frac12},\ \ \ \ \ \ 
\bgamma = (1-\bv^2 - \bar{h}_{\albe}\bar{k}^\alpha \bar{k}^\beta)^{-\frac12}.
\label{eq:normgfaf}
\eeq

With the first particle, $m$, at $\phi=0$ for $t=0$, its trajectory has coordinates  
\beq
\phi = \Omega t\ , \ \ \  \ \ \ t = \gamma \tau;
\label{eq:phase1def}
\eeq 
and the trajectory of $\bmm$ then has coordinates  
\beq\bphi = \pi+\Omega \bt\ ,\ \ \ \ \ \ 
\bt = \bgamma \btau - \pi/\Omega. 
\label{eq:phase2def}
\eeq
The positions of both particles are specified by a single parameter, 
and it is natural to choose a descriptor $\eta$ of their 
relative motion, defined by
\bea
\eta:=\bar{\phi}-\phi =\Omega(\bar{\gamma}\btau-\gamma \tau), 
\label{eq:etadef}
\eea 
where we pick $\tau=\btau=0$ when $\phi=\bphi=0$.  

A vector $R^\alpha:=x^\alpha-\bx^\alpha$ becomes 
\beqn
R^\alpha
&=& \frac1\Omega\left[(\pi-\eta)\,t^\alpha
+ (v-\bv \cos\eta)\,\varpi^\alpha - \bv\sin\eta\, \widehat\phi^\alpha\right] 
\label{eq:x-y1}\\
&=& \frac1\Omega\left[(\pi-\eta)\,t^\alpha
- (\bv- v \cos\eta)\,\bpi^\alpha - v \sin\eta\, \hbphi^\alpha\right].
\label{eq:x-y2}\eeqn

The half-retarded + half-advanced Green function 
$\dl(w)=\dl\left[(x-\bx)^2\right]$ has support on the two events that 
correspond to the roots of 
\beq
w(\eta) := (x-\bx)^2 
= \frac1{\Omega^2}\left[-(\pi-\eta)^2+v^2+\bv^2 -2v\bv\cos\eta\right]
=0. 
\label{eq:w}
\eeq
These roots are given by 
\beq 
\eta=\pi\pm\varphi, 
\label{eq:wsol}
\eeq 
with $\varphi$ the positive root of 
\beq
\varphi^2=v^2+\bv^2+2v\bv\cos\varphi. 
\label{eq:phi}
\eeq
Here $\varphi+\pi$ is the angle between $m$ and the retarded position 
of $\bar m$, and Eq. (\ref{eq:phi}) has a simple geometrical meaning, 
illustrated in Fig. \ref{fig:phi}  
\[ \varphi+\pi = \bphi_{\rm ret}- \phi,\]
in which $R$ is the distance
from $m$ to the the retarded position of $\bar m$.
 \begin{figure}[ht] 
%%% \centerline{\epsfysize=4cm \epsfbox{phi.eps}}
 \includegraphics[height=60mm,clip]{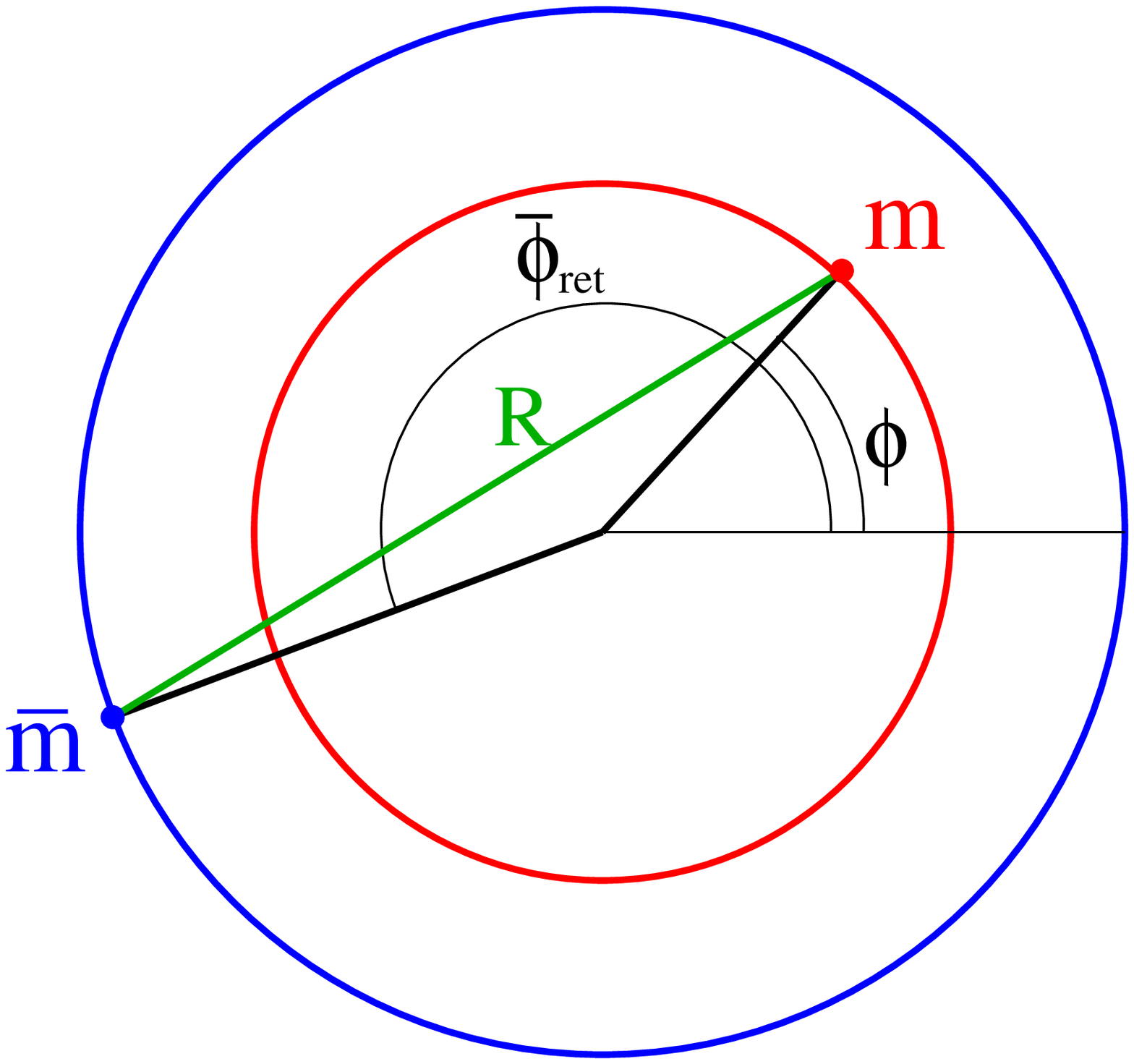}
 \caption{The figure shows mass $m$ at time $t$ and mass $\bar m$ 
 at the corresponding retarded time.  The angle   
 $\phi_{\rm ret}$ is the $\phi$ coordinate of the intersection of 
 the past light cone of $m$ with the trajectory of $\bar m$, when $m$ 
 is at $\phi$. }
 \label{fig:phi} 
 \end{figure}  
That is, using the geometric relation  
\be R^2 = a^2 + \bar a^2 -2a\bar a \cos(\bar\phi_{\rm ret} - \phi)
	= \frac1{\Omega^2} (v^2+ \bar v^2 + 2v\bar v\cos\varphi), 
\label{eq:rsq}\ee   
noting that $R = \bar t_{\rm ret} - t$, and using 
Eqs. (\ref{eq:phase1def}-\ref{eq:phase2def}), we have 
\be 
 \varphi  = \bar\phi_{\rm ret}-\pi -\phi = \Omega(\bar t_{\rm ret} - t) 
	   = \Omega R.
\ee
Then $\varphi^2 = \Omega^2  R^2$ and Eq. (\ref{eq:phi}) is immediate 
from (\ref{eq:rsq}).  Analogously, the root 
$\eta = \pi-\varphi$ corresponds to the advanced position of 
$\bmm$ defined by $-\varphi+\pi = \bphi_{\rm adv}- \phi$.

\subsection{Conserved quantities and angular velocity of circular orbits}
\label{sec:cirint}

We now rewrite the integrals appearing in Eqs.~(\ref{eq:eom1}), 
(\ref{eq:eom2}), (\ref{eq:linmom}), and (\ref{eq:angmom}) 
for a parametrization-invariant action, in terms of the parameter $\eta$. 
To evaluate the integrals, we use formulas obtained in Appendix 
\ref{sec:app1}.    
The resulting formulas for affinely parametrized formulations 
Eqs.~(\ref{eq:eomlgr1}), (\ref{eq:eomlgr2}), (\ref{eq:linmomlgr}), 
and (\ref{eq:angmomlgr}) 
are quite similar to those for the parametrization-invariant 
formulations.  
However, the fact that the action appropriate to an initial 
choice of affine parameters is not a special case of the 
parametrization-invariant action leads to definitions of a conserved 
energy that are not identical in the two formulations.

\subsubsection{Parametrization-invariant model}

For circular orbits, only the radial component of the equations of 
motion is nontrivial, and it has the form,  
\bea 
-m\gamma^{2} v \Omega
&=& \intinf d\eta\left\{\frac{\pa\Lambda}{\pa w}
\frac{2}{\bgamma\Omega^{2}}(v -\bv\cos\eta)
+\frac{\gamma}{\bgamma}\frac{\pa\Lambda}{\pa \dtx_\beta}
\hphi_{\beta}\right\} \\
&=& 4m\bmm\gamma^2\bgamma
\intinf d\eta\left\{\frac1{\Omega^2}\,\dl'(w)
(v-\bv\cos\eta)\,\tPhi(\eta,v,\bv)
\right.
\nonumber \\
&&\qquad \qquad \qquad \qquad \quad
- \dl(w)\big[(1-v\bv\cos\eta)\bv\cos\eta-\frac12 v(1-\bv^2)
- \frac12 \frac{v}{1-v^2}\tPhi(\eta,v,\bv)\big]
\left.\phantom{\frac12}\!\!\!\!\!\!\right\}, 
\label{eq:eometa1}
\eea 
\bea 
-\bar{m}\bgamma^{2}\bv\Omega 
&=& \intinf d\eta\left\{\frac{\pa\Lambda}{\pa w}
\frac{2}{\gamma\Omega^{2}}(\bv -v\cos\eta)
+\frac{\bgamma}{\gamma}\frac{\pa\Lambda}{\pa\dtbx_{\beta}}
\hbphi_{\beta}\right\}\\
&=& 4m\bmm\gamma\bgamma^2
\intinf d\eta\left\{\frac1{\Omega^2}\,\dl'(w)
(\bv-v\cos\eta)\,\tPhi(\eta,v,\bv)
\right.
\nonumber \\
&&\qquad \qquad \qquad \qquad \quad
- \dl(w)\big[(1-v\bv\cos\eta)v\cos\eta-\frac12 \bv(1-v^2)
- \frac12 \frac{\bv}{1-\bv^2}\tPhi(\eta,v,\bv)\big]
\left.\phantom{\frac12}\!\!\!\!\!\!\right\}, 
\label{eq:eometa2}
\eea 
where Eqs.~(\ref{eq:intvec1}) and (\ref{eq:intvec2}) 
in Appendix \ref{sec:app1} are used.  We have introduced here 
a function $\tPhi(\eta,v,\bv)$, defined by 
\beq
\tPhi(\eta,v,\bv):= 
\frac1{\gamma^2\bgamma^2}\big[(\dtx_\alpha\dtbx^\alpha)^2
-\frac12\dtx_\alpha\dtx^\alpha\dtbx_\beta\dtbx^\beta\big]
=(1-v\bv\cos\eta)^2-\frac12(1-v^2)(1-\bv^2). 
\label{eq:Phi}
\eeq

Since the center-of-mass frame is chosen, the only nonzero 
components of the 4-momentum and the angular momentum are 
$E := -P_\alpha(\tau,\btau) t^\alpha=-P_0(\tau,\btau)$ and 
$L := L_{12}(\tau,\btau)$.  Taking $(\tau,\btau) = (0,0)$, we have
\beqn
E = P^0(0,0)
&=& \left[m\dtx^0+\frac1{\bgamma\Omega}
\intinf d\eta\,\frac{\pa\Lambda}{\pa\dtx_0}\right](0)
+ \left[\bmm\dtbx^0+\frac1{\gamma\Omega}
\intinf d\eta\,\frac{\pa\Lambda}{\pa\dtbx_0}\right](0)
-\frac1{\gamma\bgamma\Omega^2}\intinf d\eta\,\eta
\frac{\pa\Lambda}{\pa w}2(x^0-\bx^0)
\nonumber\\
&=& m\gamma - \frac{4m\bmm\gamma\bgamma}{\Omega}\intinf d\eta\,
\dl(w)\big[(1-v\bv\cos\eta)-\frac12(1-\bv^2)
- \frac12 \frac1{1-v^2}\tPhi(\eta,v,\bv)\big]
\nonumber\\
&+& \bmm\bgamma - \frac{4m\bmm\gamma\bgamma}{\Omega}\intinf d\eta\,
\dl(w)\big[(1-v\bv\cos\eta)-\frac12(1-v^2)
- \frac12 \frac1{1-\bv^2}\tPhi(\eta,v,\bv)\big]
\nonumber\\
&&\qquad -\frac{4m\bmm\gamma\bgamma}{\Omega^3}\intinf d\eta\,
\dl'(w)\,\eta(\pi-\eta)\,\tPhi(\eta,v,\bv) , 
\label{eq:eneeta}
\eeqn
\beqn
L = L_{12}(0,0)
&=& 
\left[m(x_1\dtx_2-x_2\dtx_1)+\frac1{\bgamma\Omega}\intinf d\eta\,
\left(x_1\frac{\pa\Lambda}{\pa\dtx^2}-x_2\frac{\pa\Lambda}{\pa\dtx^1}
\right)\right](0)
\nonumber\\
&+&\left[\bmm(\bx_1\dtbx_2-\bx_2\dtbx_1)+\frac1{\gamma\Omega}\intinf d\eta\,
\left(\bx_1\frac{\pa\Lambda}{\pa\dtbx^2}-\bx_2\frac{\pa\Lambda}{\pa\dtbx^1}
\right)\right](0)
\nonumber\\
&-&\frac2{\gamma\bgamma\Omega^2}\intinf d\eta\,\eta
\left[\frac{\pa\Lambda}{\pa w}(x_2\bx_1 - x_1\bx_2)
-\frac12\left(\dtx_2\frac{\pa\Lambda}{\pa\dtx^1}
             -\dtx_1\frac{\pa\Lambda}{\pa\dtx^2}\right)\right]
\\
&=& 
\frac{m\gamma v^2}{\Omega}
-\frac{4m\bmm\gamma\bgamma v\bv}{\Omega^2}\intinf d\eta\,
\dl(w)\big[(1-v\bv\cos\eta)\cos\eta-\frac12\frac{v}{\bv}(1-\bv^2)
- \frac12 \frac{v}{\bv}\frac1{1-v^2}\tPhi(\eta,v,\bv)\big]
\nonumber\\
&+& \frac{\bmm\bgamma \bv^2}{\Omega}
-\frac{4m\bmm\gamma\bgamma v\bv}{\Omega^2}\intinf d\eta\,
\dl(w)\big[(1-v\bv\cos\eta)\cos\eta-\frac12\frac{\bv}{v}(1-v^2)
- \frac12 \frac{\bv}{v}\frac1{1-\bv^2}\tPhi(\eta,v,\bv)\big]
\nonumber\\
&&\qquad\quad
+ \frac{4m\bmm\gamma\bgamma v\bv}{\Omega^2}\intinf d\eta
\left[\frac1{\Omega^2}\dl'(w)\,\eta\sin\eta\,\tPhi(\eta,v,\bv)
+\dl(w)(1-v\bv\cos\eta)\eta\sin\eta\right],
\label{eq:angeta}
\eeqn
where we have used a formula, Eq.~(\ref{eq:intint}), derived in 
Appendix \ref{sec:app1}. 

Formulas for integrations involving a $\delta$-function 
are summarized in Appendix \ref{sec:app2}.  
First we define a function $\Phi(\varphi,v,\bv)$, related to $\tPhi$ 
by an integration: 
\beqn
\Phi(\varphi,v,\bv) 
=\frac1{\Omega^2}\intinf d\eta\,\dl(w)\,\tPhi(\eta,v,\bv)
=\frac{(1+v\bv\cos\varphi)^2 - \frac12(1-v^2)(1-\bv^2)}
{\varphi+v\bv\sin\varphi}.
\eeqn
Using this relation, we can evaluate the integrals appearing in 
the equations of motion (\ref{eq:eometa1}) and (\ref{eq:eometa2}),
obtaining an algebraic relation for the radius of each orbit (or, 
equivalently, for the velocities $v=a\Omega$ and $\bar v = \bar a \Omega$) 
in terms of the angular velocity $\Omega$: 
\beqn
-m\gamma^2 v\Omega 
&=& 4 m\bmm\gamma^2\bgamma \Omega^2
\frac1{(\varphi+v\bv\sin\varphi)^2}
\left\{(1+v\bv\cos\varphi)\bv(\varphi\cos\varphi-v^2\sin\varphi)
+\frac12 v(1-\bv^2)(\varphi+v\bv\sin\varphi)\right.
\nonumber\\
&&- \left.\frac12\big[\bv\sin\varphi(\varphi+v\bv\sin\varphi)
+ (1+v\bv\cos\varphi)(v+\bv\cos\varphi)
- \frac{v}{1-v^2}(\varphi+v\bv\sin\varphi)^2
\big]\Phi(\varphi,v,\bv)
\right\}, 
\label{eq:eom}\eeqn
\beqn
-\bmm\bgamma^2 \bv\Omega 
&=& 4 m\bmm\gamma\bgamma^2 \Omega^2
\frac1{(\varphi+v\bv\sin\varphi)^2}
\left\{(1+v\bv\cos\varphi)v(\varphi\cos\varphi-\bv^2\sin\varphi)
+\frac12 \bv(1-v^2)(\varphi+v\bv\sin\varphi)\right.
\nonumber\\
&&- \left.\frac12\big[v\sin\varphi(\varphi+v\bv\sin\varphi)
+ (1+v\bv\cos\varphi)(\bv+v\cos\varphi)
- \frac{\bv}{1-\bv^2}(\varphi+v\bv\sin\varphi)^2
\big]\Phi(\varphi,v,\bv)
\right\}.
\label{eq:eombar}\eeqn

The expression for the angular momentum (\ref{eq:angeta}) is integrated 
by substituting the equations of motion Eq.~(\ref{eq:eometa1}) 
and (\ref{eq:eometa2}) into (\ref{eq:angeta}), 
\beqn
L &=& 
\frac{4m\bmm\gamma\bgamma v\bv}{\Omega^2}\intinf d\eta\,
\dl(w)(1-v\bv\cos\eta)\eta\sin\eta
\nonumber\\
&&- 
\frac{4m\bmm\gamma\bgamma}{\Omega^4}\intinf d\eta\,
\dl'(w)\,\tPhi(\eta,v,\bv)\,
(v^2+\bv^2-2v\bv\cos\eta-v\bv\eta\sin\eta)
\\
&=& 
-\frac{2m\bmm\gamma\bgamma}{\Omega^2}\intinf d\eta\,
\left[2w\dl'(w)+\dl(w)\right]
\tPhi(\eta,v,\bv)
\label{eq:angsol1}
\\
&=& 
\frac{2m\bmm\gamma\bgamma}{\Omega^2}\intinf d\eta\,
\dl(w)\,\tPhi(\eta,v,\bv) = 2m\bmm\gamma\bgamma\,\Phi(\varphi,v,\bv).
\eeqn
Explicitly, 
\be 
 L= 2m\bmm\gamma\bgamma\,\frac{(1+v\bv\cos\varphi)^2-\frac12(1-v^2)(1-\bv^2)}
 {\varphi+v\bar v\sin\varphi},  
\label{eq:angsol}
\ee 
with $\varphi$ given by Eq. (\ref{eq:phi}).

To compute the energy $E$, we first compute $E-\Omega L$: 
\beqn
E-\Omega L &=& 
m\gamma(1-v^2)+\bmm\bgamma(1-\bv^2)
\nonumber\\
&-&\frac{4m\bmm\gamma\bgamma}{\Omega}\intinf d\eta\,
\dl(w)\left[\tPhi(\eta,v,\bv) +(1-v\bv\cos\eta)v\bv\eta\sin\eta \right]
\nonumber\\
&-&\frac{4m\bmm\gamma\bgamma}{\Omega^3}\intinf d\eta\,
\dl'(w)\,\eta\big[(\pi-\eta)+v\bv\sin\eta\big]\,\tPhi(\eta,v,\bv)
\\
&=& m\gamma(1-v^2)+\bmm\bgamma(1-\bv^2)
-\frac{2m\bmm\gamma\bgamma}{\Omega}\intinf d\eta\,
\dl(w)\,\tPhi(\eta,v,\bv)
\\
&=& m\gamma(1-v^2)+\bmm\bgamma(1-\bv^2)
-2m\bmm\gamma\bgamma\Omega\,\Phi(\varphi,v,\bv).
\label{eq:eolsol}
\eeqn
Using the definitions (\ref{eq:normgf}) of $\gamma$ and $\bar\gamma$,  
the expression for $E$ takes the simple form, 
\beq
E= \frac{m}{\gamma}+\frac{\bmm}{\bgamma}
= m(1-v^2)^{\frac12}+ \bmm(1-\bv^2)^{\frac12}. 
\label{eq:enesol}
\eeq
As noted in the introduction, this expression for $E$ is identical to 
the expression for $E$ in the {\em electromagnetic} two-body case derived 
by Schild \cite{sc63}, and to that for a scalar field, obtained in 
Appendix~\ref{sec:scalar}.  
To clarify the structure of relevant equations used in calculations 
above, their formal expressions are presented in Appendix \ref{sec:eqsEL}.

The corresponding expressions for angular momentum are each 
proportional to the potential dotted on each free index with 
the helical symmetry vector $k^\alpha$. For a scalar interaction, 
masses $m$ and $\bar m$ have scalar charges $q$ and $\bar q$. A scalar 
field $\psi$ at $m$ due to $\bar q$ satisfies 
\be
\Box \psi = -4\pi \bar q\int d\bar\tau \delta^4[x-\bar x(\bar\tau)], 
\ee
with the corresponding definition of $\bar\psi$.
Similarly, for an electromagnetic interaction with charges $e$ and $\bar e$, the 
vector potential $A_\alpha$ at $m$ due to $\bar e$ satisfies (in the Lorentz 
gauge) 
\be
\Box A_\alpha 
= -4\pi \bar e\int d\bar\tau\, \bar u_\alpha(\bar\tau)\,\delta^4[x-\bar x(\bar\tau)]. 
\ee   
With $\psi$, $A_\alpha$ and $h_{\alpha\beta}$ evaluated at $x$, 
$\bar\psi$, $\bar A_\alpha$ and $\bar h_{\alpha\beta}$ evaluated at $\bar x$, 
the angular momentum $L$ has the following forms for scalar, electromagnetic,
and post-Minkowskian gravitational interactions: 
\beqn
\mbox{Scalar charges $q$ and $\bar q$}: \qquad L 
&=&  -\frac q\gamma\psi = -\frac {\bar q}{\bar\gamma}\bar\psi
= \frac{q^2}{\gamma\bar\gamma}\frac1{\varphi+v\bar v\cos\varphi}\\
\mbox{Electromagnetic charges $e$ and -$\bar e$}: \qquad L 
&=&  -\frac e\Omega A_\alpha k^\alpha = -\frac {\bar e}\Omega \bar A_\alpha \bk^\alpha 
= e\bar e \frac {1+v\bar v\cos\varphi}{\varphi+v\bar v\cos\varphi},\\
\mbox{Post-Minkowski masses $m$ and $\bar m$}:\qquad L 
&=&  \frac {m\gamma}{2\Omega} h_{\alpha\beta} k^\alpha k^\beta
=  \frac {\bar m\bar\gamma}{2\Omega} \bar h_{\alpha\beta} \bk^\alpha \bk^\beta.
\eeqn

\subsubsection{Equations of motion, energy and angular momentum 
for the affinely parametrized model}

The analogous computations for the affinely parametrized formulation
are outlined here.  
Performing the integrals in the radial components of the equations of motions 
(\ref{eq:eomlgr1}) and (\ref{eq:eomlgr2}), we obtain   
\beqn
-m\gamma^2 v\Omega 
&=& 4 m\bmm\gamma^2\bgamma \Omega^2
\frac1{(\varphi+v\bv\sin\varphi)^2}
\left\{(1+v\bv\cos\varphi)\bv(\varphi\cos\varphi-v^2\sin\varphi)
+\frac12 v(1-\bv^2)(\varphi+v\bv\sin\varphi)\right.
\nonumber\\
&&- \left.\frac12\big[\bv\sin\varphi(\varphi+v\bv\sin\varphi)
+ (1+v\bv\cos\varphi)(v+\bv\cos\varphi)
%%- \frac{v}{1-v^2}(\varphi+v\bv\sin\varphi)^2
\big]\Phi(\varphi,v,\bv)
\right\}, 
\label{eq:eomaffine}\eeqn
\beqn
-\bmm\bgamma^2 \bv\Omega 
&=& 4 m\bmm\gamma\bgamma^2 \Omega^2
\frac1{(\varphi+v\bv\sin\varphi)^2}
\left\{(1+v\bv\cos\varphi)v(\varphi\cos\varphi-\bv^2\sin\varphi)
+\frac12 \bv(1-v^2)(\varphi+v\bv\sin\varphi)\right.
\nonumber\\
&&- \left.\frac12\big[v\sin\varphi(\varphi+v\bv\sin\varphi)
+ (1+v\bv\cos\varphi)(\bv+v\cos\varphi)
%%- \frac{\bv}{1-\bv^2}(\varphi+v\bv\sin\varphi)^2
\big]\Phi(\varphi,v,\bv)
\right\}.
\label{eq:eombaraffine}\eeqn

The angular momentum turns out to have the same form as in the 
parametrization-invariant formulation, 
\beqn
L &=& 
2m\bmm\gamma\bgamma\,\Phi(\varphi,v,\bv)
\,=\, 
2m\bmm\gamma\bgamma\,\frac{(1+v\bv\cos\varphi)^2-\frac12(1-v^2)(1-\bv^2)}
{\varphi+v\bar v\sin\varphi}. 
\label{eq:angsolaffine}
\eeqn
We find, however, that the form of $E-\Omega L$ differs from the corresponding 
Eq.~(\ref{eq:eolsol}):
\beqn
E-\Omega L = 
m\gamma(1-v^2)+\bmm\bgamma(1-\bv^2)
- 6m\bmm\gamma\bgamma\Omega\,\Phi(\varphi,v,\bv).  
\label{eq:eolsolaff}
\eeqn
The normalizations (\ref{eq:normaf}) have the explicit form  
\beqn
\left(\eta_{\albe} + h_{\albe}\right)\dtx^\alpha\dtx^\beta 
= -\gamma^2(1-v^2) + 4\bmm\gamma^2\bgamma\Omega\,\Phi(\varphi,v,\bv)
= -1,
\\
\left(\bareta_{\albe} + \bhh_{\albe}\right)\dtbx^\alpha\dtbx^\beta 
= -\bgamma^2(1-\bv^2) + 4m\gamma\bgamma^2\Omega\,\Phi(\varphi,v,\bv)
= -1.  
\eeqn
Substituting these in Eq.~(\ref{eq:eolsolaff}), 
the expression for the energy becomes  
\beq
E = \frac{m}{\gamma}+\frac{\bmm}{\bgamma}
+ 4m\bmm\gamma\bgamma\Omega\,\Phi(\varphi,v,\bv)
= m\gamma(1-v^2)+\bmm\bgamma(1-\bv^2)
- 4m\bmm\gamma\bgamma\Omega\,\Phi(\varphi,v,\bv). 
\eeq
As implied by Eqs.~(\ref{eq:angsolaffine}) and (\ref{eq:eolsolaff}), 
the form of $E$ differs from the energy (\ref{eq:enesol}) of 
the parametrization-invariant action.  Both, of course, agree in the Newtonian limit.

\section{First law of thermodynamics for binaries described by 
Fokker actions}
\label{sec:1stlaw}

The first law of thermodynamics governs nearby equilibria of conservative 
systems.  For binary systems with circular orbits, an equilibrium solution 
is a solution that is stationary in a rotating frame, a solution with a helical 
symmetry vector $k^\alpha$. When internal degrees of freedom 
(e.g., baryon number, entropy, vorticity) are fixed, the first law relates 
the change of energy to the change of angular momentum in the manner \cite{fus02}      
\be 
	\delta E = \Omega \delta L. 
\label{eq:1stlaw}\ee
In the present context, however, the presence of a radiation field whose energy 
is infinite makes the relation suspect; and the lack of a true action 
means that the simple Hamiltonian proof for point particles 
with a Newtonian potential does not hold.  But the relation (\ref{eq:1stlaw}) 
is true, and its proof is an extension of the Hamiltonian proof that uses 
the parametrization invariance of the relativistic Fokker action. 

\noindent{\sl Proposition}.  Consider a parametrization-invariant 
Fokker action of the form (\ref{eq:action}).  Suppose there is 
a family of solutions $x^\alpha(s,\tau), \bar x^\alpha(s,\bar\tau)$ 
for which the particles move in circular orbits with angular velocity 
$\Omega(s)$, with spacetime trajectories along the helical vector 
field $k^\alpha(s) = t^\alpha + \Omega(s)\phi^\alpha$.  Then
$\delta E = \Omega \delta L$, where 
\dis\delta Q:= \left.\frac {dQ}{ds}\right|_{s=0}$.    
%$\Lambda(x,\dot x, \bar x, \dot{\bar x})$.\\

\noindent{\sl Parametrization invariance and constraints}.

As a prerequisite to the proof, we begin with a brief discussion of 
parametrization invariance of an Fokker action of the form (\ref{eq:action}) 
and the constraints that follow from it. These are analogs of the 
Hamiltonian constraint (vanishing of the superhamiltonian) associated 
with a true parametrization-invariant action. A parametrization   
$\tau \rightarrow f(\tau)$ maps the path $x^\alpha(\tau)$ to the 
path $X^\alpha = x^\alpha\circ f$. Similarly,
$\bar\tau \rightarrow \bar f(\bar\tau)$ maps 
$\bar x^\alpha$ to $\bar X^\alpha = \bar x^\alpha\circ \bar f$. 
Invariance of the action, 
\[ 
I = -\int_{\tau_1}^{\tau_2}d\tau\ m(-\dtx_\alpha \dtx^\alpha)^{1/2}
    -\int_{\bar\tau_1}^{\bar\tau_2}d\bar\tau\ 
    		\bar m(-\dtbx_\alpha \dtbx^\alpha)^{1/2}
    +\int_{\tau_1}^{\tau_2}d\tau\int_{\bar\tau_1}^{\bar\tau_2}d\bar\tau 
    	\ \Lambda(x-\bx,\dot x, \dot{\bar x}),
%%koji    	\ \Lambda(w,\dot x, \dot{\bar x}),
\]
under reparametrization follows from the scaling      
\be
 \left.\Lambda(X-\bar X,\dot X^\alpha, \dot{\bar X}{}^\alpha)
%%koji \left.\Lambda[(X-\bar X)^2,\dot X^\alpha, \dot{\bar X}{}^\alpha]
	\right|_{\tau_0,\bar\tau_0}
	= \dot f(\tau_0)\dot{\bar f}(\bar\tau_0)
	  \left.\Lambda(x-\bar x,\dot x^\alpha, \dot{\bar x}{}^\alpha)
%%koji	  \left.\Lambda[(x-\bar x)^2,\dot x^\alpha, \dot{\bar x}{}^\alpha]
	  \right|_{f(\tau_0),\bar f(\bar\tau_0)}.
\ee
For $f(\tau)=k\tau$, $\bar f(\bar\tau)=\bar \tau$, we have 
\[
\left.\Lambda(X-\bar X,\dot X^\alpha, \dot{\bar X}{}^\alpha)
%%koji \left.\Lambda[(X-\bar X)^2,\dot X^\alpha, \dot{\bar X}{}^\alpha]
	\right|_{\tau_0/k,\ \bar\tau_0}
= \left.\Lambda(x-\bar x,k\dot x^\alpha, \dot{\bar x}{}^\alpha)
%%koji = \left.\Lambda[(x-\bar x)^2,k\dot x^\alpha, \dot{\bar x}{}^\alpha]
  \right|_{\tau_0,\bar\tau_0}
= k\left.\Lambda(x-\bar x,\dot x^\alpha, \dot{\bar x}{}^\alpha)
%%koji = k\left.\Lambda[(x-\bar x)^2,\dot x^\alpha, \dot{\bar x}{}^\alpha]
  \right|_{\tau_0,\bar\tau_0}.
\] 
Then the relation,
\[
\left.\frac d{dk}\left[ k\Lambda(x-\bx,\dot x,\dot {\bar x})\right]\right|_{k=1}
= \left.\frac d{dk}\left[\Lambda(x-\bx,k\dot x,\dot {\bar x})\right]\right|_{k=1},
%%koji \left.\frac d{dk}\left[ k\Lambda(w,\dot x,\dot {\bar x})\right]\right|_{k=1}
%%koji = \left.\frac d{dk}\left[\Lambda(w,k\dot x,\dot {\bar x})\right]\right|_{k=1},
\]
and its barred $\leftrightarrow$ unbarred counterpart imply 
\be 
  \Lambda = \dot x^\alpha\frac{\partial\Lambda}{\partial \dot x^\alpha}, \qquad
  \Lambda = \dtbx^\alpha\frac{\partial\Lambda}{\partial \dtbx^\alpha}.
\label{eq:homofn}
\ee

\noindent{\sl Proof of Proposition}. 

The equations of motion involve the potentials 
\be
{\cal U}_m = \int_{-\infty}^\infty d\bar\tau \Lambda,
\qquad 
{\cal U}_{\bar m} = \int_{-\infty}^\infty d\tau \Lambda.
\ee
That is, with 1-particle Lagrangians ${\cal L}_m$ and ${\cal L}_{\bar m}$ defined 
by 
\be 
{\cal L}_m = -m(-\dot x^\alpha\dot x_\alpha)^{1/2} + {\cal U}_m, 
\qquad
{\cal L}_{\bar m} = -\bar m(-\dtbx^\alpha\dtbx_\alpha)^{1/2} + {\cal U}_{\bar m},
\ee
Eqs. (\ref{eq:eom1}) and (\ref{eq:eom2}) are the equations 
of motion for the actions
\be 
I_m = \int d\tau {\cal L}_m, \qquad I_{\bar m} = \int d\bar\tau {\cal L}_{\bar m},
\ee
namely,
\be 
\frac d{d\tau}\left(\frac{\partial {\cal L}_m}{\partial \dot x^\alpha}\right)
  -\frac{\partial {\cal L}_m}{\partial x^\alpha}=0,
  \qquad
\frac d{d\bar\tau}\left(\frac{\partial {\cal L}_{\bar m}}{\partial\dot{\bar x}{}^\alpha}\right)
  -\frac{\partial {\cal L}_{\bar m}}{\partial {\bar x}^\alpha}=0. 
\ee

The one-particle 4-momentum $p_\alpha$ associated with the 
action $I_m$ is 
\be
p_\alpha = \frac{\partial {\cal L}_m}{\partial \dot x^\alpha} 
	= \frac{m\dot x_\alpha}{(-\dot x^\gamma\dot x_\gamma)^{1/2}}
	  + \frac{\partial {\cal U}_m}{\partial \dot x^\alpha}. 
\label{eq:onepmom}
\ee
Because ${\cal L}_m$ and $\dot x^\alpha$ have the same reparametrization 
scaling, namely 
\be
	{\cal L}_m \rightarrow \dot f {\cal L}_m, \qquad  
\dot x^\alpha\rightarrow \dot f\dot x^\alpha,
\ee
the momentum $p_\alpha$ is independent of the choice of 
parameter.

Let $t$ be a choice of Minkowski time, $t^\alpha$ the corresponding 
Killing vector ($\partial_t$), and $\phi^\alpha$ a rotational Killing 
vector orthogonal to $t^\alpha$.  The 1-particle energies associated 
with $I_m$ and $I_{\bar m}$ are 
\be
 E_m  = -t^\alpha p_\alpha
	=- \frac{\partial {\cal L}_m}{\partial\dot t} = \frac{m\dot t}{(-\dot x^\gamma\dot x_\gamma)^{1/2}}
	  - \frac{\partial {\cal U}_m}{\partial \dot t},
\label{eq:onepenergy1}
\ee
\be
   E_{\bar m} = -t^\alpha \bar{p}_\alpha
   		=- \frac{\partial {\cal L}_{\bar m}}{\partial\dot {\bar t}}
   	      = \frac{\bar m\dot {\bar t}}
   		{(-\dot{\bar x}{}^\gamma\dot {\bar x}_\gamma)^{1/2}}
	  	- \frac{\partial {\cal U}_{\bar m}}{\partial \dot {\bar t}};	 
\label{eq:onepenergy2}
\ee
and the 1-particle angular momenta are 
\be 
L_{\rm m} = \phi^\alpha p_\alpha
	= \frac{\partial {\cal L}_m}{\partial\dot \phi}, 
\qquad 
L_{\rm \bar m} = \phi^\alpha \bar{p}_\alpha
	= \frac{\partial {\cal L}_{\bar m}}{\partial\dot {\bar \phi}}.
\label{eq:onepangmom}
\ee
We first use the scaling relation and the equations of motion to show 
that these 1-particle momenta and angular momenta satisfy
$\delta E_m = \Omega \dl L_m$, $\delta E_{\bar m} = \Omega\delta L_{\bar m}$. 
The scaling of ${\cal L}_m$ and ${\cal L}_{\bar m}$ (parametrization invariance of $I_m$) 
implies 
\be 
{\cal L}_m = \dot x^\alpha\frac{\partial {\cal L}_m }{\partial \dot x^\alpha}, \qquad
{\cal L}_{\bar m} = \dtbx^\alpha\frac{\partial {\cal L}_{\bar m}}{\partial \dtbx^\alpha}.
\label{eq:homofnLag}
\ee
(In other words, the 1-particle superhamiltonians vanish: 
${\cal H}_m := \dot x^\alpha p_\alpha - {\cal L}_m = 0,\quad 
\bar{\cal H}_{\bar m} := \dot {\bx}^\alpha \bar p_\alpha - {\cal L}_{\bar m} = 0
$.) 
Then
\be 
  E_m = \frac{\dot x^a}{\dot t}\frac{\partial {\cal L}_m }{\partial \dot x^a}-\frac{{\cal L}_m}{\dot t}
      = v^a\frac{\partial {\cal L}_m }{\partial \dot x^a}-\frac{{\cal L}_m}{\dot t}. 	 
\ee 
(Note that ${\cal L}_m/\dot t$ is the form of the Lagrangian appropriate 
to Minkowski time: $\int dt {\cal L}_m/\dot t = \int d\tau {\cal L}_m$.)
Consider now a family of solutions to the equation of motion for
$m$ with circular orbits, each solution stationary in a comoving frame.
Neighboring orbits satisfy 
\be 
\delta E_m = \delta v^a \frac{\partial {\cal L}_m }{\partial \dot x^a} 
		+ v^a \delta \frac{\partial {\cal L}_m }{\partial \dot x^a} 
		-\delta v^a \frac{\partial {\cal L}_m }{\partial \dot x^a}
		- \delta x^a \frac{\partial {\cal L}_m }{\partial x^a}\frac1{\dot t}
	   = v^a \delta \frac{\partial {\cal L}_m }{\partial \dot x^a}
	   = \Omega \delta L_m,
\ee
where the equilibrium condition $\frac{\partial {\cal L}_m }{\partial \varpi}=0$ was 
used to infer $ \delta x^a \frac{\partial {\cal L}_m }{\partial x^a}=0$.

The total energy $E$ of Eq.~(\ref{eq:linmom}) is not, however, the sum 
of the 1-particle energies: $E_m + E_{\bar m}$ does not include 
the field energy and is not, in general, conserved.
Instead, the total energy has the form
\be
E = - P_\alpha t^\alpha 
   = E_m+ E_{\bar m} 
   -\int_{-\infty}^\tau d\tau\, t^\alpha \nabla_\alpha {\cal U}_m 
   -\int_{-\infty}^{\bar\tau} d\tau\, 
 			t^\alpha\bar\nabla_\alpha{\cal U}_{\bar m}.
\label{eq:em}\ee

The total angular momentum is similarly 
\be
L = P_\alpha\phi^\alpha 
= L_m + L_{\bar m} 
 + \int_{-\infty}^\tau d\tau\,\phi^\alpha \nabla_\alpha {\cal U}_m 
 + \int_{-\infty}^{\bar\tau} d\btau\,
 	\phi^\alpha\bar\nabla_\alpha{\cal U}_{\bar m},
\label{eq:lm}\ee
where we have used the notation 
\be 
 \phi^\alpha \nabla_\alpha {\cal U}_m
 : = \frac{\partial {\cal U}_m}{\partial\phi}
 = \frac{\partial {\cal U}_m}{\partial x^\alpha}
   \frac{\partial x^\alpha}{\partial\phi}
   + \frac{\partial {\cal U}_m}{\partial \dot x^\alpha}
   \frac{\partial \dot x^\alpha}{\partial\phi}
  = \phi^\alpha\frac{\pa\cUm}{\pa x^\alpha}
+\dtphi^\alpha\frac{\pa\cUm}{\pa \dtx^{\alpha}}. 
\eeq
(See also Eq.~(\ref{eq:gradu}) of Appendix~\ref{sec:formulaPL}.)

To recover the first law, one uses the fact that $E$ and $L$ are 
independent of $\tau$ and $\bar\tau$, setting $\tau$ and $\bar\tau$ 
to $- \infty$ to eliminate the final term: 
\begin{eqnarray}
\delta E &-& \Omega\delta L 
=\lim_{\tau,\bar\tau\rightarrow -\infty}(\delta E - \Omega\delta L)
\nonumber\\
&=&\lim_{\tau,\bar\tau\rightarrow -\infty}\left[
  \delta E_m - \Omega \delta L_m + \delta E_{\bar m} -\Omega\delta L_{\bar m}
 +\int_{-\infty}^\tau d\tau k^\alpha \nabla_\alpha \delta{\cal U}_m 
+ \int_{-\infty}^{\bar\tau} d\btau k^\alpha\bar\nabla_\alpha\delta{\cal U}_{\bar m} 
\right]\nonumber\\
&=& 0.\quad \Box
\label{eq:law1}\end{eqnarray} 
Note that the Killing symmetry, $k^\alpha\na_\alpha{\cal U}_m =0$, does not in itself 
imply $k^\alpha\na_\alpha\delta{\cal U}_m =0$, because 
$\delta k^\alpha = \phi^\alpha\delta\Omega  \neq 0$.\\  

\noindent {\sl First law for affinely parametrized action}\\

The first law also holds for the affinely parametrized action (\ref{eq:actionlgr}).  
The proof is similar to that for the parametrization-invariant action, with 
one-particle Lagrangians now defined by 
\be 
{\cal L}_m = \frac m2 \dot x^\alpha\dot x_\alpha + {\cal U}_m, 
\qquad
{\cal L}_{\bar m} = \frac{\bar m}2 \dot{\bar x}{}^\alpha\dot{\bar x}{}_\alpha + {\cal U}_{\bar m},
\ee
with the affinely parametrized $\Lambda$ replacing the parametrization-invariant 
$\Lambda$ in the definition of ${\cal U}_m$ and ${\cal U}_{\bar m}$.
Note that affine parametrization is equivalent to the conditions    
\be
  2{\cal L}_m = -m, \qquad 2{\cal L}_{\bar m}=-\bar m,
\label{eq:aflaw1}\ee
while the fact that ${\cal L}_m$ and ${\cal L}_{\bar m}$ are quadratic in the velocities $\dot x$ and $\dot{\bar x}$ 
implies
\be
2{\cal L}_m = p_\alpha \dot x^\alpha, \qquad 
2{\cal L}_{\bar m}= \bar{p}_\alpha \dot{\bar x}{}^\alpha.
\label{eq:aflaw2}\ee
The first of these relations, (\ref{eq:aflaw1}), together with the equilibrium condition $\partial{\cal L}_m/\partial\varpi=0$, implies 
\be
 \delta {\cal L}_m = \frac{\partial{\cal L}_m}{\partial \dot x^\alpha}\delta \dot x^\alpha 
		   = p_\alpha \delta \dot x^\alpha.
\label{eq:aflaw3}\ee 
From the second relation, (\ref{eq:aflaw2}), we have 
\dis 
0 =-\delta m =\delta (p_\alpha \dot x^\alpha)
 = \delta p_\alpha\, \dot x^\alpha + p_\alpha\, \delta \dot x^\alpha
$,
whence, by (\ref{eq:aflaw3}), 
\be 
 \dot x^\alpha \dl p_\alpha =0.
\ee
Again writing $E_m = -p_t$, $\dot x^a = u^t v^a$, we have 
\be 
	\delta E_m = v^a\delta p_a = \Omega \delta L_m.
\ee
Finally, the relation 
\be 
\delta E = \Omega \delta L
\ee
follows as before from Eqs. (\ref{eq:em})-(\ref{eq:law1}), which hold as 
written.

\section{Action accurate to first post-Newtonian order}
\label{sec:pn}
In a first post-Newtonian approximation, the equations of motion 
have corrections of order $v^2/c^2$ to their Newtonian terms.
As we noted above, the first post-Minkowski approximation, by including 
only terms linear in $m$ and $\bar m$, fails to be accurate to 
first post-Newtonian order. The omitted terms in the equation of motion 
are quadratic in the masses $m$ and $\bar m$, and they arise from a 
single term in the post-Newtonian Lagrangian, namely
\be
	L_{\rm PN} = -\frac{m\bar m(m+\bar m)}{2R^2}
\label{eq:lpn}\ee
We can make the Fokker action accurate to first post-Newtonian order
by adding any term that agrees to this order with (\ref{eq:lpn}), 
and we have tried two alternatives: A simplest parametrization-invariant 
choice of $\Lambda_{\rm PN}$ that reproduces (\ref{eq:lpn}) is 
\be
\Lambda _{\rm PN} 
  =-\delta \left( t-\bar{t}\right) \frac{m\bar{m}(m+\bar{m})}{2R^{2}}
     (\dtx_\gamma\dtx^\gamma\dtbx_\delta\dtbx^\delta)^{1/2},
\label{lambdapn}\ee
with value for a circular orbit given by
\be
\Lambda _{\rm PN} =-\delta \left( \eta -\pi \right) \frac{\Omega^3}2
   \frac{m\bar{m}\left(m+\bar{m}\right)}{v^{2}+\bar{v}^{2}-2v\bar{v}\cos\eta}.
\ee
A choice that is special-relativistically covariant (and parametrization
invariant) is 
\be
\Lambda_{\rm SPN} 
	=- \delta(w)\frac{m\bar{m}(m+\bar{m})}{2}  
  	  \frac{(\dtx_\gamma\dtx^\gamma
  	        \dtbx_\delta\dtbx^\delta)^{3/4}}
  	  {(R_\alpha \dtx^\alpha R_\beta \dtbx^\beta)^{1/2}}.  	   
\label{lambdaspn}\ee 

%For $N=4$, on the other hand, $\Lambda_{\rm SPN}$ is not reparametrization
%invariant, but the expression for the conserved energy retains the 
%form (AFFINE FORM?) that it has for the post-Newtonian action.

The corresponding corrections $e_{\rm PN}$ and $e_{\rm SPN}$ to the conserved 
energy and the corresponding corrections $\ell_{\rm PN}$ and $\ell_{\rm SPN}$ 
to the angular momentum are given by
\beqn
e_{\rm PN} &=& \frac12\Omega \ell_{\rm PN},
\label{epn}\\
\ell_{\rm PN} &=& -\frac{m\bar m(m+\bar m)\Omega}{\gamma\bar\gamma(v+\bar v)^2},
\label{lpn}\\
e_{\rm SPN} &=& \frac12 \Omega\ell_{\rm SPN}, 
\label{espn}\\
\ell_{\rm SPN} &=& -\frac{m\bar m(m+\bar m)\Omega}{(\gamma\bar\gamma)^{3/2}}
\frac1{(\varphi+v\bar v \sin\varphi)^2}.
\label{lspn}\eeqn

The derivation of Eqs. (\ref{epn})-(\ref{lspn}) 
and corrections to the equations of motion 
are given in Appendix \ref{sec:appn}.
 
%%%\newpage
 
\section{Numerical solution of orbital equations}
\label{sec:numer}

We have numerically solved the orbital equations, Eqs.~(\ref{eq:eom}) and
(\ref{eq:eombar}) for the parametrization-invariant model, and 
Eqs.~(\ref{eq:eomaffine}) and (\ref{eq:eombaraffine}) for the affinely 
parametrized model, finding $a(\Omega)$ 
for equal-mass particles.
Figs. \ref{fig:graph_v-rad.eps} and \ref{fig:graph_v-Ome.eps} 
show the relations for the post-Minkowski action. \\

 \begin{figure}[ht] 
%%% \centerline{\epsfysize=4cm \epsfbox{graph_v-rad.eps}}
 \includegraphics[height=50mm,clip]{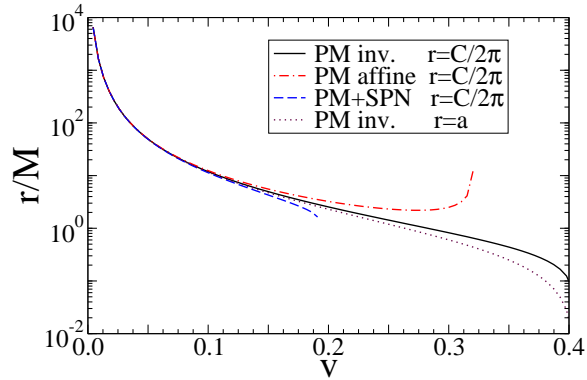}
 \caption{The velocity $v$ of a circular orbit versus the radius 
 $r$, written in the dimensionless form $r/M$, for particles 
 of equal mass $M:=m+\bmm$, where the $r$ is either the proper 
 circumferential radius $C/2\pi$ or the radial parameter $a$. 
 The motion is governed by the uncorrected parametrization-invariant 
 post-Minkowski action (PM inv.), the affine parametrized action 
 (PM affine) or the parametrization-invariant post-Minkowski action 
 with a covariant post-Newtonian correction term (\ref{lambdaspn}) 
 (PM+SPN).  For the parametrization-invariant case both $C/2\pi$ and 
 $a$ are shown. }
 \label{fig:graph_v-rad.eps} 
 \end{figure}  
 \begin{figure}[ht] 
%%% \centerline{\epsfysize=4cm \epsfbox{graph_v-Ome.eps}}
 \includegraphics[height=50mm,clip]{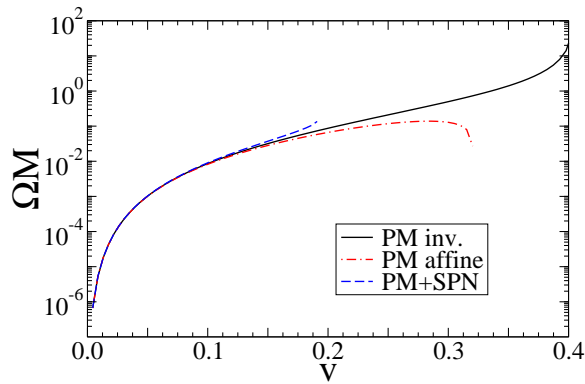}
 \caption{The corresponding relation between $\Omega$ and $v$ 
 for the orbits of Fig.~\ref{fig:graph_v-rad.eps},
 with $\Omega$ written in the dimensionless form $\Omega M$.   
 The dashed curve (PM+SPN) shows the small correction to the 
 solid curve (PM inv.) that arises 
 from adding the post-Newtonian correction term (\ref{lambdaspn}) to the 
 post-Minkowski action.}
 \label{fig:graph_v-Ome.eps} 
 \end{figure}  
 \begin{figure}[ht] 
%%% \centerline{\epsfysize=4cm \epsfbox{graph_v-Ome.eps}}
 \includegraphics[height=50mm,clip]{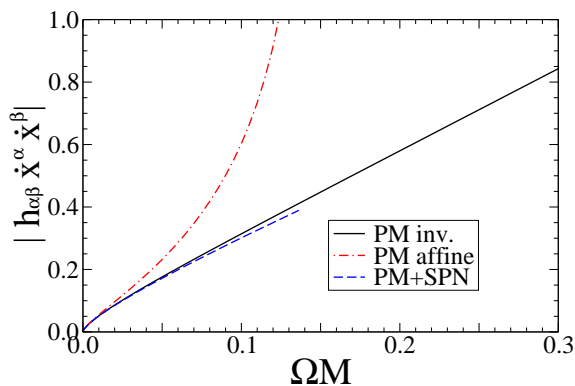}
 \caption{The corresponding relation between $\Omega$ and 
 $|h_{\alpha\beta}\dtx^\alpha\dtx^\beta|$ 
 for the orbits of Fig.~\ref{fig:graph_v-rad.eps}.   
 The dashed curve (PM+SPN) shows the first order 
 post-Newtonian correction that is nonlinear in $m$ (\ref{lambdaspn}) 
 weakens the gravitational field. }
 \label{fig:graph_Ome-h.eps} 
 \end{figure}  

As $v$ increases, relativistic beaming decreases the strength of 
the gravitational field due to $\bar m$ at the position of $m$, 
and vice-versa. In the case of scalar and electromagnetically 
bound charges, the smaller field leads to a sharply 
smaller radius for a given velocity $v$; and the same effect 
implies a larger value of $\Omega$ at fixed $v$. 
Gravity, however, has 
the competing effect that, at small radius the field is stronger 
than the Newtonian field. In the exact theory,
the result is that, for circular orbits about a fixed mass, 
the relativistic relation between $\Omega$, $r$ (and a 
velocity defined by $v\equiv \Omega r$) is identical to 
the Newtonian relation, when $r$ is taken to be the circumferential 
radius. In the post-Minkowski models, as 
Figs.~\ref{fig:graph_v-rad.eps} and \ref{fig:graph_v-Ome.eps}, 
the outcome depends on which post-Minkowski action one chooses.  
For the parametrization-invariant action, relativistic beaming 
dominates, giving a smaller radius for the same value of $v$ and 
a correspondingly larger value of $\Omega$ at fixed $v$. 
The affine action, in contrast, gives a larger value of $r$ at 
fixed $v$, an effect so pronounced for $v \gtrsim 0.3$ that  
$r$ reaches a minimum value and then increases with increasing 
$v$.

We were also surprised by the fact that the nonlinear term in 
$m$ and $\bar m$ that is absent from our first 
post-Minkowski action has sign {\em opposite} to the 
Newtonian potential term.  Fig.~\ref{fig:graph_v-Ome.eps} and 
Fig.~\ref{fig:graph_Ome-h.eps} show only a small correction 
due to the post-Newtonian term (denoted PM+SPN in the graph), and 
the correction weakens the gravitational field.

Figures \ref{fig:graph_Ome-L.eps} and \ref{fig:graph_Ome-E.eps}
display the values of the angular momentum and energy as 
functions of $\Omega$, for the parametrization-invariant post-Minkowski 
(PM inv.) action, the affine parametrized post-Minkowski action 
(PM affine), the invariant post-Minkowski action with the special 
relativistically covariant first post-Newtonian correction 
(\ref{lambdaspn}) (PM+SPN), and the invariant post-Minkowski action 
with first post-Newtonian correction (\ref{lambdapn}) (PM+PN).  
The values of Newtonian to third post-Newtonian approximations 
(0PN - 3PN) are also plotted for references.
Again relativistic beaming appears to dominate the post-Newtonian 
correction in the PM+PN and PM+SPN action, leading to a graph in which the 
energy and angular momenta have no minima.  Because the relation 
$dE=\Omega dL$ can be used to show that the innermost stable circular 
orbit (ISCO) is a minimum of $E$ and $L$, the ISCO present in 
the post-Newtonian approximation does not appear in PM inv., 
PM+SPN or PM+PN.  
Interestingly, PM affine model has a simultaneous minima in $E(\Omega)$
and $L(\Omega)$ for a sequence of circular solutions.  
Because the relation $dE=\Omega dL$ also holds for PM affine model as shown 
in Sect.~\ref{sec:1stlaw}, the circular solutions with $\Omega M$ 
larger than the value at the minima, $\Omega M = 0.0522$, 
are likely to be dynamically unstable.     
The values at the turning point of the other quantities are 
$v = 0.184, E/M = 0.988744,$ and 
$L/M^2 = 1.0279123$.  Each number above is given to computational accuracy.
 \begin{figure}[ht] 
%%% \centerline{\epsfysize=4cm \epsfbox{graph_Ome-L.eps}}
 \includegraphics[height=50mm,clip]{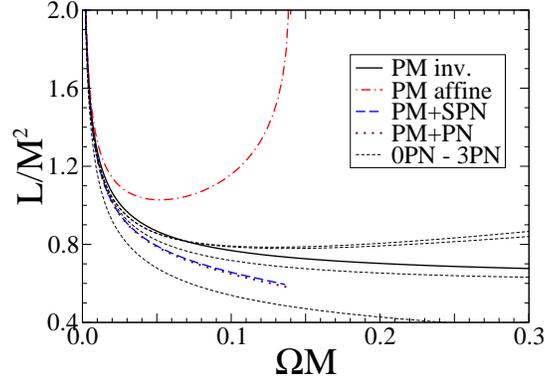}
 \caption{Angular momentum, in dimensionless form $L/M^2$, is plotted 
 against angular velocity for 8 cases. In the key labeling the curves, 
 0PN--3PN refer to models of equal-mass 
 point-particles in circular orbit in the Newtonian, 
 first post-Newtonian, second post-Newtonian, and 
 third post-Newtonian approximations from the bottom to top.}
 \label{fig:graph_Ome-L.eps} 
 \end{figure}  
 \begin{figure}[ht] 
%%% \centerline{\epsfysize=4cm \epsfbox{graph_Ome-E.eps}}
 \includegraphics[height=50mm,clip]{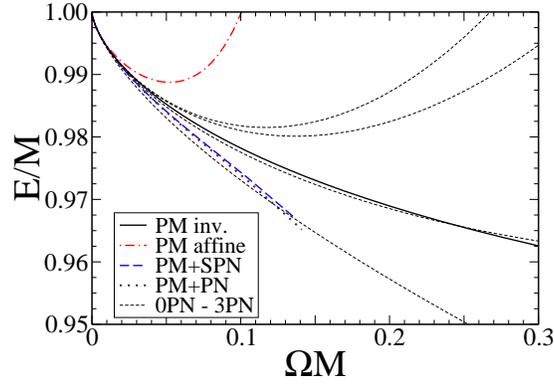}
 \caption{Energy, in dimensionless form $E/M$, is plotted against 
 angular velocity, for the models of Fig. \ref{fig:graph_Ome-L.eps}}
 \label{fig:graph_Ome-E.eps} 
 \end{figure}  

Verification of our numerical solutions can be seen from 
Fig. \ref{fig:graph_1stlaw_conv.eps}, 
in which $1-dE/\Omega dL$ is plotted against a measure $\Delta$ of the 
resolution of $dL$. As shown in Sect. \ref{sec:1stlaw}, the relation 
$dE= \Omega dL$ is exact for Fokker actions of the form of PM inv., PM affine 
PM+SPN and PM+PN, and the numerical determination finds no discrepancy.  
(The result of PM+PN is omitted in Fig.~\ref{fig:graph_1stlaw_conv.eps} 
to avoid redundancy.)\\

 \begin{figure}[ht] 
%%% \centerline{\epsfysize=5cm \epsfbox{graph_1stlaw_conv.eps}}
 \includegraphics[height=50mm,clip]{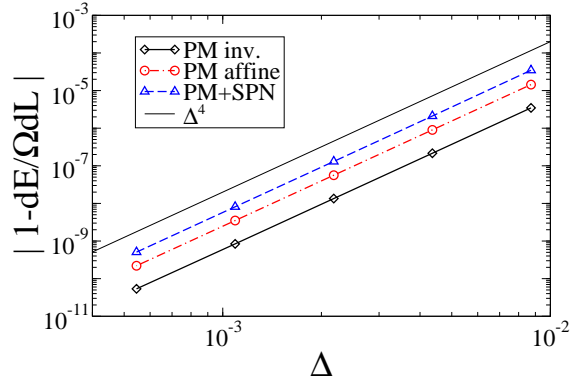}
 \caption{Accuracy of the relation $dE=\Omega dL$. Thin solid lines 
 have an inclination proportional to $O(\Delta^4)$, where 
 $\Delta$ is proportional to $dL$.  
 $dE/\Omega dL$ is evaluated at $v=0.15$ using the 4th-order 
 accurate finite difference formula (Lagrange formula).}
 \label{fig:graph_1stlaw_conv.eps} 
 \end{figure}  

\section{Discussion}
\label{sec:discussion}

In the previous section, we discussed numerical results for the 
special case of gravitationally interacting particles in circular orbits.  
Here we consider a few surprises that 
arise in our more general study of Fokker actions.  

Conserved quantities for particles with scalar, electromagnetic 
or gravitational interactions are ordinarily written as an integral 
over the field and its first time derivative on a spacelike hypersurface, 
together with sum of terms involving the position and velocity of each 
particle at its position on the hypersurface. In a Fokker formalism, 
the integral over the field is replaced by a sum of integrals over 
the trajectory of each particle.  A striking feature of the 
resulting conserved 4-momentum and angular momentum is that they 
break up into a sum of quantities that are {\em separately conserved  
for each particle}, as in Eqs.~(\ref{eq:em}) and (\ref{eq:lm}) 
(or (\ref{eq:energyapxlaw}) and (\ref{eq:angmomapxlaw})).
In these equations, the separate conservation of each contribution follows  
from the fact that the contribution associated with each particle 
depends only on the proper time of that particle.    
How is this possible, when the field energy measures 
the {\em interaction} between the fields produced by each particle; 
how is it consistent with the fact that the total 4-momentum and 
angular momentum are, in general, an exhaustive set of integrals 
of the motion?  

The answer is that the individual integrals that appear 
in the sum cannot be written as integrals over a hypersurface 
of a density locally constructed from the field and its first 
time derivative.  They therefore do not represent new integrals 
of the motion.  In fact, their existence is really a trivial 
consequence of the equation of motion satisfied by each particle:
\be
  \frac d{d\tau}\left(\frac{\pa{\cal L}_m}{\pa\dot x^\alpha}\right)
     - \frac{\pa{\cal L}_m}{\pa x^\alpha}=0
\label{eq:eomEL}
\ee 
where as before 
\be
{\cal L}_m := -m(-\dtx^\alpha\dtx_\alpha)^{1/2} + \cUm, \qquad
\cUm := \intinf d\btau\Lambda.
\ee 
Integrating the equation of motion from $\tau_1$ to $\tau_2$,
we have
\be
p_\alpha{\big |}_{\tau_1}^{\tau_2} 
   - \int_{\tau_1}^{\tau_2}d\tau \frac{\pa\cUm}{\pa x^\alpha}
   = 0,  \qquad
p_\alpha:=\frac{\pa{\cal L}_m}{\pa\dot x^\alpha}
\ee  
implying $P_{{\rm m}\alpha}$ is independent of $\tau$, where
\be
P_{{\rm m}\alpha}(\tau) :=  p_\alpha(\tau)
		- \int_{-\infty}^\tau d\tau \frac{\pa\cUm}{\pa x^\alpha}
\ee

More generally, dotting a vector $\zeta^\alpha$ 
into the equation of motion (\ref{eq:eomEL}) yields the identity 
\beq
\frac{d}{d\tau}\left[
\frac{m \dtx_\alpha\zeta^\alpha}{(-\dtx_\beta\dtx^\beta)^{\frac12}}
+\zeta^\alpha\frac{\pa\cUm}{\pa \dtx^\alpha} 
\right]
- 
\left(\zeta^\alpha\frac{\pa\cUm}{\pa x^\alpha}
+\dot \zeta^\alpha\frac{\pa\cUm}{\pa \dtx^{\alpha}}\right)
-\frac{m\dtx_\alpha \dot\zeta^\alpha}{(-\dtx_\gamma\dtx^\gamma)^{1/2}}=0.  
\eeq
Then, because a Killing vector $\zeta^\alpha$ of Minkowski space 
satisfies $\dtx^\alpha \dot\zeta_\alpha 
= \dtx^\alpha\dtx^\beta \na_\alpha \zeta_\beta = 0$, 
the quantity ${\cal Q}_m$ defined by 
\beq
{\cal Q}_m=\left[
\frac{m \dtx_\alpha\zeta^\alpha}{(-\dtx_\beta\dtx^\beta)^{\frac12}}
+\zeta^\alpha\frac{\pa\cUm}{\pa \dtx^\alpha}
\right](\tau)
- \int_{-\infty}^{\tau}d\tau
\left(\zeta^\alpha\frac{\pa\cUm}{\pa x^\alpha}
+\dot \zeta^\alpha\frac{\pa\cUm}{\pa \dtx^{\alpha}}\right)
\eeq
is independent of $\tau$. 
In particular, for $\zeta^\alpha=-t^\alpha$ and $\zeta^\alpha=\phi^\alpha$, 
${\cal Q}_m$ is 
the one-particle contribution to the energy and angular momentum, respectively.
The generalization of these relations to an $n$-particle action is immediate.

  The conserved quantities that arise from a Fokker action 
involve integrals along the particle paths whose integrands are 
not perfect time derivatives: That is, the integrals are path 
dependent.\footnote{
One can similarly construct a path dependent conserved ``momentum''
for a single Newtonian particle moving in any external field
by writing
$$ d{\bf p}/dt = - \nabla U(r) $$
$$ {\bf p} + \int_{t_0}^{t} dt \nabla U = constant.  $$
Because the integral depends on the path $r(t)$ from $t_0$ to $t$,
one does not have a first integral of the equation of motion, in the sense
of a conserved quantity that depends only on the position and
velocity of the particle.}
Although the total 4-momentum and angular momentum are also expressed 
as path-dependent integrals in the Fokker formalism, they, presumably, 
differ from their 1-particle constituents by the fact that they can 
be expressed as integrals over a hypersurface of a local function 
of the field and its first time derivative (together with the 
kinematic momentum of each particle).  The parts of these conserved 
quantities that directly involve the fields -- the sum of terms involving 
${\cal U}_m$ for each $m$ -- is not, however, the full field momentum 
or angular momentum, 
\be 
\int T_\alpha^{\ \beta} \zeta^\alpha dS_\beta.
\ee
Instead, only contributions to the field momentum (or angular momentum) 
that arise from the product of fields from different particles 
are present in the Fokker momentum (angular momentum).  
Terms quadratic in the field of a single particle must be
absorbed in the renormalized mass.

This analysis leads to a conjecture for why the conserved 
Fokker momentum and angular momentum are finite for bound systems.  
When terms quadratic in the fields of individual particles are subtracted, 
only a finite remainder survives.

Finally, we comment on ambiguities associated with the choice of
gauge and with the choice of post-Newtonian correction terms.
Solutions for two-particle circular orbits presented in this paper
are obtained in the deDonder (harmonic) gauge.  A sequence of
equal mass solutions terminates near $v\sim 0.40$ for the parametrization
invariant action, and near $v\sim 0.32$ for the affine action,
because in each case the equations of motion have no solution for $v$ 
larger than a critical value $v$.
When post-Newtonian corrections are added to the parametrization-invariant
action, we find solutions only up to
$v\sim 0.19$.  When one proceeds, for example, to higher order
calculation aiming to identify an ISCO, which may be expected to
appear near these terminal values, a choice of gauge may be
reconsidered with care.

A post-Minkowski action corrected by the terms  
$\Lambda_{\rm SPN}$ or $\Lambda_{\rm PN}$,
does not exactly reproduce the first post-Newtonian
equations of motion, because the post-Minkowski action 
contains terms of all post-Newtonian orders in the velocities.
A more systematic approximation to the higher order post-Minkowski 
Fokker action may be obtained by deriving a form of interaction term 
that agrees with a formal expansion of the stress-energy tensor
(see, for example, \cite{bel81}).

\acknowledgments
This work was supported by NSF grants PHY 0071044 and PHY 0503366.
JF thanks members of the PSW (periodic standing wave) consortium 
for relevant discussions at a meeting in Brownsville, March 12-13, 2005. 

\appendix

\section{Formulas for integration}

We present here a formalism for evaluating the integrals that arise 
in the equations of motion, and in the expressions for momentum, energy, 
and angular momentum of point particles in circular orbits.
This is closely patterned on formulas used by Schild in \cite{sc7576} 
in the case of two electrically charged particles. 

\subsection{Change of integration variables}
\label{sec:app1}

We first derive a relation for rewriting integrals with respect to 
proper time $\tau$ and $\btau$ as integrals with respect 
to the parameter $\eta$.  From Eq.~(\ref{eq:etadef}), 
\beq
\frac{d\tau}{d\eta} = - \frac{1}{\gamma\Omega}
\ \ \ \mbox{and} \ \ \
\frac{d\btau}{d\eta} = \frac{1}{\bgamma\Omega}.
\label{eq:dtdeta}
\eeq
The integral of a function $F(\eta)$ then becomes 
\beq
\intinf d\tau F(\eta) = \frac1{\gamma\Omega}\intinf d\eta\, F(\eta)
\ \ \ \mbox{and} \ \ \
\intinf d\btau F(\eta) = \frac1{\bgamma\Omega}\intinf d\eta\, F(\eta).
\label{eq:inteta}
\eeq
For a double integral of the kind that appears in the formulas for the linear momentum 
and energy, Eq.~(\ref{eq:linmom}), and the angular momentum, 
Eq.~(\ref{eq:angmom}), we may pick $\tau =\btau = 0$.  Then  
\beqn
\int_{0}^{\infty} d\tau \int_{-\infty}^{0} d\btau F(\eta)
&=&\frac1{\gamma\bgamma\Omega^2}
\int_{0}^{\infty} d\phi \int_{-\infty}^{0} d\bphi\, F(\eta)
=\frac1{\gamma\bgamma\Omega^2}
\int_{0}^{\infty} d\phi \int_{-\infty}^{-\phi} d\eta\, F(\eta)
\nonumber\\
&=&\frac1{\gamma\bgamma\Omega^2}
\int_{-\infty}^{0} d\eta\int_{0}^{-\eta} d\phi\, F(\eta)
=\frac1{\gamma\bgamma\Omega^2}
\int_{-\infty}^{0} d\eta\,(-\eta) F(\eta), 
\eeqn
and 
\beqn
-\int_{-\infty}^{0} d\tau \int_{0}^{\infty} d\btau F(\eta)
&=&-\frac1{\gamma\bgamma\Omega^2}
\int_{-\infty}^{0} d\phi \int_{0}^{\infty} d\bphi\, F(\eta)
=-\frac1{\gamma\bgamma\Omega^2}
\int_{-\infty}^{0} d\phi \int_{-\phi}^{\infty} d\eta\, F(\eta)
\nonumber\\
&=&-\frac1{\gamma\bgamma\Omega^2}
\int_{0}^{\infty} d\eta\int_{-\eta}^{0} d\phi\, F(\eta)
=-\frac1{\gamma\bgamma\Omega^2}
\int_{0}^{\infty} d\eta\,\eta F(\eta); 
\eeqn
adding the last two equalities, we have  
\beq
\left(\int_{0}^{\infty} d\tau \int_{-\infty}^{0} d\btau
-\int_{-\infty}^{0} d\tau \int_{0}^{\infty} d\btau\right) F(\eta) 
= - \frac1{\gamma\bgamma\Omega^2}\intinf d\eta\, \eta F(\eta).
\label{eq:intint}
\eeq

Our next task is to evaluate integrals of the form 
\dis\int_{-\infty}^\infty d\tau \dot {\bar X}{}^\alpha$ and 
\dis\int_{-\infty}^\infty d\bar\tau \dot X^\alpha$, 
where $X^\alpha$ and $\bar X^\alpha$ are the vectors associated with 
$m$ and $\bar m$, respectively.   In terms of components along an 
orthonormal frame, the vectors have the form
\beq
X^\alpha = X^t(\eta)\,t^\alpha + X^\varpi(\eta)\,\varpi^\alpha
+ X^{\widehat\phi}(\eta)\,\widehat\phi^\alpha + X^z(\eta)\,z^\alpha,
\eeq
\beq
\bX^\alpha = \bX^t(\eta)\,t^\alpha + \bX^\varpi(\eta)\,\bpi^\alpha
+ \bX^{\widehat\phi}(\eta)\,\widehat\phi^\alpha + \bX^z(\eta)\,z^\alpha. 
\eeq
The corresponding expressions for the derivatives are  
\beq
\dtX^\alpha = \frac{dX^\alpha}{d\tau}
=-\gamma\Omega\left(
\frac{dX^t}{d\eta}\,t^\alpha
+ \frac{dX^\varpi}{d\eta}\,\varpi^\alpha
+ \frac{dX^{\widehat\phi}}{d\eta}\,\widehat\phi^\alpha
+ \frac{dX^z}{d\eta}\,z^\alpha
+ X^{\widehat\phi}\,\varpi^\alpha
- X^\varpi\,\widehat\phi^\alpha
\right),
\eeq
\beq
\dtbX^\alpha = \frac{d\bX^\alpha}{d\btau}
=\bgamma\Omega\left(
\frac{d\bX^t}{d\eta}\,t^\alpha
+ \frac{d\bX^\varpi}{d\eta}\,\bpi^\alpha
+ \frac{d\bX^{\widehat\phi}}{d\eta}\,\hbphi^\alpha
+ \frac{d\bX^z}{d\eta}\,z^\alpha
- \bX^{\widehat\phi}\,\bpi^\alpha
+ \bX^\varpi\,\hbphi^\alpha
\right), 
\eeq
where Eq.~(\ref{eq:dtdeta}) is used.  
Then the integral of 
$\dtX^\alpha$ with respect to $\btau$, with fixed $\tau$ (and hence fixed $\phi$), is 
then given by
\beqn
\intinf d\btau \dtX^\alpha = \intinf d\btau \frac{dX^\alpha}{d\tau}
&=&-\frac{\gamma}{\bgamma}\intinf d\eta
\left[
\frac{dX^t}{d\eta}\,t^\alpha
+ \frac{dX^\varpi}{d\eta}\,\varpi^\alpha
+ \frac{dX^{\widehat\phi}}{d\eta}\,\widehat\phi^\alpha
+ \frac{dX^z}{d\eta}\,z^\alpha
+ X^{\widehat\phi}\,\varpi^\alpha
- X^\varpi\,\widehat\phi^\alpha
\right]_{\tau=\mbox{const}}
\nonumber\\
&=&
-\frac{\gamma}{\bgamma}\left\{
\big[\,X^\alpha\,\big]_{-\infty}^\infty
+ \intinf d\eta\,
\left(X^{\widehat\phi}\,\varpi^\alpha - X^\varpi\,\widehat\phi^\alpha\right)
\right\}_{\tau=\mbox{const}}
\nonumber\\
&=&
\frac{\gamma}{\bgamma}\left(
- \intinf X^{\widehat\phi} \,d\eta\,\,\varpi^\alpha
+ \intinf X^\varpi \,d\eta\,\,\widehat\phi^\alpha
\right). 
\label{eq:intvec1}
\eeqn
Similarly, the integral of 
$\dtbX^\alpha$ with respect to $\tau$, with fixed $\btau$ (and hence 
fixed $\bphi$) is 
\beqn
\intinf d\tau \dtbX^\alpha = \intinf d\tau \frac{d\bX^\alpha}{d\btau}
&=&\frac{\bgamma}{\gamma}\intinf d\eta
\left[
\frac{d\bX^t}{d\eta}\,t^\alpha
+ \frac{d\bX^\varpi}{d\eta}\,\bpi^\alpha
+ \frac{d\bX^{\widehat\phi}}{d\eta}\,\hbphi^\alpha
+ \frac{d\bX^z}{d\eta}\,z^\alpha
- \bX^{\widehat\phi}\,\bpi^\alpha
+ \bX^\varpi\,\hbphi^\alpha
\right]_{\btau=\mbox{const}}
\nonumber\\
&=&
\frac{\bgamma}{\gamma}\left\{
\big[\,\bX^\alpha\,\big]_{-\infty}^\infty
+ \intinf d\eta\,
\left(-\bX^{\widehat\phi}\,\bpi^\alpha - \bX^\varpi\,\hbphi^\alpha\right)
\right\}_{\btau=\mbox{const}}
\nonumber\\
&=&
\frac{\bgamma}{\gamma}\left(
- \intinf \bX^{\widehat\phi} \,d\eta\,\,\bpi^\alpha
+ \intinf \bX^\varpi \,d\eta\,\,\hbphi^\alpha
\right). 
\label{eq:intvec2}
\eeqn
Note that basis vectors $\varpi^\alpha$ and $\hphi^\alpha$ are functions 
of $\phi$; and $\bpi^\alpha$ and $\hbphi^\alpha$ are functions of $\bphi$.

\subsection{Integral formulas for half-retarded + half-advanced 
Green function}
\label{sec:app2}

Integrals involving the half-retarded + half-advanced Green function 
$\dl(w)$ are derived in this section.  
As mentioned in Sec.~\ref{sec:orbit}, the solutions to $w(\eta)=0$ 
are $\eta=\pi\pm\varphi$.  It is assumed that any contribution 
from $\eta = \pm\infty$ vanishes.   
{}From Eq.~(\ref{eq:w}), we have 
\beq
\left(\frac{dw}{d\eta}\right)_{\eta=\pi\pm\varphi}^{-1}
= \mp \frac{\Omega^2}{2}\frac1{\varphi +v\bv\sin\varphi},
\label{eq:dwdeta}
\eeq
and hence 
\beq
\intinf d\eta\, \dl(w)F(\eta)
=\sum_{\eta=\pi\pm\varphi}\left|\frac{dw}{d\eta}\right|^{-1} F(\eta) 
= \frac{\Omega^2}{2}\frac1{\varphi +v\bv\sin\varphi}
\left\{F(\pi+\varphi) + F(\pi-\varphi) \right\}.  
\eeq

For integrals involving a derivative of $\dl(w)$, 
\beq
\intinf d\eta\,\dl'(w)\, F(\eta)
=\intinf d\eta\,\frac{d\,\dl[w(\eta)]}{d\eta}
\left(\frac{d w}{d\eta}\right)^{-1}\, F(\eta)
=-\intinf d\eta\,\dl[w(\eta)]\,\frac{d}{d\eta}
\left[\left(\frac{dw}{d\eta}\right)^{-1}F(\eta)\right].  
\eeq
Using Eq.~(\ref{eq:dwdeta}) and its derivative, 
\beq
\frac{d}{d\eta}\left(\frac{dw}{d\eta}\right)_{\eta=\pi\pm\varphi}^{-1}
= \frac{\Omega^2}2\frac{1+v\bv\cos\varphi}{(\varphi +v\bv\sin\varphi)^2}
\label{eq:ddedwdeta}
\eeq
the integral becomes 
\beq
\intinf d\eta\,\dl'(w)\, F(\eta)
=\frac{\Omega^4}4\frac1{(\varphi +v\bv\sin\varphi)^2}
\left\{\frac{dF}{d\eta}(\pi+\varphi)-\frac{dF}{d\eta}(\pi-\varphi)
- \frac{1+v\bv\cos\varphi}{\varphi +v\bv\sin\varphi}
\big[F(\pi+\varphi)+F(\pi-\varphi)\big]
\right\}.  
\eeq

Finally, the following relation is useful for computation of 
Eq.~(\ref{eq:angsol1}) in Sec. \ref{sec:cirint}:
\beq
\intinf d\eta\,\dl'(w)\, w F(\eta)
= - \intinf d\eta\, \dl(w)\, F(\eta)
  - \intinf d\eta\, \dl(w)\, w 
\frac{d}{d\eta}\left[\frac{d\eta}{dw}F(\eta)\right]
= - \intinf d\eta\, \dl(w)\, F(\eta), 
\eeq
where, in the second term of the first equality, a derivative 
\dis \frac{d}{d\eta}\left[\frac{d\eta}{dw}F(\eta)\right]$
is assumed to be nonsingular (has an order larger than 
$O(w^{-1})$ ) in the neighborhood of $w=0$.  

%%%%%%%%%%%%%%%%%%%%%%%%%%%%%%%%%%%%%%%%%%%%%%%%%%%%%%%%%%%%%%%%%%%%%%%%%%%%%%%
\section{Post-Newtonian Corrections to E and L}
\label{sec:appn}
In presenting relations used to compute corrections from $\Lambda_{\rm PN}$
to $E$, $L$, and the equations of motion, we avoid repetition by omitting  
barred$\leftrightarrow$unbarred versions.  
The corrections to the parametrization-invariant post-Minkowski action 
are calculated, assuming the parametrization (\ref{eq:norm}).
We begin with the simpler correction term, (\ref{lambdapn}). Using 
the relations from Sect.~\ref{sec:orbit}, with $R$ in the form    
$R^{2}=\frac{1}{\Omega ^{2}}\left(v^{2}+\bar{v}^{2}-2v\bar{v}\cos\eta\right)$, 
we obtain,
\begin{eqnarray}
\Lambda _{\rm PN} 
 &=&-\delta \left( t-\bar{t}\right) \frac{m\bar{m}(m+\bar{m})}{2R^{2}}
    ( -{\dot{x}}_{\alpha }{\dot{x}}^{\alpha })^{\frac{1}{2}%
}( -\dtbx_{\beta}\dtbx^{\beta})^{\frac{1}{2}} 
\nonumber\\
&=&-\delta \left( \eta -\pi \right) \frac{\Omega ^{3}}{2}\frac{m\bar{m}\left(
m+\bar{m}\right) }{v^{2}+\bar{v}^{2}-2v\bar{v}\cos \eta }.
\end{eqnarray}
Derivatives appearing in the equations of motion and the conserved 
quantities have the form
\beqn
\frac{\partial \Lambda _{\rm PN}}{\partial \left( t-\bar{t}\right) }
&=& \delta'(\eta -\pi)\frac{\Omega ^{4}}{2}
    \frac{m\bar{m}\left(m+\bar{m}\right)}
         {v^{2}+\bar{v}^{2}-2v\bar{v}\cos\eta}, 
\\
\frac{\partial\Lambda _{\rm PN}}
	{\partial\left(x^{a}-{\bar x}^{a}\right)}
&=& \delta (\eta -\pi) \Omega^{4}\frac{m\bar{m}(m+\bar{m})}
			{(v^{2}+\bar{v}^{2}-2v\bar{v}\cos \eta )^{2}}
 \left\{(v-\bar{v}\cos\eta) \varpi_{a}-\bar{v}\sin\eta\hphi_a
 \right\},
\\
\frac{\partial \Lambda _{\rm PN}}{\partial\dot x^\beta}\hphi^{\beta}
&=&\delta (\eta-\pi) \frac{\Omega ^{3}}{2}\frac{m\bar{m}\left(
m+\bar{m}\right) \gamma v}{v^{2}+\bar{v}^{2}-2v\bar{v}\cos\eta},
\\
\frac{\partial \Lambda _{\rm PN}}{\partial \dot{x}^{0}}
&=& \frac{\partial \Lambda_{\rm PN}}{\partial \dot{x}^{\alpha }}t^{\alpha}
= -\delta (\eta-\pi) \frac{\Omega^{3}}{2}\frac{m\bar{m}
    (m+\bar{m}) \gamma }{v^{2}+\bar{v}^{2}-2v\bar{v}\cos\eta}.
\eeqn
The index $a$ of $x^a$ is spatial.  
The post-Newtonian correction term to the $\varpi$-component of the 
equation of motion is then
\be 
\int_{-\infty }^{\infty }d\btau \left[ \frac{\partial \Lambda _{\rm PN}}
{\partial R^{\alpha }}-\frac{d}{d\tau}\frac{\partial \Lambda
_{\rm PN}}{\partial \dot{x}^{\alpha }}\right] \varpi ^{\alpha }=m\bar{m}\left( m+%
\bar{m}\right) \frac{\Omega ^{3}}{\bar{\gamma}\left( v+\bar{v}\right) ^{3}}%
\left[ 1+\frac{1}{2}\gamma ^{2}v(v+\bar{v})\right]. 
\ee
The correction to the angular momentum is 
\begin{eqnarray*}
\ell_{\rm PN} &=&-\frac{1}{\gamma \bar{\gamma}\Omega ^{3}}
\int_{-\infty }^{\infty}d\eta \left( 
v\frac{\pa \Lambda _{\rm PN}}{\pa R^a}\varpi^a
-\bar{v}\frac{\pa \Lambda _{\rm PN}}{\pa R^a}\bpi^a\right) 
-\frac{v}{\gamma \bar{\gamma}\Omega ^{3}}
\int_{-\infty}^{\infty }d\eta \,\eta
\frac{\partial \Lambda _{\rm PN}}{\partial R^{a}}\hphi^a
\nonumber\\
&=&
\frac{m\bmm(m+\bmm)\Omega}{\gamma \bar{\gamma}}
\int_{-\infty }^{\infty }d\eta\, \delta(\eta-\pi)\left[
- \frac{1}{v^2+\bv^2-2v\bv\cos\eta}
+ \frac{v\bv\sin\eta}{\left(v^2+\bv^2-2v\bv\cos\eta\right)^2}\right]
\nonumber\\
&=&-\frac{m\bmm(m+\bmm)\Omega}{\gamma\bgamma(v+\bv)^2}, 
\end{eqnarray*}
and the corresponding correction to the energy is
\begin{eqnarray*}
e_{\rm PN} -\Omega\, \ell_{\rm PN}
&=& 
-\frac{1}{\bgamma \Omega}\intinf d\eta 
\frac{\pa\Lambda_{\rm PN}}{\pa \dtx^\alpha}k^\alpha
-\frac{1}{\gamma \Omega }\intinf d\eta 
\frac{\pa\Lambda_{\rm PN}}{\pa \dtbx^\alpha}\bk^\alpha
+\frac1{\gamma\bgamma\Omega^2}\intinf d\eta\,
\eta\frac{\pa\Lambda_{\rm PN}}{\pa R^\alpha}k^\alpha
\nonumber\\
&=&
\frac{m\bmm(m+\bmm)\Omega^2}{\gamma\bgamma}
\intinf d\eta \left\{\dl(\eta-\pi)\left[
\frac1{v^2+\bv^2-2v\bv\cos\eta}
-\frac{v\bv\,\eta\sin\eta}{\left(v^2+\bv^2-2v\bv\cos\eta\right)^2}
\right]\right.
\nonumber\\
&&\hspace{3.5cm} \left.
+ \frac{d\delta (\eta-\pi)}{d\eta}\frac12
\frac{\eta}{v^2+\bv^2-2v\bv\cos\eta}\right\}
\\
&=&\frac{m\bmm(m+\bmm)\Omega^2}{2\gamma\bgamma\left(v+\bv\right)^{2}}
\,=\,-\frac12\Omega\, \ell_{\rm PN}.
\end{eqnarray*}
Finally, 
\beq
e_{\rm PN}=\frac12\Omega\, \ell_{\rm PN}
=-\frac{m\bmm(m+\bmm)\Omega^2}{2\gamma\bgamma(v+\bv)^2}.
\eeq

The analogous corrections from the special-relativistically covariant
post-Newtonian correction are as follows.\\
Correction to the equation of motion:
\begin{eqnarray} 
&& \hspace{-1.5cm}
\intinf d\bar\tau \left[ 
\frac{\pa \Lambda _{\rm SPN}}{\pa R^\alpha}
-\frac{d}{d\tau}\frac{\pa}{\pa\dtx^\alpha}\right]\varpi^\alpha 
= \frac{1}{\bgamma\Omega}\intinf d\eta
  \frac{\pa\Lambda_{\rm SPN}}{\pa R^\alpha}\varpi^\alpha
 +\frac{\gamma}{\bgamma}\intinf d\eta 
  \frac{\pa \Lambda_{\rm SPN}}{\pa \dtx^\alpha}\hphi^\alpha
\nonumber \\
&=&-\frac{m\bmm(m+\bmm) \Omega }{\bar{\gamma}\left( \gamma 
\bar{\gamma}\right) ^{1/2}}\int_{-\infty }^{\infty }d\eta 
\left\{ \frac1{\Omega^2}\delta^{\prime }(w) 
	\sigma\frac{v-\bar{v}\cos\eta}{\eta -\pi -v\bar{v}\sin \eta}\right.
\nonumber\\
&&\hspace{4cm}
\left.+\delta (w)
\sigma \left[ \frac{\frac{1}{2}\bar{v}\sin \eta }{(\eta -\pi -v\bar{v}\sin
\eta )^{2}}-\frac{\frac34\gamma ^{2}v}{\eta -\pi -v\bar{v}\sin \eta }%
\right] \right\} \nonumber \\
&=&\frac{m\bar{m}(m+\bar m) \Omega ^{3}}{\bgamma \left( \gamma 
\bar{\gamma}\right) ^{1/2}}\frac{1}{\left( \varphi +v\bar{v}\sin \varphi
\right) ^{2}}\left\{ \frac34\gamma ^{2}v+\frac{\bar{v}\sin \varphi }{%
\varphi +v\bar{v}\sin \varphi }+\frac{\left( 1+v\bar{v}\cos \varphi \right)
\left( v+\bar{v}\cos \varphi \right) }{\left( \varphi +v\bar{v}\sin \varphi
\right) ^{2}}\right\} , 
\end{eqnarray}
where $\sigma=\pm 1$ for $\eta-\pi-v\bv\sin\eta \gtrless 0$, respectively.  

Correction to the angular momentum:
\begin{eqnarray}
\ell_{\rm SPN} 
&=&
-\frac1{\gamma \bgamma\Omega^3}\intinf d\eta
\left( v\frac{\pa\Lambda_{\rm SPN}}{\pa R^\alpha}\varpi^\alpha
-\bv\frac{\pa \Lambda_{\rm SPN}}{\pa R^\alpha}\bpi^\alpha\right) 
-\frac1{\gamma\bgamma\Omega^2}\intinf d\eta \,\eta
\left( 
\frac{\pa \Lambda_{\rm SPN}}{\pa R^\alpha}\phi^\alpha
- \frac{\pa\Lambda_{\rm SPN}}{\pa \dtx^\alpha}\dot{\phi}^\alpha 
\right)
\nonumber\\
&=&\frac{m\bar{m}(m+\bar m) }{\left( \gamma \bar{\gamma}\right)
^{3/2}\Omega }\int_{-\infty }^{\infty }d\eta \left( \frac{1}{\Omega ^{2}}%
\delta ^{\prime }(w) \sigma \frac{v^{2}+\bar{v}^{2}-2v\bar{%
v}\cos \eta -v\bar{v}\eta \sin \eta }{\eta -\pi -v\bar{v}\sin \eta }+\delta
(w) \sigma \frac{\frac{1}{2}v\bar{v}\left( \sin \eta -\eta
\cos \eta \right) }{\left( \eta -\pi -v\bar{v}\sin \eta \right) ^{2}}\right) 
\nonumber\\
&=&-\frac{m\bmm(m+\bmm)\Omega}{\left(\gamma\bgamma\right)^{3/2}}
\frac{1}{\left( \varphi + v\bv\sin \varphi \right)^2}.
\end{eqnarray}
Correction to the energy:
\begin{eqnarray}
e_{\rm SPN}-\Omega\, \ell_{\rm SPN} 
&=& 
-\frac{1}{\bgamma \Omega}\intinf d\eta 
\frac{\pa\Lambda_{\rm SPN}}{\pa \dtx^\alpha}k^\alpha
-\frac{1}{\gamma \Omega }\intinf d\eta 
\frac{\pa\Lambda_{\rm SPN}}{\pa \dtbx^\alpha}\bk^\alpha
+\frac1{\gamma\bgamma\Omega^2}\intinf d\eta\,\eta\left(
\frac{\pa\Lambda_{\rm SPN}}{\pa R^\alpha}k^\alpha
+\frac{\pa \Lambda _{\rm SPN}}{\pa \dtx^{\alpha}}\Omega\dot{\phi}^\alpha
\right)
\nonumber\\
&=&-\frac{m\bar{m}(m+\bar m) }{\left( \gamma \bar{%
\gamma}\right) ^{3/2}}\int_{-\infty }^{\infty }d\eta \left[ \frac{1}{\Omega
^{2}}\delta ^{\prime }(w) \sigma \eta +\delta \left(
w \right) \sigma \frac{\eta \left( 1-v\bar{v}\cos \eta \right) }{%
2\left( \eta -\pi -v\bar{v}\sin \eta \right) ^{2}}
-\delta(w)\sigma\frac1{\eta -\pi -v\bar{v}\sin \eta }
\right]  
\nonumber\\
&=&\frac12\frac{m\bar{m}(m+\bar m) \Omega ^{2}}{\left(
\gamma \bar{\gamma}\right) ^{3/2}}\frac{1}{\left( \varphi +v\bar{v}\sin
\varphi \right)^{2}} 
=-\frac12 \Omega \ell_{\rm SPN}.
\end{eqnarray}
Finally,
\be
e_{\rm SPN}=\frac12 \Omega \ell_{\rm SPN}
=-\frac12\frac{m\bar{m}(m+\bar m)
\Omega ^{2}}{\left( \gamma \bar{\gamma}\right) ^{3/2}}\frac{1}{\left(
\varphi +v\bar{v}\sin \varphi \right) ^{2}}.
\ee
%%\newpage
%%%%%%%%%%%%%%%%%%%%%%%%%%%%%%%%%%%%%%%%%%%%%%%%%%%%%%%%%%%%%%%%%%%%%%%%%%%%%

\section{Scalar Interaction}
\label{sec:scalar} 
A scalar field is described by a Fokker action of the form (\ref{eq:action}),
with 
\be
\Lambda =q{\bar q}\delta(w)\left( -\dot{x}_{\alpha }\dtx^\alpha \right)^{1/2}
\left( -\dtbx_\alpha\dtbx^\alpha \right)^{1/2}.
\ee
The equation of motion for two particles in circular orbit has, for 
$m$, the form,
\begin{eqnarray*}
\frac{d}{d\tau}\frac{m\dot{x}_{\alpha }}{\left( -\dot{x}_{\alpha }\dot{x}%
^{\alpha }\right) ^{1/2}}\varpi ^{\alpha } &=&-m\gamma ^{2}v\Omega
=\int_{-\infty }^{\infty }d\eta \left\{ \frac{\partial \Lambda }{\partial
w}\frac{2}{\bar{\gamma}\Omega ^{2}}\left( v-\bar{v}\cos \eta \right) +%
\frac{\gamma }{\bar{\gamma}}\frac{\partial \Lambda }{\partial \dot{x}%
^{\alpha }}\hphi ^\alpha\right\}  \\
&=&\frac{2q{\bar q}}{\bar{\gamma}}\int_{-\infty }^{\infty }d\eta \left\{ 
\frac{1}{\Omega ^{2}}\delta ^{\prime }(w) \left( v-\bar{v}%
\cos \eta \right) -\delta(w)\frac{1}{2}\gamma
^{2}v\right\}  \\
&=&\frac{q{\bar q}\Omega ^{2}}{\bar{\gamma}^{2}}\frac{1}{\varphi +v\bar{v}%
\sin \varphi }\left\{ \gamma ^{2}v+\frac{\bar{v}\sin \varphi }{\varphi +v%
\bar{v}\sin \varphi }+\frac{\left( 1+v\bar{v}\cos \varphi \right) \left( v+%
\bar{v}\cos \varphi \right) }{\varphi +v\bar{v}\sin \varphi }\right\}, 
\end{eqnarray*}
with the corresponding barred$\leftrightarrow$unbarred equation for 
$\bar m$.\footnote{An equivalent solution has been obtained independently by 
Jean-Philippe Bruneton and Gilles Esposito-Far\`ese\cite{be05}.}

The formalism developed earlier quickly leads to expressions for 
angular momentum and energy: 
\beqn
L &=&-\int_{-\infty }^{\infty }d\eta \frac{\partial \Lambda }{\partial w}
	\frac{2}{\gamma \bar{\gamma}\Omega ^{4}}\left( v^{2}+\bar{v}^{2}-2v%
\bar{v}\cos \eta \right) -\frac{2}{\gamma \bar{\gamma}\Omega ^{2}}%
\int_{-\infty }^{\infty }d\eta \eta \left[ \frac{\partial \Lambda }{\partial w}
\left( x_{2}\bx_{1}-x_{1}\bx_{2}\right) -\frac{1}{2}\left( \dot{x}_{2}%
\frac{\partial \Lambda }{\partial \dot{x}^{1}}-\dot{x}_{1}\frac{\partial
\Lambda }{\partial \dot{x}^{2}}\right) \right]  
\nonumber\\
&=&-\frac{2q{\bar q}}{\gamma \bar{\gamma}\Omega ^{2}}\int_{-\infty }^{\infty
}d\eta \frac{1}{\Omega ^{2}}\delta ^{\prime }(w) \left(
v^{2}+\bar v^{2}-2v\bar{v}\cos \eta -v\bar{v}\eta\sin \eta \right)  
\nonumber\\
&=&\frac{q{\bar q}}{\gamma \bar{\gamma}}\frac{1}{\varphi +v\bar{v}\sin\varphi}.
\eeqn

\beqn
E-\Omega L 
&=&
-\left[ m\dtx_\alpha k^{\alpha } +\frac1{\Omega \bgamma}\intinf d\eta 
\frac{\pa \Lambda}{\pa \dtx^\alpha} k^\alpha \right]
-\left[ \bmm\dtbx_\alpha \bk^\alpha +\frac1{\Omega\gamma}
\intinf d\eta \frac{\pa \Lambda}{\pa \dtbx^\alpha}\bk^{\alpha } \right]
%%%\nonumber\\
%%%&&
+\frac{2}{\gamma \bgamma}\Omega^2\intinf d\eta\,
\eta \frac{\pa \Lambda}{\pa w}R_{\alpha } k^\alpha 
\nonumber\\
&=& m \gamma \left( 1-v^{2}\right) - \frac{q{\bar q}}{\Omega \bar{\gamma}}%
\int_{-\infty }^{\infty }d\eta \delta(w)\gamma \left(
1-v^{2}\right) +\bar{m}\bar{\gamma}\left( 1-\bar{v}^{2}\right) 
 -\frac{q{\bar q}}{\Omega \gamma }\int_{-\infty}^{\infty}d\eta \delta(w)
	\bar{\gamma}\left( 1-\bar{v}^{2}\right)  
\nonumber\\
&&+\frac{2q{\bar q}}{\gamma \bar{\gamma}\Omega ^{3}}\int_{-\infty }^{\infty
}d\eta \delta ^{\prime }(w) \eta \left( \eta -\pi -v\bar{v}%
\sin \eta \right) 
\nonumber\\
&=&\frac{m}{\gamma }+\frac{\bar{m}}{\bar{\gamma}}
   -\frac{2q{\bar q}}{\gamma\bar{\gamma}\Omega}
   \int_{-\infty }^{\infty }d\eta\delta(w) 
   -\frac{q{\bar q}}{\gamma \bar{\gamma}\Omega }
   \int_{-\infty }^{\infty}d\eta \delta'(w) \eta \frac{dw}{d\eta} 
\nonumber\\
&=&\frac{m}{\gamma }+\frac{\bar{m}}{\bar{\gamma}}-\frac{q{\bar q}}{\gamma 
\bar{\gamma}\Omega }\int_{-\infty }^{\infty }d\eta \delta
(w)  
\nonumber\\
&=&\frac{m}{\gamma }+\frac{\bar{m}}{\bar{\gamma}}-\frac{q{\bar q}\Omega }{%
\gamma \bar{\gamma}}\frac1{\varphi +v\bar{v}\sin \varphi } 
\nonumber\\
&=&\frac{m}{\gamma }+\frac{\bar{m}}{\bar{\gamma}}-\Omega L.
\eeqn
Then
\beqn
E&=&\frac{m}{\gamma}+\frac{\bar m}{\bar\gamma},\\
L&=&\frac{q\bar{q}}{\gamma \bar\gamma}\frac1{\varphi+v\bar v\sin\varphi}.
\eeqn

%%\newpage

%%%%%%%%%%%%%%%%%%%%%%%%%%%%%%%%%%%%%%%%%%%%%%%%%%%%%%%%%%%%%%%%%%%%%%%%%%%%%

\section{Derivations of the momentum and angular momentum formulas}
\label{sec:formulaPL} 

In this appendix, we present calculations and useful expressions relating 
to the momentum and the angular momentum formulas, for an arbitrary 
interaction term $\Lambda(x-\bx,\dtx,\dtbx)$.  

\subsection{Derivations of Eqs.~(\ref{eq:linmom}) and (\ref{eq:angmom})}

Dettman and Schild \cite{ds54} derived the momentum and the angular 
momentum formulas from Poincar\'e invariance of the general Fokker action.
We follow their calculation for the case of our  
action for two point particles to derive Eqs.~(\ref{eq:linmom}) 
and (\ref{eq:angmom}), from the varied action (\ref{eq:dlaction}).  

We first rewrite Eq.~(\ref{eq:dlaction}) for $\dl I$,   
substituting the equation of motion for each particle, Eqs.~(\ref{eq:eom1})
and (\ref{eq:eom2}):
\bea
\dl I(\tau_1, \tau_2, \btau_{1},\btau_{2})
&=&\left[\frac{m \dtx_{\alpha}}{(-\dtx_{\gamma}\dtx^{\gamma})^{\frac{1}{2}}}
+\intinf d\btau
\frac{\pa\Lambda}{\pa \dtx^{\alpha}}\right]\dl x^{\alpha}
\left.\phantom{\frac12}\!\!\!\!\!\right|_{\tau_{1}}^{\tau_{2}} 
%%%{\Big |}_{\tau_{1}}^{\tau_{2}} 
%
+\int_{\tau_{1}}^{\tau_{2}}d\tau \frac{d}{d\tau}\left[\dl x^{\alpha}
\left(\int_{\btau_{1}}^{\btau_{2}}- \intinf\right)d\btau 
\frac{\pa\Lambda}{\pa \dtx^{\alpha}}\right]
\nonumber\\
&+&\int_{\tau_1}^{\tau_2}d\tau\,\dl x^{\alpha}
\left(\int_{\btau_{1}}^{\btau_{2}}-\intinf\right)d\btau
\left[\frac{\pa\Lambda}{\pa R^{\alpha}}
-\frac{d}{d\tau}\frac{\pa\Lambda}{\pa\dtx^{\alpha}}\right]
\nonumber\\
&+&\left[\frac{\bar{m}\dtbx_{\alpha}}
{(-\dtbx_{\gamma}\dtbx^{\gamma})^{\frac{1}{2}}}
+\intinf d\tau\frac{\pa\Lambda}{\pa\dtbx^{\alpha}}\right]
\dl\bx^{\alpha}
\left.\phantom{\frac12}\!\!\!\!\!\right|_{\btau_{1}}^{\btau_{2}} 
%%%{\Big |}_{\btau_{1}}^{\btau_{2}}
%
+\int_{\btau_{1}}^{\btau_{2}}d\btau \frac{d}{d\btau}\left[\dl\bx^{\alpha}
\left(\int_{\tau_1}^{\tau_2}-\intinf\right)d\tau
\frac{\pa\Lambda}{\pa\dtbx^{\alpha}}\right]
\nonumber\\
&+&\int_{\btau_{1}}^{\btau_{2}}d\btau\,\dl \bx^{\alpha}
\left(\int_{\tau_1}^{\tau_2}-\intinf\right)d\tau
\left[\frac{\pa\Lambda}{\pa\bR^{\alpha}}
-\frac{d}{d\btau}\frac{\pa\Lambda}{\pa\dtbx^{\alpha}}\right]
\nonumber\\ 
\nonumber\\
&=&\left[\frac{m \dtx_{\alpha}}{(-\dtx_{\gamma}\dtx^{\gamma})^{\frac{1}{2}}}
+\intinf d\btau
\frac{\pa\Lambda}{\pa \dtx^{\alpha}}\right]\dl x^{\alpha}
\left.\phantom{\frac12}\!\!\!\!\!\right|_{\tau_{1}}^{\tau_{2}} 
%%%{\Big |}_{\tau_{1}}^{\tau_{2}}
\nonumber\\
&+&\int_{\tau_1}^{\tau_2}d\tau
\left(\int_{\btau_{1}}^{\btau_{2}}-\intinf\right)d\btau
\left[\,\dl x^{\alpha}\frac{\pa\Lambda}{\pa R^{\alpha}}
+\dl \dtx^\alpha\frac{\pa\Lambda}{\pa \dtx^{\alpha}}
\right]
\nonumber\\
&+&\left[\frac{\bar{m}\dtbx_{\alpha}}
{(-\dtbx_{\gamma}\dtbx^{\gamma})^{\frac{1}{2}}}
+\intinf d\tau\frac{\pa\Lambda}{\pa\dtbx^{\alpha}}\right]
\dl\bx^{\alpha}
\left.\phantom{\frac12}\!\!\!\!\!\right|_{\btau_{1}}^{\btau_{2}} 
%%%{\Big |}_{\btau_{1}}^{\btau_{2}}
\nonumber\\
&+&\int_{\btau_{1}}^{\btau_{2}}d\btau
\left(\int_{\tau_1}^{\tau_2}-\intinf\right)d\tau
\left[\,\dl \bx^{\alpha}\frac{\pa\Lambda}{\pa\bR^{\alpha}}
+\dl \dtbx^\alpha\frac{\pa\Lambda}{\pa\dtbx^{\alpha}}
\right],   
\label{eq:dlI}
\eea
where 
\be
\dl \dtx^\alpha := \frac{d\,\dl x^{\alpha}}{d\tau}, \qquad
\dl \dtbx^\alpha := \frac{d\,\dl\bx^{\alpha}}{d\btau}.
\ee

We next restrict the variations of the trajectory of each particle, 
$\dl x^\alpha$ and $\dl\bx^\alpha$, to infinitesimal Poincar\'e  
transformations (which we will subsequently take to have the forms 
(\ref{eq:infsp}) and (\ref{eq:infrot})).   
Since the Fokker action, and hence the interaction $\Lambda$, 
is Poincar\'e invariant, the interaction term $\Lambda$ satisfies 
the identity 
\beq
\Lambda(x+\dl x - \bx - \dl\bx,\dtx+\dl\dtx,\dtbx+\dl\dtbx)
=\Lambda(x-\bx,\dtx,\dtbx). 
\label{eq:Lorentzinv}
\eeq
Expanding the identity to first order in $\delta x$ and $\delta\bar x$, we have 
\beq
\dl x^\alpha\frac{\pa\Lambda}{\pa R^\alpha}
+\dl \bx^\alpha\frac{\pa\Lambda}{\pa \bR^\alpha}
+\dl \dtx^\alpha\frac{\pa\Lambda}{\pa\dtx^\alpha}
+\dl \dtbx^\alpha\frac{\pa\Lambda}{\pa\dtbx^\alpha}=0.  
\label{eq:Lorentzrel}
\eeq
Using this relation, the variation of the action (\ref{eq:dlI}) becomes 
\bea
\dl I(\tau_1, \tau_2, \btau_{1},\btau_{2})
&=&\left[\frac{m \dtx_{\alpha}}{(-\dtx_{\gamma}\dtx^{\gamma})^{\frac{1}{2}}}
+\intinf d\btau
\frac{\pa\Lambda}{\pa \dtx^{\alpha}}\right]\dl x^{\alpha}
\left.\phantom{\frac12}\!\!\!\!\!\right|_{\tau_{1}}^{\tau_{2}} 
%%%{\Big |}_{\tau_{1}}^{\tau_{2}}
%
+\left[\frac{\bar{m}\dtbx_{\alpha}}
{(-\dtbx_{\gamma}\dtbx^{\gamma})^{\frac{1}{2}}}
+\intinf d\tau\frac{\pa\Lambda}{\pa\dtbx^{\alpha}}\right]
\dl\bx^{\alpha}
\left.\phantom{\frac12}\!\!\!\!\!\right|_{\btau_{1}}^{\btau_{2}} 
%%%{\Big |}_{\btau_{1}}^{\btau_{2}}
\nonumber\\
&&+\left(\intinf d\tau\int_{\btau_{1}}^{\btau_{2}}d\btau - 
\int_{\tau_1}^{\tau_2}d\tau\intinf d\btau\right)
\left[\,\dl x^{\alpha}\frac{\pa\Lambda}{\pa R^{\alpha}}
+\dl \dtx^\alpha\frac{\pa\Lambda}{\pa \dtx^{\alpha}}
\right].
\label{eq:dlI2}
\eea
The double integral in Eq.~(\ref{eq:dlI2}) 
is rearranged to separate the contribution from $(\tau_1,\btau_1)$ and 
$(\tau_2,\btau_2)$ explicitly as follows, 
\beqn
\intinf \int_{\btau_{1}}^{\btau_{2}}
- \int_{\tau_1}^{\tau_2}\intinf 
&=&
\int_{-\infty}^{\tau_{1}}
\left(\int_{\btau_{1}}^{\infty}-\int_{\btau_{2}}^{\infty}\right)
+ \int_{\tau_{1}}^{\infty} 
\left(\int_{-\infty}^{\btau_{2}}-\int_{-\infty}^{\btau_{1}}\right)
\nonumber\\
&-& \left(\int_{\tau_1}^{\infty} - \int_{\tau_2}^{\infty} \right)
\int_{-\infty}^{\btau_{2}}
- \left(\int_{-\infty}^{\tau_2}-\int_{-\infty}^{\tau_1} \right)
\int_{\btau_{2}}^{\infty} 
\nonumber\\
&=&
\left(\int_{\tau_2}^{\infty} \int_{-\infty}^{\btau_{2}}
-\int_{-\infty}^{\tau_2}\int_{\btau_{2}}^{\infty}  \right)
- \left(\int_{\tau_{1}}^{\infty} \int_{-\infty}^{\btau_{1}}
-\int_{-\infty}^{\tau_{1}}\int_{\btau_{1}}^{\infty}\right).
\eeqn
Finally, the variation $\dl I$ with respect to an infinitesimal 
Poincar\'e transformation takes the form 
\bea
\dl I(\tau_1, \tau_2, \btau_{1},\btau_{2})
&=&\left[\frac{m \dtx_{\alpha}}{(-\dtx_{\gamma}\dtx^{\gamma})^{\frac{1}{2}}}
+\intinf d\btau
\frac{\pa\Lambda}{\pa \dtx^{\alpha}}\right]\dl x^{\alpha}
\left.\phantom{\frac12}\!\!\!\!\!\right|_{\tau_{1}}^{\tau_{2}} 
%%%{\Big |}_{\tau_{1}}^{\tau_{2}}
%
+\left[\frac{\bar{m}\dtbx_{\alpha}}
{(-\dtbx_{\gamma}\dtbx^{\gamma})^{\frac{1}{2}}}
+\intinf d\tau\frac{\pa\Lambda}{\pa\dtbx^{\alpha}}\right]
\dl\bx^{\alpha}
\left.\phantom{\frac12}\!\!\!\!\!\right|_{\btau_{1}}^{\btau_{2}} 
%%%{\Big |}_{\btau_{1}}^{\btau_{2}}
\nonumber\\
&&+\left(\int_{\tau}^{\infty} \int_{-\infty}^{\btau}
-\int_{-\infty}^{\tau}\int_{\btau}^{\infty}  \right)
\left[\,\dl x^{\alpha}\frac{\pa\Lambda}{\pa R^{\alpha}}
+\dl \dtx^\alpha\frac{\pa\Lambda}{\pa \dtx^{\alpha}}
\right]d\tau d\btau
\left.\phantom{\frac12}\!\!\!\!\!\right|_{\btau_{1}}^{\btau_{2}} .
%%%{\Big |}_{\btau_{1}}^{\btau_{2}}.
\label{eq:deltaI}
\eea

To derive the conserved momentum, Eq. (\ref{eq:linmom}), a translation 
by a constant vector (\ref{eq:infsp}) is substituted 
in (\ref{eq:deltaI}):
\bea
\frac{\dl I}{\dl a^\alpha}
&=&
\left[\frac{m \dtx_{\alpha}}
{(-\dtx_{\gamma}\dtx^{\gamma})^{\frac{1}{2}}}
+\intinf d\btau\frac{\pa\Lambda}{\pa \dtx^{\alpha}}
\right]_{\tau_{1}}^{\tau_{2}} 
+\left[\frac{\bar{m}\dtbx_{\alpha}}
{(-\dtbx_{\gamma}\dtbx^{\gamma})^{\frac{1}{2}}}
+\intinf d\tau\frac{\pa\Lambda}{\pa\dtbx^{\alpha}}
\right]_{\btau_{1}}^{\btau_{2}} 
\nonumber\\
&&
+\left(\int_{\tau}^{\infty} \int_{-\infty}^{\btau}
-\int_{-\infty}^{\tau}\int_{\btau}^{\infty}  \right)
\frac{\pa\Lambda}{\pa R^{\alpha}}d\tau d\btau
\left.\phantom{\frac12}\!\!\!\!\!\right|_{\btau_{1}}^{\btau_{2}} ,
%%%{\Big |}_{\btau_{1}}^{\btau_{2}}, 
\label{eq:dlIdla}
\eea
where we used $\dl\dtx^\alpha = da^\alpha/d\tau = 0$.

The angular momentum similarly corresponds to an infinitesimal rotation 
(\ref{eq:infrot}): 
\bea
2\frac{\dl I}{\dl \epsilon^{\beta\alpha}}
&=&\left[\frac{m \left(x_\alpha\dtx_{\beta}-x_\beta\dtx_{\alpha}\right)}
{(-\dtx_{\gamma}\dtx^{\gamma})^{\frac{1}{2}}}
+\intinf d\btau
\left(x_\alpha\frac{\pa\Lambda}{\pa \dtx^{\beta}}
-x_\beta\frac{\pa\Lambda}{\pa \dtx^{\alpha}}\right)
\right]_{\tau_{1}}^{\tau_{2}} 
\nonumber\\
&+&\left[\frac{\bar{m}\left(\bx_\alpha\dtbx_{\beta}-\bx_\beta\dtbx_{\alpha}\right)}
{(-\dtbx_{\gamma}\dtbx^{\gamma})^{\frac{1}{2}}}
+\intinf d\tau
\left(\bx_\alpha\frac{\pa\Lambda}{\pa\dtbx^{\beta}}
-\bx_\beta\frac{\pa\Lambda}{\pa\dtbx^{\alpha}}\right)
\right]_{\btau_{1}}^{\btau_{2}} 
\nonumber\\
&+&\left(\int_{\tau}^{\infty} \int_{-\infty}^{\btau}
-\int_{-\infty}^{\tau}\int_{\btau}^{\infty}  \right)
\left[\,x_{\alpha}\frac{\pa\Lambda}{\pa R^{\beta}}
-x_{\beta}\frac{\pa\Lambda}{\pa R^{\alpha}}
+\dtx_\alpha\frac{\pa\Lambda}{\pa \dtx^{\beta}}
-\dtx_\beta\frac{\pa\Lambda}{\pa \dtx^{\alpha}}
\right]d\tau d\btau
\left.\phantom{\frac12}\!\!\!\!\!\right|_{\btau_{1}}^{\btau_{2}}, 
%%%{\Big |}_{\btau_{1}}^{\btau_{2}}, 
\label{eq:dlIdleps}
\eea
where $\dl \dtx^\alpha = \epsilon^{\albe}\dtx_\beta$ is used.  

An analogous derivation of the momentum and angular momentum for 
the affinely parametrized action results in the replacements 
\beq
\frac{m\dtx_{\alpha}}
{(-\dtx_{\gamma}\dtx^{\gamma})^{\frac{1}{2}}}
\rightarrow m\dtx_{\alpha},\qquad
\frac{\bar{m}\dtbx_{\alpha}}
{(-\dtbx_{\gamma}\dtbx^{\gamma})^{\frac{1}{2}}}
\rightarrow \bar{m}\dtbx_{\alpha}, 
\label{eq:rplmom}
\eeq
in the momentum equation (\ref{eq:dlIdla}), and 
\beq
\frac{m \left(x_\alpha\dtx_{\beta}-x_\beta\dtx_{\alpha}\right)}
{(-\dtx_{\gamma}\dtx^{\gamma})^{\frac{1}{2}}}
\rightarrow 
m \left(x_\alpha\dtx_{\beta}-x_\beta\dtx_{\alpha}\right),
\qquad
\frac{\bar{m}\left(\bx_\alpha\dtbx_{\beta}-\bx_\beta\dtbx_{\alpha}\right)}
{(-\dtbx_{\gamma}\dtbx^{\gamma})^{\frac{1}{2}}}
\rightarrow 
\bar{m}\left(\bx_\alpha\dtbx_{\beta}-\bx_\beta\dtbx_{\alpha}\right), 
\label{eq:rplangmom}
\eeq
in the angular momentum equation (\ref{eq:dlIdleps}).  

The double integrals in Eqs.(\ref{eq:deltaI})-(\ref{eq:dlIdleps}) have a 
useful alternative form:
\bea
\left(\int_{\tau}^{\infty}\int_{-\infty}^{\btau} 
- \int_{-\infty}^{\tau}\int_{\btau}^{\infty}\right)d\tau d\btau
&=&
\left(\intinf -\int_{-\infty}^{\tau}\right)\int_{-\infty}^{\btau}
d\tau d\btau
-\int_{-\infty}^{\tau}\left(\intinf-\int_{-\infty}^{\btau}\right)
d\tau d\btau
\nonumber \\
&=&
-\int_{-\infty}^{\tau}d\tau\intinf d\btau
+ \int_{-\infty}^{\btau}d\btau\intinf d\tau.  
\label{eq:dblint}
\eea
The terms depending on the proper time of 
each path $\tau$ and $\btau$ can be separated by writing the 
double integral in the above form, (\ref{eq:dblint}), as seen in 
Eqs.~(\ref{eq:energyapxlaw}) and (\ref{eq:angmomapxlaw}).  
These are used in the proof of the first law in 
Sec.~\ref{sec:1stlaw}.

\subsection{Formulas for $E$, $L$, and $E-\Omega L$}
\label{sec:eqsEL}

Next, we derive expressions for the 
nonzero components of the angular momentum $L := L_{12}(\tau,\btau)$, 
the 4-momentum $E := -P_\alpha(\tau,\btau) t^\alpha$,
and a combination $E-\Omega L$ for parametrization 
invariant action.  The formulas for the affinely parametrized 
action can be obtained by replacements analogous to 
Eqs.~(\ref{eq:rplmom}) and (\ref{eq:rplangmom}), 
which are not repeated.  

We begin with the angular momentum: Since the basis $\phi^\alpha$ and 
$\bphi^\alpha$ at positions of particles $\{m,x\}$ and $\{\bmm,\bx\}$ 
have components in Cartesian coordinate 
$\phi^\alpha = (-x^2,x^1,0) = (-x_2,x_1,0)$ and 
$\bphi^\alpha = (-\bx^2,\bx^1,0) = (-\bx_2,\bx_1,0)$, one has relations 
\beq
x_1 A_2 - x_2 A_1 = A_\alpha\phi^\alpha \ \mbox{ and } \ 
\bx_1 \bA_2 - \bx_2 \bA_1 = \bA_\alpha\bphi^\alpha,   
\label{eq:dualrelation}
\eeq
where 
$A_\alpha$ and $\bA_\alpha$ are vectors at positions $m$ and $\bmm$.  
Applying the relations (\ref{eq:dualrelation}) to the $L_{12}$ component of 
the angular momentum  equation (\ref{eq:angmom}) yields   
\beqn
L
&=&\left[
\frac{m \dtx_\alpha\phi^\alpha}{(-\dtx_\beta\dtx^\beta)^{\frac12}}
+\intinf d\btau \frac{\pa\Lambda}{\pa \dtx^\alpha}\phi^\alpha
\right](\tau) 
\,+\,\left[
\frac{\bmm \dtbx_\alpha\bphi^\alpha}{(-\dtbx_\beta\dtbx^\beta)^{\frac12}}
+\intinf d\tau \frac{\pa\Lambda}{\pa\dtbx^\alpha}\bphi^\alpha
\right](\btau)
\nonumber\\
&&+ \left(\int_{\tau}^{\infty}\int_{-\infty}^{\btau}
-\int_{-\infty}^{\tau}\int_{\btau}^{\infty}\right)
\left(\frac{\pa\Lambda}{\pa R^\alpha}\phi^\alpha
+\frac{\pa\Lambda}{\pa \dtx^{\alpha}}\dtphi^\alpha
\right)d\tau d\btau,   
\label{eq:angmomapx}
\eeqn
where we define $\dtphi^\alpha = \frac{d}{d\tau}\phi^\alpha$. 
From the definition of the linear momentum 
(\ref{eq:linmom}), the energy $E =- P_\alpha t^\alpha$ 
has the corresponding form
\beqn
E
&=&-\left[\frac{m\dtx_{\alpha}t^\alpha}
{(-\dtx_{\gamma}\dtx^{\gamma})^{\frac{1}{2}}}
+\intinf d\btau
\frac{\pa\Lambda}{\pa \dtx^{\alpha}}t^\alpha\right](\tau)
-\left[\frac{\bar{m}\dtbx_{\alpha}t^\alpha}
{(-\dtbx_{\gamma}\dtbx^{\gamma})^{\frac{1}{2}}}
+ \intinf d\tau
\frac{\pa\Lambda}{\pa\dtbx^{\alpha}}t^\alpha\right](\btau)
\nonumber \\
&&-\left(\int_{\tau}^{\infty}\int_{-\infty}^{\btau}
-\int_{-\infty}^{\tau}\int_{\btau}^{\infty}\right)
\frac{\pa\Lambda}{\pa R^{\alpha}}t^\alpha d\tau d\btau. 
\label{eq:energyapx}
\eeqn
The combination $E-\Omega L$ then becomes 
\beqn
E - \Omega L
&=&-\left[
\frac{m \dtx_\alpha k^\alpha}{(-\dtx_\beta\dtx^\beta)^{\frac12}}
+\intinf d\btau \frac{\pa\Lambda}{\pa \dtx^\alpha} k^\alpha
\right](\tau) 
\,-\,\left[
\frac{\bmm \dtbx_\alpha\bk^\alpha}{(-\dtbx_\beta\dtbx^\beta)^{\frac12}}
+\intinf d\tau \frac{\pa\Lambda}{\pa\dtbx^\alpha}\bk^\alpha
\right](\btau)
\nonumber\\
&&- \left(\int_{\tau}^{\infty}\int_{-\infty}^{\btau}
-\int_{-\infty}^{\tau}\int_{\btau}^{\infty}\right)
\left(\frac{\pa\Lambda}{\pa R^\alpha} k^\alpha
+\frac{\pa\Lambda}{\pa \dtx^{\alpha}}\Omega\dtphi^\alpha
\right)d\tau d\btau .
\label{eq:eomelapx}
\eeqn

One can write the energy (\ref{eq:energyapx}) and 
the angular momentum (\ref{eq:angmomapx}) in a form 
related to the one particle energy and angular momentum 
defined in Sec.~\ref{sec:1stlaw}, Eqs.~(\ref{eq:onepmom}), 
and (\ref{eq:onepenergy1})--(\ref{eq:onepangmom}), 
together with the relation for the double 
integral (\ref{eq:dblint}).  
Interestingly, the resulting formulas for the energy and 
angular momentum depend {\em separately} on the 
proper time of each path, $\tau$ and $\btau$.  
Using the property 
\beq
\frac{\pa \Lambda}{\pa R^\alpha}
= \frac{\pa \Lambda}{\pa x^\alpha}
= -\frac{\pa \Lambda}{\pa \bx^\alpha}, 
\label{eq:Lampro}
\eeq
we can rewrite the energy (\ref{eq:energyapx}) in terms of 
the one-particle potential $\cUm$ and $\bcUm$, 
\beqn
E
&=&
-\left[\frac{m\dtx_{\alpha}t^\alpha}
{(-\dtx_{\gamma}\dtx^{\gamma})^{\frac{1}{2}}}
+t^\alpha\frac{\pa}{\pa \dtx^{\alpha}}\intinf d\btau\Lambda
\right](\tau)
-\left[\frac{\bar{m}\dtbx_{\alpha}t^\alpha}
{(-\dtbx_{\gamma}\dtbx^{\gamma})^{\frac{1}{2}}}
+ t^\alpha\frac{\pa}{\pa\dtbx^{\alpha}}\intinf d\tau\Lambda
\right](\btau)
\nonumber \\
&&
+\left(
\int_{-\infty}^{\tau}d\tau\intinf d\btau
- \int_{-\infty}^{\btau}d\btau\intinf d\tau
\right)
\frac{\pa\Lambda}{\pa R^{\alpha}}t^\alpha  
\nonumber \\
&=&
-\left[\frac{m\dtx_{\alpha}t^\alpha}
{(-\dtx_{\gamma}\dtx^{\gamma})^{\frac{1}{2}}}
+t^\alpha\frac{\pa\cUm}{\pa\dtx^\alpha}
\right](\tau)
+\int_{-\infty}^{\tau}d\tau\, t^\alpha \frac{\pa\cUm}{\pa x^\alpha}
\nonumber \\
&&
-\left[\frac{\bar{m}\dtbx_{\alpha}t^\alpha}
{(-\dtbx_{\gamma}\dtbx^{\gamma})^{\frac{1}{2}}}
+ t^\alpha\frac{\pa\bcUm}{\pa\dtbx^\alpha}
\right](\btau)
+\int_{-\infty}^{\btau}d\btau\, t^\alpha \frac{\pa\bcUm}{\pa\bx^\alpha}.  
\label{eq:energyapxlaw}
\eeqn

To relate the angular momentum (\ref{eq:angmomapx}) 
to one-particle angular momentum, we notice that 
the relation (\ref{eq:Lorentzrel}), together with properties 
(\ref{eq:dualrelation}) and (\ref{eq:Lampro}), implies
\beq
\phi^\alpha\frac{\pa\Lambda}{\pa x^\alpha}
+\bphi^\alpha\frac{\pa\Lambda}{\pa \bx^\alpha}
+\dtphi^\alpha\frac{\pa\Lambda}{\pa\dtx^\alpha}
+\dtbphi{}^\alpha\frac{\pa\Lambda}{\pa\dtbx^\alpha}=0.  
\eeq
Like the energy, the angular momentum can be written in terms of the  
potentials $\cUm$ and $\bcUm$, 
\beqn
L
&=&\left[
\frac{m \dtx_\alpha\phi^\alpha}{(-\dtx_\beta\dtx^\beta)^{\frac12}}
+\phi^\alpha\frac{\pa}{\pa \dtx^\alpha} \intinf d\btau \Lambda
\right](\tau) 
\,+\,\left[
\frac{\bmm \dtbx_\alpha\bphi^\alpha}{(-\dtbx_\beta\dtbx^\beta)^{\frac12}}
+\bphi^\alpha\frac{\pa}{\pa\dtbx^\alpha}\intinf d\tau \Lambda
\right](\btau)
\nonumber\\
&&- 
\left(\int_{-\infty}^{\tau}d\tau\intinf d\btau
- \int_{-\infty}^{\btau}d\btau\intinf d\tau\right)
\left(\frac{\pa\Lambda}{\pa R^\alpha}\phi^\alpha
+\frac{\pa\Lambda}{\pa \dtx^{\alpha}}\dtphi^\alpha
\right)
\nonumber\\
&=&\left[
\frac{m \dtx_\alpha\phi^\alpha}{(-\dtx_\beta\dtx^\beta)^{\frac12}}
+\phi^\alpha\frac{\pa\cUm}{\pa \dtx^\alpha} 
\right](\tau) 
- \int_{-\infty}^{\tau}d\tau
\left(\phi^\alpha\frac{\pa\cUm}{\pa x^\alpha}
+\dtphi^\alpha\frac{\pa\cUm}{\pa \dtx^{\alpha}}\right)
\nonumber\\
&+&\left[
\frac{\bmm \dtbx_\alpha\bphi^\alpha}{(-\dtbx_\beta\dtbx^\beta)^{\frac12}}
+\bphi^\alpha\frac{\pa\bcUm}{\pa\dtbx^\alpha}
\right](\btau)
- \int_{-\infty}^{\btau}d\btau
\left(\bphi^\alpha\frac{\pa\bcUm}{\pa \bx^\alpha}
+\dtbphi^\alpha\frac{\pa\bcUm}{\pa \dtbx^{\alpha}}\right).
\label{eq:angmomapxlaw}
\eeqn
The terms in the brackets in the above Eqs.~(\ref{eq:energyapxlaw}) 
and (\ref{eq:angmomapxlaw}) are the one-particle energy and 
angular momentum; the contribution from each particle which is moving 
in the field produced by the other particle (see Sec.~\ref{sec:1stlaw}).

It is also noticeable that the formulas for $E$, $L$ and $E-\Omega L$, 
Eqs.~(\ref{eq:energyapx}), (\ref{eq:angmomapx}), and (\ref{eq:eomelapx}), 
can be written in a common form, because of a property 
$\dot{t}^\alpha:=dt^\alpha/d\tau = 0$, 
$\dot{k}^\alpha=\Omega\dtphi^\alpha$ accordingly, and 
those for corresponding barred quantities.  
Writing $\cal Q$ to represent these conserved quantities, 
$E$, $L$, and $E-\Omega L$, 
and $\zeta^\alpha$ for the associated vectors, $t^\alpha$, 
$\phi^\alpha$, and $k^\alpha$, we have 
\newcommand{\dtzeta}{\dot{\zeta}}
\newcommand{\bzeta}{\bar{\zeta}}
\newcommand{\dtbzeta}{\dot{\bar{\zeta}}{}}
\bea
\cal Q
&=&\left[
\frac{m \dtx_\alpha\zeta^\alpha}{(-\dtx_\beta\dtx^\beta)^{\frac12}}
+\intinf d\btau \frac{\pa\Lambda}{\pa \dtx^\alpha}\zeta^\alpha
\right](\tau) 
\,+\,\left[
\frac{\bmm \dtbx_\alpha\bzeta^\alpha}{(-\dtbx_\beta\dtbx^\beta)^{\frac12}}
+\intinf d\tau \frac{\pa\Lambda}{\pa\dtbx^\alpha}\bzeta^\alpha
\right](\btau)
\nonumber\\
&&+ \left(\int_{\tau}^{\infty}\int_{-\infty}^{\btau}
-\int_{-\infty}^{\tau}\int_{\btau}^{\infty}\right)
\left(\frac{\pa\Lambda}{\pa R^\alpha}\zeta^\alpha
+\frac{\pa\Lambda}{\pa \dtx^{\alpha}}\dtzeta^\alpha
\right)d\tau d\btau.   
\label{eq:qapx}
\eea
Also for the expressions in terms of the one-particle potentials, 
we have
\beqn
\cal Q&=&\left[
\frac{m \dtx_\alpha\zeta^\alpha}{(-\dtx_\beta\dtx^\beta)^{\frac12}}
+\zeta^\alpha\frac{\pa\cUm}{\pa \dtx^\alpha} 
\right](\tau) 
- \int_{-\infty}^{\tau}d\tau
\left(\zeta^\alpha\frac{\pa\cUm}{\pa x^\alpha}
+\dtzeta^\alpha\frac{\pa\cUm}{\pa \dtx^{\alpha}}\right)
\nonumber\\
&+&\left[
\frac{\bmm \dtbx_\alpha\bzeta^\alpha}{(-\dtbx_\beta\dtbx^\beta)^{\frac12}}
+\bzeta^\alpha\frac{\pa\bcUm}{\pa\dtbx^\alpha}
\right](\btau)
- \int_{-\infty}^{\btau}d\btau
\left(\bzeta^\alpha\frac{\pa\bcUm}{\pa \bx^\alpha}
+\dtbzeta^\alpha\frac{\pa\bcUm}{\pa \dtbx^{\alpha}}\right).
\label{eq:qapxlaw}
\eeqn
In Eqs.~(\ref{eq:em})--(\ref{eq:law1}) in Sec.~\ref{sec:1stlaw}, 
the integrand of Eq.~(\ref{eq:qapxlaw}) is written in a short form 
defined by 
\beq
\zeta^\alpha\na_\alpha\cUm:=\zeta^\alpha\frac{\pa\cUm}{\pa x^\alpha}
+\dtzeta^\alpha\frac{\pa\cUm}{\pa \dtx^{\alpha}}, \qquad 
\bzeta^\alpha\bar{\na}_\alpha\bcUm:=\bzeta^\alpha\frac{\pa\bcUm}{\pa \bx^\alpha}
+\dtbzeta^\alpha\frac{\pa\bcUm}{\pa \dtbx^{\alpha}}.  
\label{eq:gradu}\eeq

As seen in the above expression (\ref{eq:qapxlaw}), 
contributions to the total energy and angular momentum 
from time $ \tau $ and $ \bar{\tau} $ are separated.
It means that each piece conserves independently; 
$E$ $L$ and $E -\Omega L$, are a sum of two conserved quantities
associated with $\tau$ and $\bar{\tau}$.
\\

For circular motion, the velocities are given by  
$\dtx^\alpha = \gamma k^\alpha$ and $\dtbx^\alpha = \gamma \bk^\alpha$, 
and the parametrization-invariant interaction term satisfies  
Eq.~(\ref{eq:homofn}) (Euler's relation for a homogeneous function of 
degree one in $\dtx^\alpha$ and $\dtbx^\alpha$); we then have
\beqn
E - \Omega L
&=&-\left[
\frac{m \dtx_\alpha k^\alpha}{(-\dtx_\beta\dtx^\beta)^{\frac12}}
+\frac1{\gamma}\intinf d\btau \Lambda\right](\tau) 
\,-\,\left[
\frac{\bmm \dtbx_\alpha\bk^\alpha}{(-\dtbx_\beta\dtbx^\beta)^{\frac12}}
+\frac1{\bgamma}\intinf d\tau \Lambda\right](\btau)
\nonumber\\
&&- \left(\int_{\tau}^{\infty}\int_{-\infty}^{\btau}
-\int_{-\infty}^{\tau}\int_{\btau}^{\infty}\right)
\left(\frac{\pa\Lambda}{\pa R^\alpha} k^\alpha
+\frac{\pa\Lambda}{\pa \dtx^{\alpha}}\Omega\dtphi^\alpha
\right)d\tau d\btau.    
\label{eq:eomelapx2}
\eeqn

Only the radial components of the equations of motion, (\ref{eq:eom1}) 
and (\ref{eq:eom2}), are nontrivial, and they take the form 
\be
\frac{d}{d\tau}\left[\frac{m\dtx_{\alpha}}
{(-\dtx_{\gamma}\dtx^{\gamma})^{\frac{1}{2}}}\right]\varpi^\alpha
=\intinf d\btau \frac{\pa\Lambda}{\pa R^{\alpha}}\varpi^\alpha
+\frac{\gamma\Omega^2}{v}\intinf d\btau
\frac{\pa\Lambda}{\pa \dtx^{\alpha}}\phi^\alpha
\label{eq:eom1apx}
\ee
\be
\frac{d}{d\btau}\left[\frac{\bmm\dtbx_{\alpha}}
{(-\dtbx_{\gamma}\dtbx^{\gamma})^{\frac{1}{2}}}\right]\bpi^\alpha
=\intinf d\tau\frac{\pa\Lambda}{\pa\bR^{\alpha}}\bpi^\alpha
+\frac{\bgamma\Omega^2}{\bv}
\intinf d\tau\frac{\pa\Lambda}{\pa\dtbx^{\alpha}}\bphi^\alpha.
\label{eq:eom2apx}
\ee
In the above we used Eqs.~(\ref{eq:intvec1}) and (\ref{eq:intvec2})
in Appendix \ref{sec:app1}, to eliminate the $\tau$ derivative in 
the $\btau$ integral (and respectively $\btau$ derivative 
in the $\tau$ integral), then transformed the integration variable $\eta$ back to 
$\tau$ $(\btau)$ using (\ref{eq:inteta}).

The radial component of the accelerations 
$\ddtx_{\alpha}\varpi^\alpha=-\gamma^2 v \Omega$ 
and $\ddtbx_{\alpha}\bpi^\alpha=-\bgamma^2 \bv \Omega$ 
are related to $\dtx_\alpha\phi^\alpha=\gamma v^2/\Omega$ and 
$\dtbx_\alpha\bphi^\alpha=\bgamma\bv^2/\Omega$, respectively, by
\beq
\frac{m\dtx_{\alpha}\phi^\alpha}
{(-\dtx_{\gamma}\dtx^{\gamma})^{\frac{1}{2}}}
=-\frac{v}{\gamma\Omega^2}
\frac{d}{d\tau}\left[\frac{m\dtx_{\alpha}}
{(-\dtx_{\gamma}\dtx^{\gamma})^{\frac{1}{2}}}\right]\varpi^\alpha
\ \mbox{ and }\ 
\frac{\bmm\dtbx_{\alpha}\bphi^\alpha}
{(-\dtbx_{\gamma}\dtbx^{\gamma})^{\frac{1}{2}}}
=-\frac{\bv}{\bgamma\Omega^2}
\frac{d}{d\btau}\left[\frac{\bmm\dtbx_{\alpha}}
{(-\dtbx_{\gamma}\dtbx^{\gamma})^{\frac{1}{2}}}\right]\bpi^\alpha, 
\label{eq:accelapx}
\eeq
where $\dtx_\alpha\ddtx^\alpha = 0 = \dtbx_\alpha\ddtbx^\alpha $ are used.  
With this relation (\ref{eq:accelapx}), the radial equations of motion 
(\ref{eq:eom1apx}) and (\ref{eq:eom2apx}) are substituted to further simplify 
the angular momentum formula (\ref{eq:angmomapx}):
\beqn
L
&=&-\frac{v}{\gamma\Omega^2}
\left[\intinf d\btau \frac{\pa\Lambda}{\pa R^{\alpha}}\varpi^\alpha
\right](\tau) 
\,-\,\frac{\bv}{\bgamma\Omega^2}
\left[\intinf d\tau \frac{\pa\Lambda}{\pa \bR^{\alpha}}\bpi^\alpha
\right](\btau)
\nonumber\\
&&+ \left(\int_{\tau}^{\infty}\int_{-\infty}^{\btau}
-\int_{-\infty}^{\tau}\int_{\btau}^{\infty}\right)
\left(\frac{\pa\Lambda}{\pa R^\alpha}\phi^\alpha
+\frac{\pa\Lambda}{\pa \dtx^{\alpha}}\dtphi^\alpha
\right)d\tau d\btau .   
\label{eq:angmomapx2}
\eeqn

\subsection{Conserved quantities in Post-Minkowski gravity 
in terms of $h_{\albe}$ and $\bhh_{\albe}$}
\label{sec:eqshij} 

As shown in Sec.~\ref{sec:PMinv}, the parametrization invariant 
interaction (\ref{eq:intgeo}) leads to 
equations of motion, Eqs.~(\ref{eq:hsol}) and (\ref{eq:bhsol}),
that describe point particles in a post-Minkowski approximation 
with half-advanced + half-retarded fields. 
 
The momentum and angular momentum formulas, 
in the form of a sum of one-particle contributions, 
can be written in terms of $h_{\albe}$ and $\bhh_{\albe}$.  
We present these formulas valid for arbitrary particle trajectories  
$m$ and $\bmm$.

We begin by rewriting the variation of $I$ with respect to 
an infinitesimal Poincar\'e transformation, (\ref{eq:dlI2}), 
in terms of one-particle potentials $\cUm$ and $\bcUm$,
\bea
\dl I(\tau_1, \tau_2, \btau_{1},\btau_{2})
&=&\left[\frac{m \dtx_{\alpha}}{(-\dtx_{\gamma}\dtx^{\gamma})^{\frac{1}{2}}}
+\intinf d\btau
\frac{\pa\Lambda}{\pa \dtx^{\alpha}}\right]\dl x^{\alpha}
\left.\phantom{\frac12}\!\!\!\!\!\right|_{\tau_{1}}^{\tau_{2}} 
%%%{\Big |}_{\tau_{1}}^{\tau_{2}}
%
-\int_{\tau_1}^{\tau_2}d\tau\intinf d\btau 
\left[\,\dl x^{\alpha}\frac{\pa\Lambda}{\pa R^{\alpha}}
+\dl \dtx^\alpha\frac{\pa\Lambda}{\pa \dtx^{\alpha}}
\right]
\nonumber\\
&+&\left[\frac{\bar{m}\dtbx_{\alpha}}
{(-\dtbx_{\gamma}\dtbx^{\gamma})^{\frac{1}{2}}}
+\intinf d\tau\frac{\pa\Lambda}{\pa\dtbx^{\alpha}}\right]
\dl\bx^{\alpha}
\left.\phantom{\frac12}\!\!\!\!\!\right|_{\btau_{1}}^{\btau_{2}} 
%%%{\Big |}_{\btau_{1}}^{\btau_{2}}
%
-\int_{\btau_1}^{\btau_2}d\btau \intinf d\tau
\left[\,\dl \bx^{\alpha}\frac{\pa\Lambda}{\pa \bR^{\alpha}}
+\dl \dtbx^\alpha\frac{\pa\Lambda}{\pa \dtbx^{\alpha}}
\right]
\\
&=&\left[\frac{m \dtx_{\alpha}}{(-\dtx_{\gamma}\dtx^{\gamma})^{\frac{1}{2}}}
+ \frac{\pa\cUm}{\pa \dtx^{\alpha}}\right]\dl x^{\alpha}
\left.\phantom{\frac12}\!\!\!\!\!\right|_{\tau_{1}}^{\tau_{2}} 
%%%{\Big |}_{\tau_{1}}^{\tau_{2}}
%
-\int_{\tau_1}^{\tau_2}d\tau
\left[\,\dl x^{\alpha}\frac{\pa\cUm}{\pa x^{\alpha}}
+\dl \dtx^\alpha\frac{\pa\cUm}{\pa \dtx^{\alpha}}
\right]
\nonumber\\
&+&\left[\frac{\bar{m}\dtbx_{\alpha}}
{(-\dtbx_{\gamma}\dtbx^{\gamma})^{\frac{1}{2}}}
+\frac{\pa\bcUm}{\pa\dtbx^{\alpha}}\right]
\dl\bx^{\alpha}
\left.\phantom{\frac12}\!\!\!\!\!\right|_{\btau_{1}}^{\btau_{2}} 
%%%{\Big |}_{\btau_{1}}^{\btau_{2}}
%
-\int_{\btau_1}^{\btau_2}d\btau 
\left[\,\dl \bx^{\alpha}\frac{\pa\bcUm}{\pa \bx^{\alpha}}
+\dl \dtbx^\alpha\frac{\pa\bcUm}{\pa \dtbx^{\alpha}}
\right] .
\label{eq:dlI3}
\eea
The conserved momentum and angular momentum are then derived as follows,
\bea
P_\alpha(\tau,\btau)
&=&\left[\frac{m \dtx_{\alpha}}{(-\dtx_{\gamma}\dtx^{\gamma})^{\frac{1}{2}}}
+ \frac{\pa\cUm}{\pa \dtx^{\alpha}}\right](\tau)
-\int_{-\infty}^{\tau}d\tau\frac{\pa\cUm}{\pa x^{\alpha}}
\nonumber\\
&+&\left[\frac{\bar{m}\dtbx_{\alpha}}
{(-\dtbx_{\gamma}\dtbx^{\gamma})^{\frac{1}{2}}}
+ \frac{\pa\bcUm}{\pa\dtbx^{\alpha}}\right](\btau)
-\int_{-\infty}^{\btau}d\btau 
\frac{\pa\bcUm}{\pa \bx^{\alpha}}, 
\label{eq:linmomonep}
\eea
and 
\bea
L_{\albe}(\tau,\btau)
&=&\left[\frac{m \left(x_\alpha\dtx_{\beta}-x_\beta\dtx_{\alpha}\right)}
{(-\dtx_{\gamma}\dtx^{\gamma})^{\frac{1}{2}}}
+
\left(x_\alpha\frac{\pa\cUm}{\pa \dtx^{\beta}}
-x_\beta\frac{\pa\cUm}{\pa \dtx^{\alpha}}\right)
\right](\tau)
\nonumber\\
&-&\int_{-\infty}^{\tau}d\tau 
\left[\,x_{\alpha}\frac{\pa\cUm}{\pa x^{\beta}}
-x_{\beta}\frac{\pa\cUm}{\pa x^{\alpha}}
+\dtx_\alpha\frac{\pa\cUm}{\pa \dtx^{\beta}}
-\dtx_\beta\frac{\pa\cUm}{\pa \dtx^{\alpha}}
\right]
\nonumber\\
&+&\left[\frac{\bar{m}\left(\bx_\alpha\dtbx_{\beta}-\bx_\beta\dtbx_{\alpha}\right)}
{(-\dtbx_{\gamma}\dtbx^{\gamma})^{\frac{1}{2}}}
+\left(\bx_\alpha\frac{\pa\bcUm}{\pa\dtbx^{\beta}}
-\bx_\beta\frac{\pa\bcUm}{\pa\dtbx^{\alpha}}\right)
\right](\btau)
\nonumber\\
&-&\int_{-\infty}^{\btau}d\btau 
\left[\,\bx_{\alpha}\frac{\pa\bcUm}{\pa \bx^{\beta}}
-\bx_{\beta}\frac{\pa\bcUm}{\pa \bx^{\alpha}}
+\dtbx_\alpha\frac{\pa\bcUm}{\pa \dtbx^{\beta}}
-\dtbx_\beta\frac{\pa\bcUm}{\pa \dtbx^{\alpha}}
\right].  
\label{eq:angmomonep}
\eea

For the interaction of Eq.~(\ref{eq:intgeo}) and corresponding fields 
(\ref{eq:hsol}) and (\ref{eq:bhsol}), the one-particle potentials 
are related to the fields $h_{\albe}$ and $\bhh_{\albe}$ by
\beq
\cUm = \intinf d\btau\Lambda = 
\frac12 m\, h_{\albe}
\frac{\dtx^\alpha\dtx^\beta}{(-\dtx_\gamma\dtx^\gamma)^\frac12}, 
\qquad
\bcUm = \intinf d\tau\Lambda = 
\frac12 \bmm\, \bhh_{\albe}
\frac{\dtbx^\alpha\dtbx^\beta}{(-\dtbx_\gamma\dtbx^\gamma)^\frac12}; 
\label{eq:oneph1}
\eeq
and their derivatives with respect to position and velocity 
take the form 
\bea
&&
\frac{\pa\cUm}{\pa x^\alpha} = 
\frac12 m\, \frac{\pa h_{\gade}}{\pa x^\alpha}
\frac{\dtx^\gamma\dtx^\delta}{(-\dtx_\ep\dtx^\ep)^\frac12}, 
\qquad
\frac{\pa\cUm}{\pa \dtx^\alpha} = 
 m\left[\,h_{\albe}
 + \frac12\eta_{\albe} \frac{h_{\gade}\dtx^\gamma\dtx^\delta}
 {(-\dtx_\ep\dtx^\ep)}\,\right]
  \frac{\dtx^\beta}{(-\dtx_\gamma\dtx^\gamma)^\frac12}, 
\label{eq:oneph2}
\\&&
\frac{\pa\bcUm}{\pa \bx^\alpha} = 
\frac12 \bmm\, \frac{\pa \bhh_{\gade}}{\pa \bx^\alpha}
\frac{\dtbx^\gamma\dtbx^\delta}{(-\dtbx_\ep\dtbx^\ep)^\frac12}, 
\qquad
\frac{\pa\bcUm}{\pa \dtbx^\alpha} = 
 \bmm\left[\,\bhh_{\albe}
 + \frac12\bareta_{\albe}\frac{\bhh_{\gade}\dtbx^\gamma\dtbx^\delta}
 {(-\dtbx_\ep\dtbx^\ep)}\,\right]
 \frac{\dtbx^\beta}{(-\dtbx_\gamma\dtbx^\gamma)^\frac12}. 
\label{eq:oneph3}
\eea
Substituting relations (\ref{eq:oneph2}) and (\ref{eq:oneph3}) in  
Eqs.~(\ref{eq:linmomonep}) and (\ref{eq:angmomonep}), we obtain  
explicit expressions for the momentum and angular momentum in terms of
$h_{\albe}$ and $\bhh_{\albe}$, 
\bea
P_\alpha(\tau,\btau)
&=&
m\left[\,\eta_{\albe} + h_{\albe}
+\frac12\eta_{\albe}\frac{h_{\gade}\dtx^\gamma\dtx^\delta}
{(-\dtx_{\ep}\dtx^{\ep})}\,
\right]
\frac{ \dtx^{\beta}}{(-\dtx_{\gamma}\dtx^{\gamma})^{\frac{1}{2}}}(\tau)
-\frac12 m\int_{-\infty}^{\tau}d\tau\, 
\frac{\pa h_{\gade}}{\pa x^\alpha}
\frac{\dtx^\gamma\dtx^\delta}{(-\dtx_\ep\dtx^\ep)^\frac12}
\nonumber\\
&+&
\bmm\left[\,\eta_{\albe} + \bhh_{\albe}
+\frac12\eta_{\albe}\frac{\bhh_{\gade}\dtbx^\gamma\dtbx^\delta}
{(-\dtbx_{\ep}\dtbx^{\ep})}\,
\right]
\frac{ \dtbx^{\beta}}{(-\dtbx_{\gamma}\dtbx^{\gamma})^{\frac{1}{2}}}(\btau)
-\frac12 \bmm\int_{-\infty}^{\btau}d\btau\, 
\frac{\pa \bhh_{\gade}}{\pa \bx^\alpha}
\frac{\dtbx^\gamma\dtbx^\delta}{(-\dtbx_\ep\dtbx^\ep)^\frac12}, 
\label{eq:onepalinmom}
\eea
and 
\bea
L_{\albe}(\tau,\btau)
&=&
m\left\{\frac{ \left(x_\alpha\dtx_{\beta}-x_\beta\dtx_{\alpha}\right)}
{(-\dtx_{\gamma}\dtx^{\gamma})^{\frac{1}{2}}}
\left[
1+ \frac12 \frac{h_{\gade}\dtx^\gamma\dtx^\delta}
 {(-\dtx_\ep\dtx^\ep)}
\right]
+
\frac{\left(x_\alpha h_{\bega}-x_\beta h_{\alpha\gamma}\right)\dtx^\gamma}
{(-\dtx_\delta\dtx^\delta)^\frac12}
\right\}(\tau)
\nonumber\\
&-& m \int_{-\infty}^{\tau}d\tau \left[\,
\frac12 \left(x_\alpha \frac{\pa h_{\gade}}{\pa x^\beta}
-x_\beta \frac{\pa h_{\gade}}{\pa x^\alpha}\right)
\frac{\dtx^\gamma\dtx^\delta}{(-\dtx_\ep\dtx^\ep)^\frac12}
+
\frac{\left(\dtx_\alpha h_{\bega}-\dtx_\beta h_{\alpha\gamma}\right)\dtx^\gamma}
{(-\dtx_\delta\dtx^\delta)^\frac12}
\right]
\nonumber\\
&+&
\bmm\left\{\frac{ \left(\bx_\alpha\dtbx_{\beta}-\bx_\beta\dtbx_{\alpha}\right)}
{(-\dtbx_{\gamma}\dtbx^{\gamma})^{\frac{1}{2}}}
\left[
1+ \frac12 \frac{\bhh_{\gade}\dtbx^\gamma\dtbx^\delta}
 {(-\dtbx_\ep\dtbx^\ep)}
\right]
+
\frac{\left(\bx_\alpha \bhh_{\bega}-\bx_\beta \bhh_{\alpha\gamma}\right)\dtbx^\gamma}
{(-\dtbx_\delta\dtbx^\delta)^\frac12}
\right\}(\btau)
\nonumber\\
&-& \bmm \int_{-\infty}^{\btau}d\btau \left[\,
\frac12 \left(\bx_\alpha \frac{\pa \bhh_{\gade}}{\pa \bx^\beta}
-\bx_\beta \frac{\pa \bhh_{\gade}}{\pa \bx^\alpha}\right)
\frac{\dtbx^\gamma\dtbx^\delta}{(-\dtbx_\ep\dtbx^\ep)^\frac12}
+
\frac{\left(\dtbx_\alpha \bhh_{\bega}-\dtbx_\beta \bhh_{\alpha\gamma}\right)\dtbx^\gamma}
{(-\dtbx_\delta\dtbx^\delta)^\frac12}
\right].
\label{eq:onepaangmom}
\eea

For completeness, we give the corresponding form of the conserved 
quantity $\cal Q$ associated with a Killing vector $\zeta^\alpha$ 
of Minkowski space, rewriting Eq.~(\ref{eq:qapxlaw}), 
in terms of $h_{\albe}$ and 
$\bhh_{\albe}$.  Applying a property of $\zeta^\alpha$, 
$\dtx^\alpha \dot\zeta_\alpha 
= \dtx^\alpha\dtx^\beta \na_\alpha \zeta_\beta = 0$, we have
\beqn
\cal Q&=&
m 
\left[
\eta_{\albe}+ h_{\albe}
+\frac12 \eta_{\albe}\frac{h_{\gade}\dtx^\gamma\dtx^\delta}
 {(-\dtx_\ep\dtx^\ep)}
\right]
\frac{\zeta^\alpha\dtx^\beta}{(-\dtx_\gamma\dtx^\gamma)^{\frac12}}(\tau) 
\nonumber\\
&-& m \int_{-\infty}^{\tau}d\tau \left[\,
\frac12 \zeta^\alpha \frac{\pa h_{\bega}}{\pa x^\alpha}
\frac{\dtx^\beta\dtx^\gamma}{(-\dtx_\dl\dtx^\dl)^\frac12}
+
\frac{h_{\albe}\dtzeta^\alpha \dtx^\beta}
{(-\dtx_\gamma\dtx^\gamma)^\frac12}
\right]
\nonumber\\
&+&
\bmm 
\left[
\eta_{\albe}+ \bhh_{\albe}
+ \frac12 \eta_{\albe}\frac{\bhh_{\gade}\dtbx^\gamma\dtbx^\delta}
 {(-\dtbx_\ep\dtbx^\ep)}
\right]
\frac{\bzeta^\alpha\dtbx^\beta}{(-\dtbx_\gamma\dtbx^\gamma)^{\frac12}}(\btau) 
\nonumber\\
&-& \bmm \int_{-\infty}^{\btau}d\btau \left[\,
\frac12 \bzeta^\alpha \frac{\pa \bhh_{\bega}}{\pa \bx^\alpha}
\frac{\dtbx^\beta\dtbx^\gamma}{(-\dtbx_\dl\dtbx^\dl)^\frac12}
+
\frac{\bhh_{\albe}\dtbzeta^\alpha \dtbx^\beta}
{(-\dtbx_\gamma\dtbx^\gamma)^\frac12}
\right].
\label{eq:onepaqapx}
\eeqn
\\

For the affinely parametrized interaction (\ref{eq:interac}), 
we obtain analogous expressions for momentum and angular momentum 
in terms of $h_{\albe}$ and $\bhh_{\albe}$.  Since the 
form (\ref{eq:interac}) relates to the solutions 
\bea
h_{\albe}(x) &=& 4\bmm\intinf d\btau\,\dl(w)
\left(\dtbx_\alpha\dtbx_\beta-\frac12\eta_{\albe}\dtbx_\gamma\dtbx^\gamma\right), 
\\
\bhh_{\albe}(\bx) &=& 4m\intinf d\tau\,\dl(w)
\left(\dtx_\alpha\dtx_\beta-\frac12\eta_{\albe}\dtx_\gamma\dtx^\gamma\right), 
\eea
the one-particle potentials for the affinely parametrized interaction
are written, 
\beq
\cUm = \intinf d\btau\Lambda = 
\frac12 m\, h_{\albe}\dtx^\alpha\dtx^\beta, 
\qquad
\bcUm = \intinf d\tau\Lambda = 
\frac12 \bmm\, \bhh_{\albe}\dtbx^\alpha\dtbx^\beta, 
\label{eq:oneph1aff}
\eeq
and their derivatives with respect to the position and to the velocity 
become 
\bea
&&
\frac{\pa\cUm}{\pa x^\alpha} = 
\frac12 m\, \frac{\pa h_{\gade}}{\pa x^\alpha}
\dtx^\gamma\dtx^\delta, 
\qquad
\frac{\pa\cUm}{\pa \dtx^\alpha} = 
 m\,h_{\alpha\gamma}\dtx^\gamma
\label{eq:oneph2aff}
\\&&
\frac{\pa\bcUm}{\pa \bx^\alpha} = 
\frac12 \bmm\, \frac{\pa \bhh_{\gade}}{\pa \bx^\alpha}
\dtbx^\gamma\dtbx^\delta, 
\qquad
\frac{\pa\bcUm}{\pa \dtbx^\alpha} = 
 \bmm\,\bhh_{\alpha\gamma}
 \dtbx^\gamma. 
\label{eq:oneph3aff}
\eea

Replacements (\ref{eq:rplmom}), (\ref{eq:rplangmom}) and 
substitution of (\ref{eq:oneph2aff}) and (\ref{eq:oneph3aff}) 
in Eqs.~(\ref{eq:linmomonep}) and (\ref{eq:angmomonep})
yield formulas for the momentum and angular momentum the expressions, 
\bea
P_\alpha(\tau,\btau)
&=&
m\left(\eta_{\albe} + h_{\albe}\right)\dtx^{\beta}(\tau)
-\frac12 m\int_{-\infty}^{\tau}d\tau\, 
\frac{\pa h_{\gade}}{\pa x^\alpha}
\dtx^\gamma\dtx^\delta
\nonumber\\
&+&
\bmm\left(\eta_{\albe} + \bhh_{\albe}\right)\dtbx^{\beta}(\btau)
-\frac12 \bmm\int_{-\infty}^{\btau}d\btau\, 
\frac{\pa \bhh_{\gade}}{\pa \bx^\alpha}
\dtbx^\gamma\dtbx^\delta, 
\label{eq:onepalinmomaff}
\eea
and 
\bea
L_{\albe}(\tau,\btau)
&=&
m\Big[\left(x_\alpha\dtx_{\beta}-x_\beta\dtx_{\alpha}\right)
+\left(x_\alpha h_{\bega}-x_\beta h_{\alpha\gamma}\right)\dtx^\gamma
\Big](\tau)
\nonumber\\
&-& m \int_{-\infty}^{\tau}d\tau \left[\,
\frac12 \left(x_\alpha \frac{\pa h_{\gade}}{\pa x^\beta}
-x_\beta \frac{\pa h_{\gade}}{\pa x^\alpha}\right)
\dtx^\gamma\dtx^\delta
+
\left(\dtx_\alpha h_{\bega}-\dtx_\beta h_{\alpha\gamma}\right)\dtx^\gamma
\right]
\nonumber\\
&+&
\bmm\Big[\left(\bx_\alpha\dtbx_{\beta}-\bx_\beta\dtbx_{\alpha}\right)
+\left(\bx_\alpha \bhh_{\bega}-\bx_\beta \bhh_{\alpha\gamma}\right)\dtbx^\gamma
\Big](\btau)
\nonumber\\
&-& \bmm \int_{-\infty}^{\btau}d\btau \left[\,
\frac12 \left(\bx_\alpha \frac{\pa \bhh_{\gade}}{\pa \bx^\beta}
-\bx_\beta \frac{\pa \bhh_{\gade}}{\pa \bx^\alpha}\right)
\dtbx^\gamma\dtbx^\delta
+
\left(\dtbx_\alpha \bhh_{\bega}-\dtbx_\beta \bhh_{\alpha\gamma}\right)\dtbx^\gamma
\right].
\label{eq:onepaangmomaff}
\eea

Finally, the charge $\cal Q$ has for the affinely parametrized interaction the form 
\bea
\cal Q&=&
m \left(\eta_{\albe} + h_{\albe}\right)
\zeta^\alpha\dtx^\beta(\tau) 
- m \int_{-\infty}^{\tau}d\tau \left(
\frac12 \zeta^\alpha \frac{\pa h_{\bega}}{\pa x^\alpha}
\dtx^\beta\dtx^\gamma
+ h_{\albe}\dtzeta^\alpha \dtx^\beta
\right)
\nonumber\\
&+&
\bmm \left(\eta_{\albe} + \bhh_{\albe}\right)
\bzeta^\alpha\dtbx^\beta(\btau) 
- \bmm \int_{-\infty}^{\btau}d\btau \left(
\frac12 \bzeta^\alpha \frac{\pa \bhh_{\bega}}{\pa \bx^\alpha}
\dtbx^\beta\dtbx^\gamma
+ \bhh_{\albe}\dtbzeta^\alpha \dtbx^\beta
\right).
\label{eq:onepaqaff}
\eea

%%\newpage
%%%%%%%%%%%%%%%%%%%%%%%%%%%%%%%%%%%%%%%%%%%%%%%%%%%%%%%%%%%%%%%%%%%%%%%%%%%%%


\begin{thebibliography}{99}

\bibitem{bd92}
J. K. Blackburn and S. Detweiler, Phys. Rev. D {\bf 46}, 2318 (1992).
S. Detweiler, Phys. Rev. D {\bf 50}, 4929 (1994).
\bibitem{bct98} P. R. Brady, J. D. E. Creighton, and K. S. Thorne, Phys. Rev. 
D {\bf 58}, 061501 (1998). 
\bibitem{whelan00} J.T. Whelan, W. Krivan, and R.H. Price, Class. Quant.
Grav. {\bf 17}, 4895 (2000)
\bibitem{whelan02} J.T. Whelan, C. Beetle, W. Krivan, and 
R.H. Price, Class. Quant. Grav. {\bf 19}, 1285 (2002).
\bibitem{fus02} J. L. Friedman, K. Ury\=u and M. Shibata, 
Phys. Rev. D {\bf 65}, 064035 (2002).
\bibitem{price04} R. H. Price, Class. Quant. Grav. {\bf 21}, S281 (2004)
\bibitem{andrade04} Z. Andrade et al. Phys. Rev. D {\bf 70}, 
064001 (2004).
\bibitem{torre03} C. G. Torre, J. Math. Phys. {\bf 44}, 6223 (2003).
\bibitem{bromley05} B. Bromley, R. Owen, and R.H. Price, 
Phys. Rev. D {\bf 71}, 104017 (2005).
\bibitem{klein05} C. Klein, Phys.Rev. D {\bf70} 124026 (2004).
\bibitem{sc63} A. Schild, Phys. Rev. {\bf 131}, 2762 (1963).
\bibitem{schonberg46} M. Sch\"{o}nberg, Phys. Rev. {\bf 69}, 211 (1946).
\bibitem{fo29} A. D. Fokker, Zeits. f. Physik {\bf 58}, 386 (1929).
\bibitem{wf4549} J. A. Wheeler and R. P. Feynman, Rev. Mod. Phys. 
{\bf 17}, 157, (1945); {\bf 21}, 425, (1949)
\bibitem{yo06}
S. Yoshida, B. C. Bromley, J. S. Read, K. Ury\=u, and 
J. L. Friedman, 
{\it Models of helically symmetric binary systems}, 
Class. Quant. Grav. in press (2006).
\bibitem{bbp06}
C. Beetle, B. C. Bromley, R.H. Price, 
{\it The periodic standing-wave approximation: eigenspectral 
computations for linear gravity and nonlinear toy models}, gr-qc0602027, 
Phys. Rev. D submitted (2006).
\bibitem{uryu06} K. Ury\=u et.~al., in preparation.
\bibitem{ds54} J. W. Dettman and A. Schild, Phys. Rev. {\bf 95}, 1057 (1954).
\bibitem{hg62} P. Havas and J. N. Goldberg, Phys. Rev. {\bf 128}, 398 (1962)
\bibitem{ra73} P. Ramond, Phys. Rev. D {\bf 7}, 449 (1973)
\bibitem{bel81} L. Bel, T. Damour, N. Deruelle, J. Iba\~nez,
 and J. Martin, Gen. Rel. Grav. {\bf 13}, 963 (1981)
\bibitem{sc7576} A. Schild, Ann. Phys. {\bf 93}, 88 (1975); 
{\bf 99}, 434 (1976).  
\bibitem{be05} J.P. Bruneton and G. Esposito-Far\`ese, 
%%%`The two-body problem in Nordstr\"om's scalar gravity', 
in preparation. 

\end{thebibliography}
\end{document}